\documentclass[prb,reprint,aps,superscriptaddress,longbibliography]{revtex4-2}
\usepackage{graphicx}% Include figure files
\usepackage{dcolumn}% Align table columns on decimal point
\usepackage{bm}% bold math
\usepackage{braket}
\usepackage{amsmath}
\usepackage{amssymb}
\usepackage{feynmp}
\usepackage{graphics}
\usepackage{enumerate}
\usepackage{CJKutf8}
\usepackage[dvipsnames]{xcolor}

\usepackage{comment}
\usepackage[normalem]{ulem} % sout 
\usepackage[hidelinks]{hyperref}% add hypertext capabilities

\everymath{\displaystyle}

\begin{document}

\title{Non-uniform quantum geometry stabilizes generalized Wigner crystals
% \TS{A Lindemann-type criterion for generalized Wigner crystals in Aharanav-Casher bands}
}
%Generalized Wigner crystals and fractional Chern insulators in bands with non-uniform quantum geometry

\author{Nicol\'as Morales-Dur\'an}
\email{nmoralesduran@flatironinstitute.org}
\affiliation{Center for Computational Quantum Physics, Flatiron Institute, New York, New York 10010, USA}

\author{Tobias M. R. Wolf}
\affiliation{Department of Physics, The University of Texas at Austin, Austin, Texas 78712, USA}

\author{Jingtian Shi}
\affiliation{Materials Science Division, Argonne National Laboratory, Lemont, Illinois 60439, USA}

\author{Tomohiro Soejima (\begin{CJK*}{UTF8}{bsmi}副島智大\end{CJK*})}
\affiliation{Center for Computational Quantum Physics, Flatiron Institute, New York, New York 10010, USA}
\affiliation{Center for Quantum Phenomena, Department of Physics, New York University, 726 Broadway, New York, New York 10003, USA}

\author{Andrew J. Millis}
\affiliation{Center for Computational Quantum Physics, Flatiron Institute, New York, New York 10010, USA}
\affiliation{Department of Physics, Columbia University, New York, New York 10027, USA}

\author{Jennifer Cano}
\affiliation{Center for Computational Quantum Physics, Flatiron Institute, New York, New York 10010, USA}
\affiliation{Department of Physics and Astronomy, Stony Brook University, Stony Brook, New York 11794, USA}

\date{\today}

\begin{abstract}
Moiré materials host fractional Chern insulators and electron crystals in close proximity, but the mechanism selecting between them remains an open question. We address this competition in Chern bands with ideal but momentum-dependent 
quantum geometry -- Aharonov-Casher bands. We present an ansatz wave function for generalized Wigner crystals and, by comparing its energy to that of the competing Laughlin-like state, map out the phase diagram at filling fraction $\nu=1/m$ as a function of the degree of geometric non-uniformity. Our work identifies quantum geometry-controlled zero point fluctuations of the charge density of the generalized Wigner crystal as the mechanism controlling its relative stability, implying a kind of quantum Lindemann criterion for the crystal-liquid phase boundary. 
\end{abstract}

\maketitle

{\em Introduction --}
Narrow topological bands can host distinct forms of correlated order, including fractional Chern insulator (FCI) states, which preserve translation symmetry while developing topological order~\cite{Wen_FCI,DasSarma_Sun_FCI,Neupert_FCI,Sheng_FCI,Bernevig_Regnault_FCI}, and generalized Wigner crystals \cite{Wigner1934WC}, which spontaneously break the discrete translation symmetry of the host lattice.
Remarkably, recent experiments in moiré platforms, including twisted transition metal dichalcogenide (TMD) homobilayers and rhombohedral graphene heterostructures, have reported signatures of both fractionalized \cite{cai2023signatures, zeng2023thermodynamic, park2023observation, xu2023observation,lu2024fractional} and crystalline states \cite{xu2025superconductivity,sun2026twistangle,Lu2025Extended,Aronson2025Displacement}, with the two phases appearing in close proximity in their phase diagrams.
Understanding the mechanisms that select between these competing states remains an important question.

A key ingredient in these platforms is the presence of nontrivial and non-uniform quantum geometry. While it plays a crucial role in stabilizing fractional states~\cite{roy2014band,jackson2015geometric,Parameswaran2013Fractional,ledwith2023vortexability,ledwith2020fractional,wang2021exact,cano2026ideal}, quantum geometry can also stabilize crystals~\cite{valenti2025quantum}, significantly modify their properties~\cite{Joy2026ChiralWigner,kim2026exchange}, and enable entirely new crystalline states all together~\cite{Joy2026ChiralWigner,Zverevich2026Spin, Zhou2025NewClasses,dong2024anomalous,zeng2024sublattice,crepel2025efficient,Soejima2025Lambda,Desrochers2026Electronic,Dong2024Stability,Patri2024Extended,Tan2024Parent,Tan2025Variational,valenti2025quantum,Soejima2024Anomalous,Zhou2024Fractional,kim2026exchange,Su2025Moire}. These observations necessitate a broader investigation of crystalline states beyond those supported on trivial bands~\cite{Drummond2009WC,Smith2024WC,Kim2022Interstitials,Kim2024Dynamical,Valenti2025Gate,Esterlis2025Bilayer,Esterlis2025MagnetismMoire,Zhou2021BilayerCrystal,Smolenski2021WignerCrystal,Sung2025Electronic, CornellWigner,CaliforniaWigner,Berkeley,CornellWignerStripe,STM_Wigner,ContinuousWigner} or on Landau levels with uniform quantum geometry~\cite{Yoshioka_Fukuyama,Yoshioka_Lee,maki1983static,MacDonald1985Broken,Platzman1993QuantumFreezing}.

Aharonov-Casher (AC) bands~\cite{aharonov1979ground} provide a convenient platform to study the effect of non-uniform quantum geometry on the FCI-crystal competition. They exhibit a flat electronic dispersion and, crucially, a non-uniform quantum geometry that can be tuned while preserving the vanishing bandwidth and ideal band condition, thus isolating the effects of quantum geometry.
AC bands also model Chern bands that emerge in semiconductor moiré materials \cite{Duran2024Magic,Shi2024Adiabatic}. The ideal geometry of AC bands renders fractional Chern insulator states exact zero-energy ground states of short-range interactions~\cite{ledwith2023vortexability,ledwith2020fractional,wang2021exact,cano2026ideal}. This property allows one to construct Laughlin-type wave functions analytically and evaluate their energies efficiently even in the presence of long-ranged interactions. By contrast, no comparable analytical description of the competing crystalline states has been available. 
Consequently, although the destabilization of Laughlin states by geometric non-uniformity is well understood~\cite{shi2026effects,Wu2012Adiabatic,jackson2015geometric,Moitra2026Instability,Parameswaran2013Fractional}, a framework for understanding the energetic advantage of the competing crystalline states has been lacking.

In this work, we introduce a simple variational ansatz for generalized Wigner crystals in AC bands: a Slater determinant of localized orbitals arranged on a periodic lattice. A multipole expansion shows that the minimum-spread orbital yields the lowest variational energy, which decreases sharply with increasing quantum-geometric non-uniformity. Comparing this energy with that of Laughlin-type states, we obtain phase diagrams at fillings $\nu=1/3$, $1/7$, and $1/9$, demonstrating that strong non-uniformity favors crystallization, consistent with previous studies of FCI gaps~\cite{shi2026effects}. We find the Laughlin energy is nearly insensitive to non-uniformity of quantum geometry, such that the phase boundary is almost entirely determined by the crystal's zero-point spread, giving a quantum analogue of the Lindemann criterion.
We further find that the rotational symmetry of the optimal crystal depends on the sign of the non-uniformity, revealing a purely quantum-geometric mechanism for symmetry breaking. These results show how intra-unit-cell geometric structure enhances orbital localization and drives the transition from a fractionalized liquid to an electron crystal.

{\em Aharonov-Casher bands: a model for moiré materials --} 
In twisted TMD homobilayers such as tMoTe$_2$, the complex interlayer tunneling produces a relatively flat topmost valence band with Chern number $C=1$ \cite{Wu2019Topological, Yu2020Giant, Zhang2025Experimental,Thompson2025Microscopic}, whose geometric properties resemble those of the lowest Landau level \cite{devakul2021magic,Duran2023Pressure} and provide a favorable platform for the emergence of correlated topological phases.
Here, we study a simplified model -- the Aharonov-Casher band -- that captures the key features of the topmost moiré band of tMoTe$_2$: flat dispersion, Chern number $C=1$ and ideal quantum geometry. The adiabatic approximation \cite{Duran2024Magic,Shi2024Adiabatic} provides a precise connection between tMoTe$_2$ and the AC model. Generalizations of this mapping have also been developed for rhombohedral graphene \cite{tan2025ideal}.

\begin{figure}
    \centering
    \includegraphics[width=0.48\textwidth]{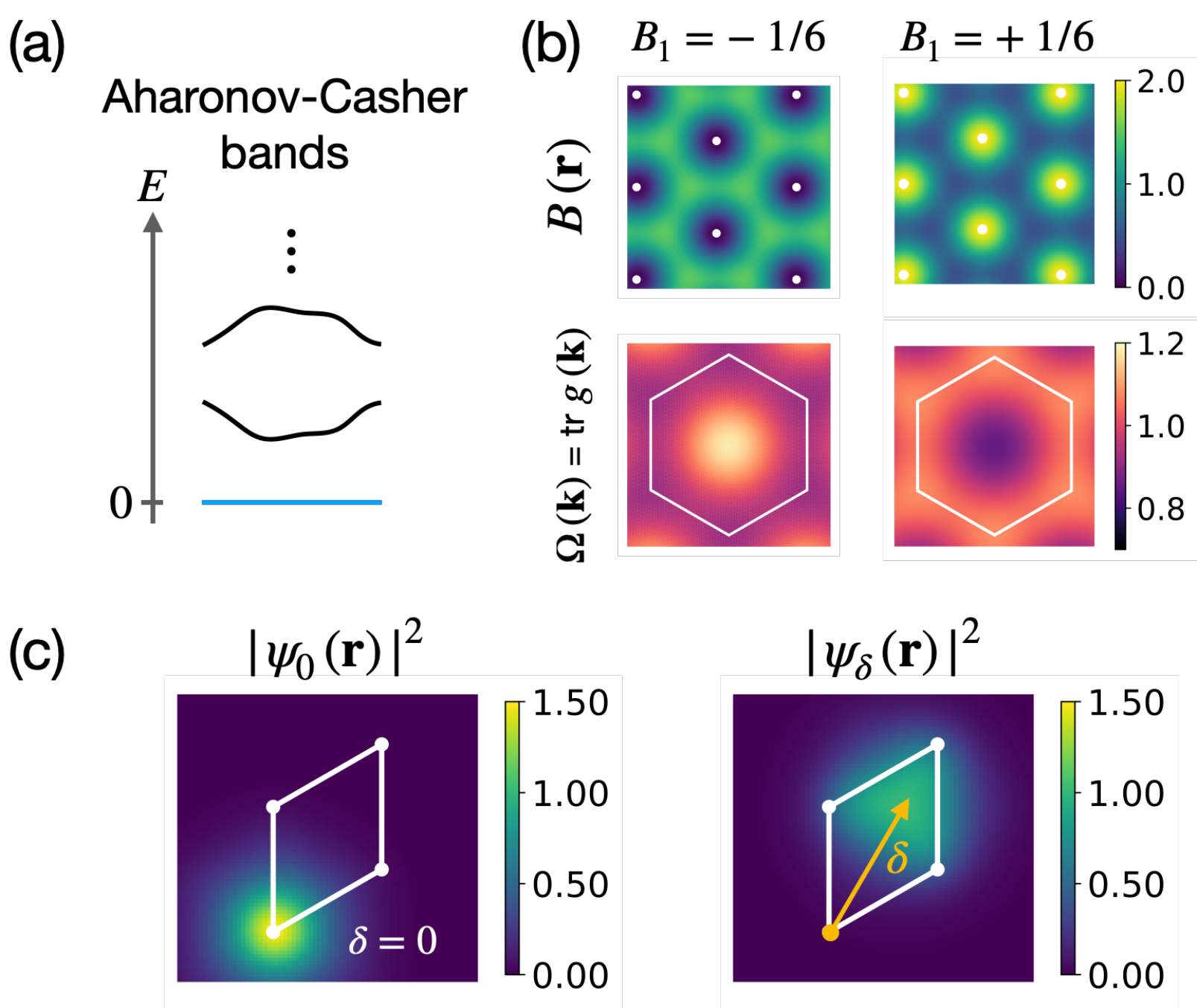}
    
    \caption{(a) Schematic representation of the bands resulting from diagonalizing the Hamiltonian given by Eq.~\eqref{eq:AC_Hamiltonian}. In this work we focus on the lowest band (indicated in blue), which we refer to as the AC band. This band is flat and has ideal quantum geometry. (b) Examples of magnetic field distributions generating the AC bands and their corresponding Berry curvature, $\Omega_{\bm k}$, with negative (left column) and positive (right column) harmonic $B_1$. Because the AC band is ideal, it satisfies $\Omega_{\bm k}=\text{tr}\,g_{\bm k}$, where $g_{\bm k}$ is the quantum metric. (c) Charge densities of two localized orbitals centered at different positions $\bm \delta$ within the moiré unit cell. We have used $B_1=-1/6$.}
    \label{fig:Schematics}
\end{figure}

An AC band is the lowest band of the Hamiltonian 
\begin{align}
    H_{\text{AC}}=\frac{1}{2m^*}\left[ \bm \Pi^2+B({\bm r})\right],
    \label{eq:AC_Hamiltonian}
\end{align}
which describes a single spin sector of a non-relativistic particle of effective mass $m^*$, subject to a periodic magnetic field. The kinetic-momentum operator ${\bm \Pi}$ is defined as
\begin{align}
    \bm \Pi={\bm p}+{\bm A}(\bm r), \quad\text{with}\quad \nabla \times {\bm A}({\bm r})=B({\bm r}).
    \label{eq:defpiB}
\end{align}
As schematically indicated in Fig.~\ref{fig:Schematics}(a), the lowest energy band of the Hamiltonian defined by Eq. \eqref{eq:AC_Hamiltonian} is flat for any choice of $B({\bm r})$ \cite{aharonov1979ground}. Here we choose the magnetic field $B({\bm r})$ to be invariant under translation by lattice vectors $\{ {\bm R}^0\}$ with lattice constant $a_M$, whose corresponding reciprocal lattice vectors are $\{ \bm G\}$. Motivated by experiments in moir\'e systems, we specialize to the triangular lattice and refer to the lattice defined by $\{ {\bm R}^0\}$  as the moir\'e lattice.
  
We split the magnetic field into its average and a varying piece,
\begin{align}
    B({\bm r})=B_0+\delta B({\bm r}).
    \label{eq:MagneticFieldSplit}
\end{align}
Following the adiabatic mapping, we fix $B_0$ so that each moir\'e unit cell is pierced by one magnetic flux quantum $\Phi_0$ \cite{Duran2024Magic,Shi2024Adiabatic}, and define the associated magnetic length as $\ell^2=\hbar/(eB_0)$.

For simplicity, we take $\delta B (\mathbf{r})$ to be the sum of first harmonics:
\begin{align}
    \delta B({\bm r})=B_1 \sum_{{\bm G} \in \text{1st shell}}e^{\,i\,{\bm G}\cdot {
    \bm r
    }}.
    \label{eq:BG_harmonics}
\end{align}
$B_1$ has units of $\ell^{-2}$ and determines the momentum-space profile of the band's Berry curvature~\cite{wang2021exact}. In Fig.~\ref{fig:Schematics}(b) we show two examples of magnetic field distributions studied in this work, together with the Berry curvatures of the resulting AC bands. 

The ideal quantum geometry of AC bands implies that their wave functions take the form \cite{wang2021exact,ledwith2020fractional} 
\begin{align}
    \psi^{AC}({\bm r})=e^{\chi({\bm r})}\,\psi^{LLL}({\bm r}),
    \label{eq:AC_wavefunction}
\end{align}
where $\psi^{LLL}({\bm r})$ is a lowest Landau level (LLL) wave function and $\chi({\bm r})$ is the K\"ahler potential that satisfies $\nabla^2 \chi({\bm r})=\delta B({\bm r})$ \cite{wang2021exact,ledwith2020fractional}. In the limit $B_1=0$ the AC band reduces to the LLL, where the competition between fractionalized and crystalline phases has been extensively studied \cite{MacDonald1985Broken,MacDonald1984Influence,Girvin_Lam,Yi1998Laughlin, Yoshioka_Lee, Yoshioka_Fukuyama,maki1983static,Platzman1993QuantumFreezing,DaSilva2016Fractional,Levesque1984Crystallization}.\\

{\em  Candidate ground states of Aharonov-Casher bands --}
\begin{figure*}
    \centering
    \includegraphics[width=0.95\textwidth]{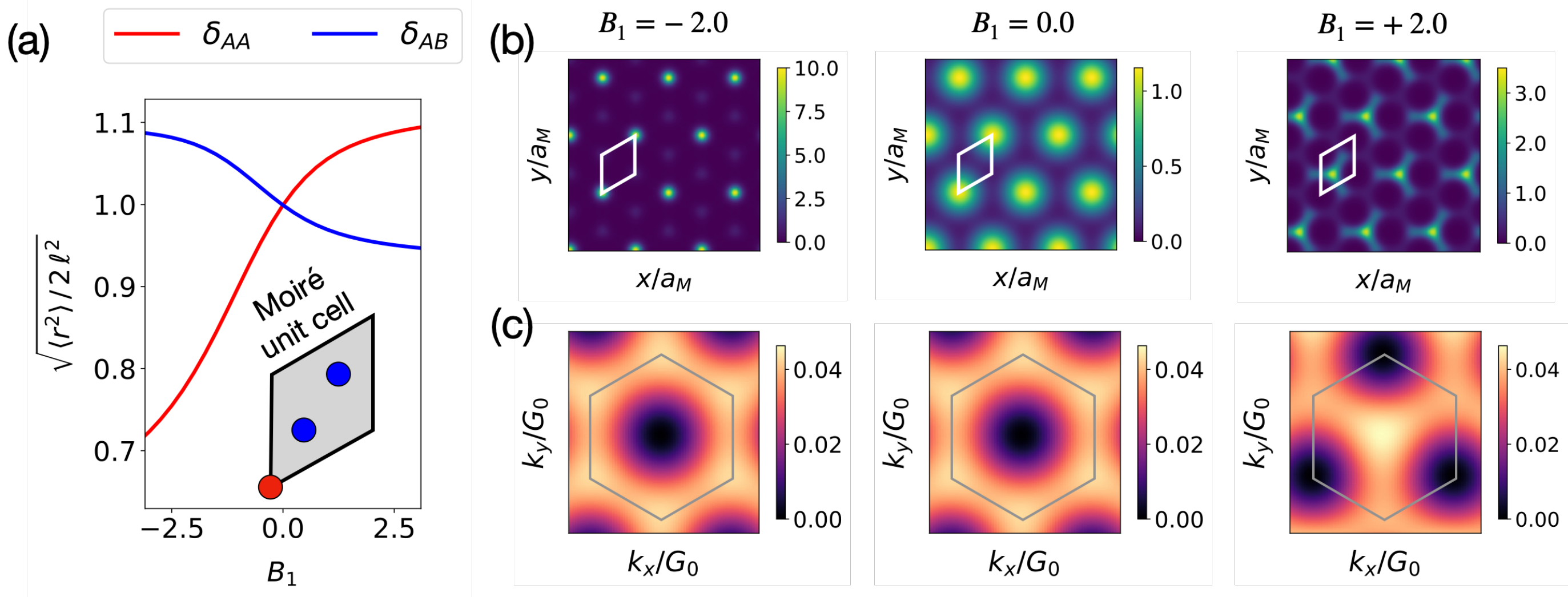}
    
    \caption{(a) Spread of most localized AC orbitals centered at high-symmetry points within the moiré unit cell, as a function of $B_1$. The high-symmetry points, labeled ${\bm\delta}_{\text{AA}}$ (red) and ${\bm \delta}_{\text{AB}}/{\bm \delta}_{\text{BA}}$ (blue) are indicated in the inset. (b) Charge density of the crystal state at filling $\nu=1/3$ for different values of $B_1$. The moiré unit cell is indicated. (c) Brillouin zone occupation $|c({\bm k})|^2$ for an orbital from each crystal in (b). For $B_1=+2$ there is an equivalent crystal with orbitals centered at $\bm \delta_{\text{BA}}$, whose momentum distribution is peaked at the other three corners of the Brillouin zone, $G_0=4\pi/\sqrt{3}a_M$ is the magnitude of the first-shell reciprocal lattice vectors.}
    \label{fig:Crystal_Densities}
\end{figure*}

We now consider the candidate ground states of AC bands as a function of $B_1$ for electron densities corresponding to a filling fraction $\nu=1/m$ per moiré unit cell, where $m$ is an odd integer.
For short-range interactions, the following Laughlin-like state is a zero energy state ~\cite{wang2021exact,ledwith2020fractional}
\begin{align}
    \ket{\Psi_{\text{L}}^{\text{AC}}}=\left( \prod_i e^{\chi(\bm r_i)} \right)\ket{\Psi_{\text{L}}^{\text{LLL}}},
    \label{eq:Laughlin_State}
\end{align}
where the product is taken over all particle indices $i$ and $\ket{\Psi_{L}^{\text{LLL}}}$ is the Laughlin ground state for a spatially uniform magnetic field \cite{Laughlin1983Anomalous}. For the Coulomb interaction, the energy of the Laughlin-like state is given, to second order in $B_1$, by
\begin{align}
    \frac{E_{\text L}^{\text{AC}}}{N}\approx\frac{E_{\text{L}}^{\text{LLL}}}{N}+\left( \frac{9\,\bar{f}_{\bm G_0}}{8\pi^2}\right)\,B_1^2,
    \label{eq:AC_Laughlin_Energy}
\end{align}
where $N$ is the electron number and $\bar{f}_{\bm G_0}\equiv \langle\bar{\rho}^{\dagger}_{\bm G_0}\left[H,\bar{\rho}_{\bm G_0}\right]\rangle$ is the single-mode approximation to the  oscillator strength of the uniform Laughlin state \cite{girvin1986magneto} evaluated at the first shell of reciprocal lattice vectors; see Appendix \ref{Appendix:LaughlinPerturbation} for a derivation. In addition, the analytic form of Eq.~\eqref{eq:Laughlin_State} allows one to compute its energy for arbitrary values of $B_1$ via plasma analogy Monte Carlo calculations \cite{Wolf2025Intraband, Kousa2025Magnetoroton}; see Appendix \ref{Appendix:MonteCarlo} for further details.

The main state competing with the Laughlin liquid is a generalized  Wigner crystal with periodicity commensurate with the moiré lattice, $a_c=\nu^{-1/2}\,a_M$. Following Maki and Zotos \cite{maki1983static}, we propose a Slater-determinant variational wave function
\begin{align}
    \Psi_{\text{C}}^{\text{AC}}(\{ {\bm r}_i \})=\left( \frac{1}{N!}\right)^{1/2} \det |T_{{\bm R}_j}\psi^{\text{AC}}({\bm r}_i)|,
    \label{eq:MZ_Ansatz}
\end{align}
where $\{ {\bm r}_i\}$ are the electron coordinates, $\{ {\bm R}_j\} $ are the crystal site centers, $\psi^{\text{AC}}({\bm r})$ is a localized single-electron wave function with weight entirely in the AC band, and $T_{{\bm R}_j}$ is the magnetic translation operator. We assume that the crystal sites form a triangular lattice commensurate with the magnetic field, so that $\{{\bm R}_j\}$ are invariant under translations generated by a subset of the $\{ {\bm R}^0\}$.

For dilute electron density the orbital overlaps, and therefore the exchange energies, are negligible and the energy is determined by the direct Coulomb interaction term between different orbitals (there is no kinetic energy, because the AC band is flat; see Fig.~\ref{fig:Schematics}(a)). The minimum energy of the crystalline state in Eq.~\eqref{eq:MZ_Ansatz} is determined by minimizing the spread of $\psi^\text{AC}(\bm r)$. An accurate heuristic is the second cumulant $\braket{r^2}_c$ defined by
\begin{align}
    \braket{r^2}_{c}=\int d^2r\,({\bm r}-\langle{\bm r}\rangle)^2 |\psi^{\text{AC}}({\bm r})|^2,
    \label{eq:r2c}
\end{align} 
where $\langle \bm r \rangle$ is the center of mass of $\psi^{\text AC}$. Indeed combining the standard multipole expansion of the direct Coulomb energy with approximate $C_3$-symmetry and Gaussianity of the wave functions gives (see Appendix \ref{Appendix:Multi_pole})

\begin{align}
    \frac{E_{\text{C}}^{\text{AC}} [\psi^{\text{AC}}]}{N}\approx \frac{e^2}{2\ell}\sum_{i}\left[\frac{1}{R_{i}}+\frac{\braket{r^2}_{c}}{2R_{i}^3}+\frac{9\, \braket{r^2}_{c}^2}{8R_{i}^5}\right]{- E_{\text{bg}}}
    ,
    \label{eq:AC_Crystal_Energy}
\end{align}
where $R_{i}=|{\bm R}_i|$ runs over all crystal sites, $E_\text{bg}$ is the contribution from the neutralizing background,  and  terms of higher order in $R^{-1}$ are not explicitly written.

To find the orbital $\psi^{\text{AC}}$ that minimizes Eq.~(\ref{eq:r2c}), it is useful to introduce an auxiliary variable $\bm \delta$ and to rewrite the minimization of Eq.~(\ref{eq:r2c})  as
\begin{equation}
    \min_{\psi} \langle r^2 \rangle_c = \min_{\psi, \mathbf{\delta}} \langle ( {\bm r}-{\bm \delta})^2 \rangle = \min_{\mathbf{\delta}} \langle ({\bm r}-{\bm \delta})^2 \rangle^{\text{opt}},
\end{equation}
where the first equality follows because for a fixed $\psi^\text{AC}$,
the optimal ${\bm \delta}$ satisfies ${\bm \delta}  = \langle {\bm r} \rangle$ and
$\langle ({\bm r} - {\bm \delta})^2 \rangle = \langle r^2 \rangle_c$.  Introducing  $\langle ({\bm r}-{\bm \delta})^2 \rangle^{\text{opt}} = \min\nolimits_{\psi} \langle ({\bm r}-{\bm \delta})^2 \rangle$ as the minimal (optimized) wave function spread for fixed $\bm \delta$, the second equality then shows that by minimizing the spread over $\bm{\delta}$ we accomplish the needed minimization over $\psi^{\text{AC}}$. 

We can obtain $\langle (\bm r - \bm \delta)^2 \rangle^{\text{opt}}$ by solving for the ground state of an effective Hamiltonian in momentum space. To see this, we decompose the AC orbital into a sum over the quasi-Bloch states of the band~\cite{ferrari1990twodimensional,haldane2018modularinvariant}, 
\begin{align}
    \psi^{\text{AC}}({\bm r})=\int d^2 k\,c({\bm k})\,\psi_{\bm k}^{\text{AC}}({\bm r}).
    \label{eq:Orbital_expansion}
\end{align}
This decomposition yields an effective Hamiltonian in momentum space \cite{Price2015Artificial,claassen2015position,okuma2024constructing,li2024constraints} 

\begin{align}
    \braket{(\bm r-\bm \delta)^2}=\int d^2 k\,c^*({\bm k})\left[({\bm\Pi}_{\bm k} - {\bm \delta})^2+\text{tr}\,g_{\bm k}\right]c({\bm k}),
    \label{eq:Effective_H}
\end{align}
where ${\bm \Pi}_{\bm k}=-i{\bm \partial}_{\bm k} +{\bm A}_{\bm k}$ is a vector-valued operator, with ${\bm A}_{\bm k}$ the Berry connection, and $g_{\bm k}$ is the quantum metric of the AC band. We have used the notation that for a vector ${\bm A}$, $({\bm A})^2 = A_x^2 + A_y^2$. We denote the optimally localized orbital by
\begin{equation}
    \psi_{\bm \delta}^{\text{AC}} (\bm r)= \mathrm{argmin} \langle (\bm r-\bm \delta)^2 \rangle.
\end{equation}
Due to the non-uniform magnetic field at $B_1 \neq 0$, both the optimal spread and the charge density distribution depend on $\bm \delta$. In Fig.~\ref{fig:Schematics}(c) we show examples of AC orbitals localized at different positions within the unit cell. We find that the optimal orbitals, obtained by varying over $\bm \delta$, are localized at the high-symmetry points of the moiré unit cell, that we label ${\bm \delta}_{\text{AA}}, {\bm \delta}_{\text{AB/BA}}$ (see Appendix~\ref{Appendix:Gaussian_Orbital}). Specifically, for $B_1<0$, the optimally localized AC orbital is centered at ${\bm \delta}_{\text{AA}}$, while for $B_1>0$, the optimally localized AC orbital is centered at ${{\bm \delta}_{\text{AB}/\text{BA}}}$, as shown in Fig.~\ref{fig:Crystal_Densities}(a). These locations correspond to a minimum of $B({\bm r})$ and a maximum of $\chi({\bm r})$.

The change in the location of the optimal orbital between $B_1 < 0$ and $B_1 > 0$ results in crystals with different $n$-fold rotational symmetries, denoted by $C_n$. 
In Fig.~\ref{fig:Crystal_Densities}(b) we show the charge densities of crystals given by Eq.~\eqref{eq:MZ_Ansatz} with optimal orbitals.
For $B_1 > 0$, the orbitals are centered at ${\bm \delta}_{\text{AB}/\text{BA}}$ and break the $C_6$-symmetry of $B({\bm r})$ down to $C_3$, while for $B_1 < 0$, the crystal remains $C_6$-symmetric. Following Ref.~\cite{mishra2026charge}, we refer to these crystals as the $C_3$-crystal and $C_6$-crystal, respectively.

The momentum space structure of the optimal orbitals reveals the quantum geometric origin of their real space behavior.
In Fig.~\ref{fig:Crystal_Densities}(c) we show $|c({\bm k})|^2$ of the optimal orbitals at $B_1 = -2, 0, +2$.
%\TS{\sout{The optimal orbital results from a balance between minimizing the terms in the effective Hamiltonian in Eq.~\eqref{eq:Effective_H}: each orbital is peaked in regions where $\text{tr}g_{\bm k}$ is small, while retaining a degree of delocalization determined by the competing ``kinetic'' term.}}
At $B_1=0$, the momentum space distribution is given in terms of a Jacobi theta function~\cite{ishikawa1997flux, ishikawa1998duality}, which minimizes the ``kinetic'' part of the effective Hamiltonian. The honeycomb structure of $|c({\bm k})|^2$ matches the honeycomb minima of $\text{tr}\,g_{\bm{k}}$ at $B_1 < 0$. As a result, $|c({\bm{k}})|^2$ remains nearly identical. At $B_1 > 0$, however, $|c({\bm{k}})|^2$ becomes modified to take advantage of the triangular lattice minima of $\text{tr}\,g_{\bm{k}}$, and the optimal orbitals form a balance between the ``kinetic'' and $\text{tr}\,g_{\bm{k}}$ terms.

The behavior of optimized orbitals can be captured by a natural analytical ansatz:
We multiply the maximally localized coherent state at $B_1 = 0$ with the K\"ahler potential factor $e^{\chi(r)}$:
\begin{align}
    \psi^{\text{AC}}_{\bm \delta}({\bm r})
    &\approx \left(\frac{1}{2\pi \ell^2\mathcal{N}_{{\bm \delta}}^2}\right)^{1/2}\, e^{-\frac{|{\bm r}-{\bm \delta}|^2}{4\ell^2}+i\frac{({\bm \delta}\times {\bm r})_z}{2\ell^2}+\chi({\bm r})}, 
    \label{eq:AC_GaussianOrbital}
\end{align}
where $\mathcal{N}_{\bm \delta}$ is a normalization constant.
For a uniform quantum metric, $B_1=0$, Eq.~\eqref{eq:AC_GaussianOrbital} is a Gaussian in the LLL \cite{maki1983static} and minimizes Eq.~\eqref{eq:Effective_H} with $\braket{r^2}_c=2\ell^2$.
At nonzero $B_1$, the $e^{\chi({\bm r})}$ factor changes the spatial profile of the orbital to enhance the weight on the region with small $B({\bm r})$, consistent with the profile observed in Fig.~\ref{fig:Crystal_Densities}(b).
In the Appendix \ref{Appendix:Gaussian_Orbital}, we show the quantitative agreement between the optimal orbitals and analytical orbitals given by Eq.~\eqref{eq:AC_GaussianOrbital}, and compute the spread and the crystal energy analytically.

\begin{figure}
    \centering
    \includegraphics[width=0.48\textwidth]{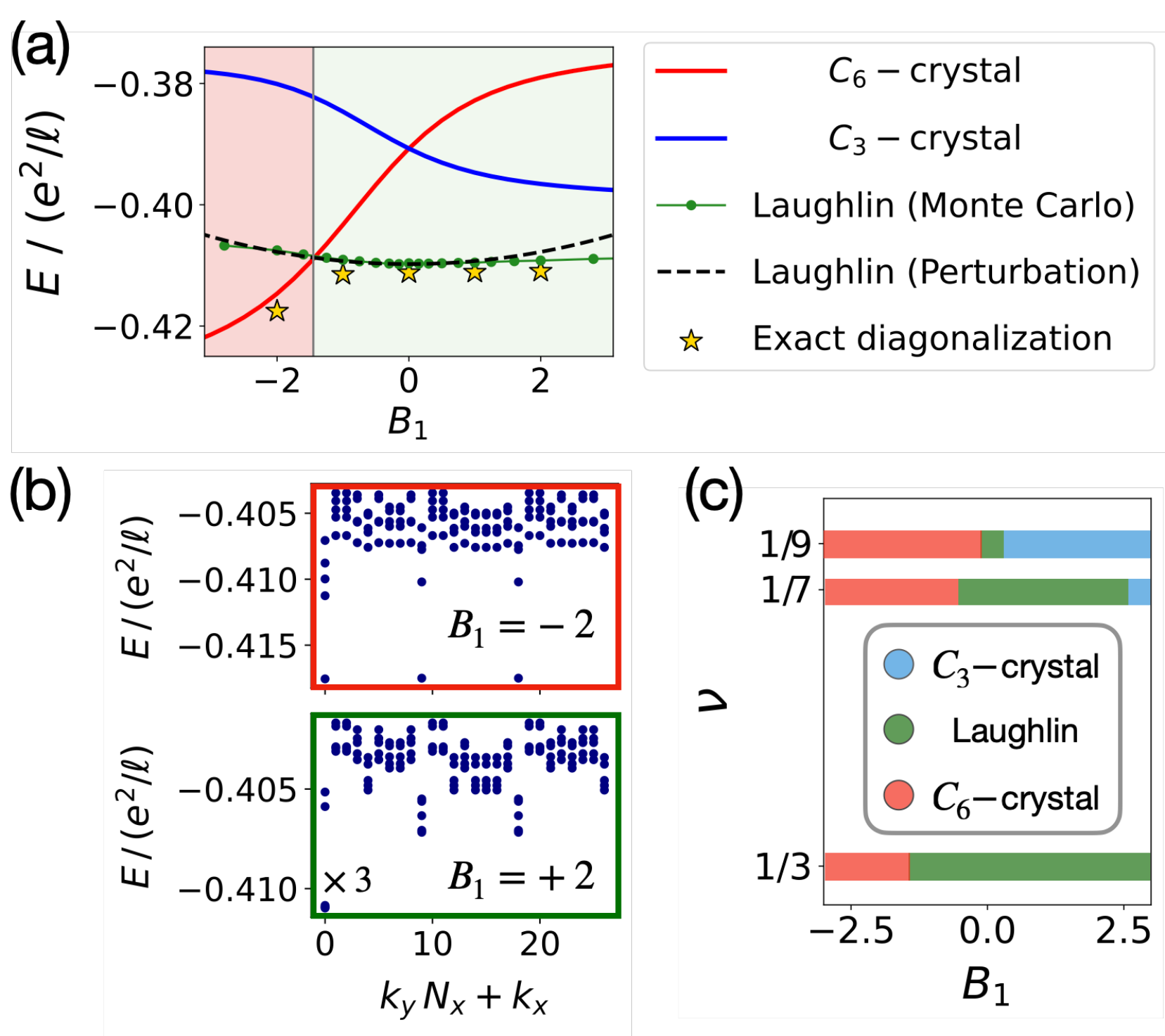}
    
    \caption{(a) Energy per particle of crystal and liquid states at filling $\nu=1/3$ of AC bands, as a function of $B_1$. The golden stars indicate exact diagonalization energies for a system with $N=9$ particles. (b) Exact diagonalization spectra for two AC bands: For $B_1=-2$ the ground state breaks the discrete translation symmetry of the magnetic field, forming a generalized Wigner crystal with $\sqrt{3}\times \sqrt{3}$ order. For $B_1=+2$ the ground state degeneracy is consistent with an FCI Laughlin-like ground state. (c) Phase diagram of AC bands as a function of $\nu$ and $B_1$.}
    \label{fig:PhaseDiagram}
\end{figure}

{\em Phase diagram of Aharonov-Casher bands at fractional fillings --}
We now compare the energy of candidate variational states. In Fig.~\ref{fig:PhaseDiagram}(a) we show how the energies of the $C_3$-crystal, the $C_6$-crystal, and the Laughlin-like liquid depend on $B_1$, at $\nu=1/3$. We consider $B_1 \in [-3,+3]$. The Laughlin energy is computed using Eq.~\eqref{eq:AC_Laughlin_Energy} (black dashed line) and through plasma analogy Monte Carlo (green line), while the crystalline energies are computed from the optimal orbital spread at ${\bm\delta}_{\text{AA}}$ and ${\bm \delta}_{AB/BA}$.
The energy of the Laughlin state is weakly affected by the geometry non-uniformity and the perturbative result reproduces the Monte Carlo result very well, even beyond $|B_1|\ll1$. On the other hand, the crystalline energy depends strongly on $B_1$, especially at $B_1 < 0$.

As a result, the energy of the $C_6$-crystal crosses the Laughlin state energy at around $B_1\approx -1.45$, while the $C_3$-crystal does not cross the Laughlin state energy for all the values of $B_1>0$ considered in our study. This observation is quantitatively confirmed by ED calculations on a finite system with $N=9$ particles. Representative many-body spectra for $B_1=-2$ and $B_1=+2$ are shown in Fig.~\ref{fig:PhaseDiagram}(b), and the corresponding ED ground state energies are indicated as golden stars in Fig.~\ref{fig:PhaseDiagram}(a). The ED spectrum for $B_1=-2$ is consistent with a $\sqrt{3}\times\sqrt{3}$ generalized Wigner crystal, while the ED spectrum for $B_1=+2$ has three-fold topological degeneracy at $\bm \gamma$, expected for an FCI in the studied finite size system. We repeat this procedure for other commensurate fillings, and obtain the phase diagram in Fig.~\ref{fig:PhaseDiagram}(c), which reveals that the $C_3$-crystal can be stabilized (see Appendix \ref{Appendix_Other_fillings} for how these energies behave).

{\em Discussion --}
In narrow Chern bands, the intra-unit-cell spatial structure induced by quantum geometry modifies the wave functions of translation-symmetry-breaking insulators, and therefore can influence ground state energetics. Here we studied a flat Chern band model with ideal but non-uniform quantum geometry -- the AC band. Making use of its special geometric structure, we uncover the central mechanism governing the ground state competition: non-uniform quantum geometry favors electron crystallization by enabling greater orbital localization, while imposing an in practice weak energetic cost on the Laughlin state, whose correlations must accommodate the resulting real-space structure.
Increasing the quantum geometry non-uniformity ultimately favors a crystalline ground state, while the global structure of the quantum geometry determines the symmetry and preferred site centers of the lowest-energy crystal.

The special geometric structure of the AC band allows us to compute the Laughlin state energy for large systems, which enabled a mapping of the phase diagram as a function of quantum geometry non-uniformity and filling fraction.
However, the underlying mechanism by which quantum geometry controls orbital localization does not rely on ideal geometry. This motivates future studies to address the applicability of our results beyond the ideal limit.

Our results have an interesting connection to Lindemann's criterion \cite{Lindemann1910}, according to which a solid melts when the root mean square fluctuation of atomic positions about the lattice sites becomes a fraction of the interatomic distance. The weak dependence of the Laughlin energy on quantum geometric effects allows us to approximate it at $B_1 \neq 0$ by its value at $B_1 = 0$; the crystal melts when the zero point fluctuations become large enough such that $E^{\text{AC}}_\text{L}(B_1 = 0) = E^{\text{AC}}_\text{C}(\langle r^2 \rangle_c)$. We find that at melting, the ratio between the orbital spread and the crystal lattice constant, $\sqrt{\langle r^2\rangle_c}/a_c$, is approximately $ 0.25,0.19,0.17$ for $\nu=1/3, 1/7, 1/9$, respectively. These values are not universal; they depend on $\nu$ because the energy of the FCI liquid is density-dependent, but they are of the order of the values found in other 2D and 3D melting transitions \cite{Khrapak2020Lindemann,Bedanov1985modified,Ross1969Generalized}. We expect this criterion to be robust even if we consider more general AC bands by adding higher harmonics.

We did not address crystalline ground states whose periodicity is incommensurate with that of the magnetic field, as such states are not captured by our ansatz. This situation arises, for instance, at filling $\nu=1/5$, where an incommensurate crystal as well as a stripe phase will compete with the Laughlin liquid. We leave this interesting scenario for future work. All the crystalline phases considered in this work are topologically trivial, but ans\"atze for Wigner crystals with non-zero Chern number, which are energetically competitive at higher filling fractions \cite{MacDonald1984Influence,MacDonald1985Broken,mishra2026charge}, could also be constructed. 
Thus, our approach can be generalized to the large variety of crystalline states that have recently been proposed \cite{kim2026exchange,Joy2026ChiralWigner,Zhou2025NewClasses,Zverevich2026Spin}. In addition, the framework presented here provides a controlled platform to investigate the role of quantum geometry on the collective modes of both topological liquids \cite{Kousa2025Magnetoroton,Paul2026Shining,shi2026effects} and crystalline states \cite{Dong2025Phonons,Desrochers2026Elastic} on the same footing.

The AC band we study is experimentally relevant, as it can approximate the topmost band of twisted TMD moiré materials~\cite{Duran2024Magic,Shi2024Adiabatic,shi2026effects}. Most experiments on tMoTe$_2$ have observed a trivial insulating state at filling fraction $\nu=1/3$, consistent with a crystalline phase. However, a recent experiment \cite{Pan2026optical} has reported a weak FCI state. While the experimental band structure has a nonzero band width and a finite band gap to the next band, our result may serve as a starting point for understanding this delicate competition. Near filling fraction $\nu=2/3$, experiments on tMoTe$_2$ have reported a rich interplay of multiple phases, such as FCIs, re-entrant integer quantum anomalous Hall (RIQAH) states and possible signatures of superconductivity \cite{xu2025superconductivity}. Additionally, ED and density matrix renormalization group studies suggest that superconductivity can emerge close to $\nu=2/3$ in AC bands \cite{Guerci_AC_SC} and related models \cite{wang2025chiral}. A detailed investigation of the AC band phase diagram at $\nu=2/3$ is left for future work.

Finally, we expect that removing the projection to the lowest-energy band, which gave us analytical and numerical control over the problem, will only quantitatively change the picture. 
In fact, band-mixing allows further localization of the electronic orbitals \cite{MacDonald1984Influence,mishra2026charge,moralesduran2026bandmixing}, thus favoring crystalline states  over Laughlin liquids and decreasing the magnitude of critical values for $B_1$ where the FCI to crystal transition occurs.\\

{\it Acknowledgements--} The Flatiron Institute is a division of the Simons Foundation. We thank Conor Smith for related collaborations and Lei Chen, Eslam Khalaf, Bishoy M. Kousa, Allan H. MacDonald, Sparsh Mishra, Daniel Muñoz-Segovia, and Nicolas Regnault for helpful discussions. JS acknowledges support from the US Department of Energy, Office of Science, Basic Energy Sciences, Materials Sciences and Engineering Division.
This work was partly supported by JST PRESTO, Japan, Grant Number JPMJPR2455 to T.S.
JC acknowledges support from the Air Force Office of Scientific Research under Grant No. FA9550-24-1-0222. This work was performed in part at the Aspen Center for Physics, which is supported by National Science Foundation grant PHY-2210452 and by a grant from the Alfred P Sloan Foundation (G-2026-79536).

\bibliography{references}

\begin{thebibliography}{104}%
\makeatletter
\providecommand \@ifxundefined [1]{%
 \@ifx{#1\undefined}
}%
\providecommand \@ifnum [1]{%
 \ifnum #1\expandafter \@firstoftwo
 \else \expandafter \@secondoftwo
 \fi
}%
\providecommand \@ifx [1]{%
 \ifx #1\expandafter \@firstoftwo
 \else \expandafter \@secondoftwo
 \fi
}%
\providecommand \natexlab [1]{#1}%
\providecommand \enquote  [1]{``#1''}%
\providecommand \bibnamefont  [1]{#1}%
\providecommand \bibfnamefont [1]{#1}%
\providecommand \citenamefont [1]{#1}%
\providecommand \href@noop [0]{\@secondoftwo}%
\providecommand \href [0]{\begingroup \@sanitize@url \@href}%
\providecommand \@href[1]{\@@startlink{#1}\@@href}%
\providecommand \@@href[1]{\endgroup#1\@@endlink}%
\providecommand \@sanitize@url [0]{\catcode `\\12\catcode `\$12\catcode
  `\&12\catcode `\#12\catcode `\^12\catcode `\_12\catcode `\%12\relax}%
\providecommand \@@startlink[1]{}%
\providecommand \@@endlink[0]{}%
\providecommand \url  [0]{\begingroup\@sanitize@url \@url }%
\providecommand \@url [1]{\endgroup\@href {#1}{\urlprefix }}%
\providecommand \urlprefix  [0]{URL }%
\providecommand \Eprint [0]{\href }%
\providecommand \doibase [0]{https://doi.org/}%
\providecommand \selectlanguage [0]{\@gobble}%
\providecommand \bibinfo  [0]{\@secondoftwo}%
\providecommand \bibfield  [0]{\@secondoftwo}%
\providecommand \translation [1]{[#1]}%
\providecommand \BibitemOpen [0]{}%
\providecommand \bibitemStop [0]{}%
\providecommand \bibitemNoStop [0]{.\EOS\space}%
\providecommand \EOS [0]{\spacefactor3000\relax}%
\providecommand \BibitemShut  [1]{\csname bibitem#1\endcsname}%
\let\auto@bib@innerbib\@empty
%</preamble>
\bibitem [{\citenamefont {Tang}\ \emph {et~al.}(2011)\citenamefont {Tang},
  \citenamefont {Mei},\ and\ \citenamefont {Wen}}]{Wen_FCI}%
  \BibitemOpen
  \bibfield  {author} {\bibinfo {author} {\bibfnamefont {E.}~\bibnamefont
  {Tang}}, \bibinfo {author} {\bibfnamefont {J.-W.}\ \bibnamefont {Mei}},\ and\
  \bibinfo {author} {\bibfnamefont {X.-G.}\ \bibnamefont {Wen}},\ }\bibfield
  {title} {\bibinfo {title} {High-temperature fractional quantum {H}all
  states},\ }\href {https://doi.org/10.1103/PhysRevLett.106.236802} {\bibfield
  {journal} {\bibinfo  {journal} {Phys. Rev. Lett.}\ }\textbf {\bibinfo
  {volume} {106}},\ \bibinfo {pages} {236802} (\bibinfo {year}
  {2011})}\BibitemShut {NoStop}%
\bibitem [{\citenamefont {Sun}\ \emph {et~al.}(2011)\citenamefont {Sun},
  \citenamefont {Gu}, \citenamefont {Katsura},\ and\ \citenamefont
  {Das~Sarma}}]{DasSarma_Sun_FCI}%
  \BibitemOpen
  \bibfield  {author} {\bibinfo {author} {\bibfnamefont {K.}~\bibnamefont
  {Sun}}, \bibinfo {author} {\bibfnamefont {Z.}~\bibnamefont {Gu}}, \bibinfo
  {author} {\bibfnamefont {H.}~\bibnamefont {Katsura}},\ and\ \bibinfo {author}
  {\bibfnamefont {S.}~\bibnamefont {Das~Sarma}},\ }\bibfield  {title} {\bibinfo
  {title} {Nearly flatbands with nontrivial topology},\ }\href
  {https://doi.org/10.1103/PhysRevLett.106.236803} {\bibfield  {journal}
  {\bibinfo  {journal} {Phys. Rev. Lett.}\ }\textbf {\bibinfo {volume} {106}},\
  \bibinfo {pages} {236803} (\bibinfo {year} {2011})}\BibitemShut {NoStop}%
\bibitem [{\citenamefont {Neupert}\ \emph {et~al.}(2011)\citenamefont
  {Neupert}, \citenamefont {Santos}, \citenamefont {Chamon},\ and\
  \citenamefont {Mudry}}]{Neupert_FCI}%
  \BibitemOpen
  \bibfield  {author} {\bibinfo {author} {\bibfnamefont {T.}~\bibnamefont
  {Neupert}}, \bibinfo {author} {\bibfnamefont {L.}~\bibnamefont {Santos}},
  \bibinfo {author} {\bibfnamefont {C.}~\bibnamefont {Chamon}},\ and\ \bibinfo
  {author} {\bibfnamefont {C.}~\bibnamefont {Mudry}},\ }\bibfield  {title}
  {\bibinfo {title} {Fractional quantum {H}all states at zero magnetic field},\
  }\href {https://doi.org/10.1103/PhysRevLett.106.236804} {\bibfield  {journal}
  {\bibinfo  {journal} {Phys. Rev. Lett.}\ }\textbf {\bibinfo {volume} {106}},\
  \bibinfo {pages} {236804} (\bibinfo {year} {2011})}\BibitemShut {NoStop}%
\bibitem [{\citenamefont {Sheng}\ \emph {et~al.}(2011)\citenamefont {Sheng},
  \citenamefont {Gu}, \citenamefont {Sun},\ and\ \citenamefont
  {Sheng}}]{Sheng_FCI}%
  \BibitemOpen
  \bibfield  {author} {\bibinfo {author} {\bibfnamefont {D.~N.}\ \bibnamefont
  {Sheng}}, \bibinfo {author} {\bibfnamefont {Z.-C.}\ \bibnamefont {Gu}},
  \bibinfo {author} {\bibfnamefont {K.}~\bibnamefont {Sun}},\ and\ \bibinfo
  {author} {\bibfnamefont {L.}~\bibnamefont {Sheng}},\ }\bibfield  {title}
  {\bibinfo {title} {Fractional quantum {H}all effect in the absence of
  {L}andau levels},\ }\href {https://doi.org/10.1038/ncomms1380} {\bibfield
  {journal} {\bibinfo  {journal} {Nature Communications}\ }\textbf {\bibinfo
  {volume} {2}},\ \bibinfo {pages} {389} (\bibinfo {year} {2011})}\BibitemShut
  {NoStop}%
\bibitem [{\citenamefont {Regnault}\ and\ \citenamefont
  {Bernevig}(2011)}]{Bernevig_Regnault_FCI}%
  \BibitemOpen
  \bibfield  {author} {\bibinfo {author} {\bibfnamefont {N.}~\bibnamefont
  {Regnault}}\ and\ \bibinfo {author} {\bibfnamefont {B.~A.}\ \bibnamefont
  {Bernevig}},\ }\bibfield  {title} {\bibinfo {title} {Fractional {C}hern
  insulator},\ }\href {https://doi.org/10.1103/PhysRevX.1.021014} {\bibfield
  {journal} {\bibinfo  {journal} {Phys. Rev. X}\ }\textbf {\bibinfo {volume}
  {1}},\ \bibinfo {pages} {021014} (\bibinfo {year} {2011})}\BibitemShut
  {NoStop}%
\bibitem [{\citenamefont {Wigner}(1934)}]{Wigner1934WC}%
  \BibitemOpen
  \bibfield  {author} {\bibinfo {author} {\bibfnamefont {E.}~\bibnamefont
  {Wigner}},\ }\bibfield  {title} {\bibinfo {title} {On the interaction of
  electrons in metals},\ }\href {https://doi.org/10.1103/PhysRev.46.1002}
  {\bibfield  {journal} {\bibinfo  {journal} {Phys. Rev.}\ }\textbf {\bibinfo
  {volume} {46}},\ \bibinfo {pages} {1002} (\bibinfo {year}
  {1934})}\BibitemShut {NoStop}%
\bibitem [{\citenamefont {Cai}\ \emph {et~al.}(2023)\citenamefont {Cai},
  \citenamefont {Anderson}, \citenamefont {Wang}, \citenamefont {Zhang},
  \citenamefont {Liu}, \citenamefont {Holtzmann}, \citenamefont {Zhang},
  \citenamefont {Fan}, \citenamefont {Taniguchi}, \citenamefont {Watanabe},
  \citenamefont {Ran}, \citenamefont {Cao}, \citenamefont {Fu}, \citenamefont
  {Xiao}, \citenamefont {Yao},\ and\ \citenamefont {Xu}}]{cai2023signatures}%
  \BibitemOpen
  \bibfield  {author} {\bibinfo {author} {\bibfnamefont {J.}~\bibnamefont
  {Cai}}, \bibinfo {author} {\bibfnamefont {E.}~\bibnamefont {Anderson}},
  \bibinfo {author} {\bibfnamefont {C.}~\bibnamefont {Wang}}, \bibinfo {author}
  {\bibfnamefont {X.}~\bibnamefont {Zhang}}, \bibinfo {author} {\bibfnamefont
  {X.}~\bibnamefont {Liu}}, \bibinfo {author} {\bibfnamefont {W.}~\bibnamefont
  {Holtzmann}}, \bibinfo {author} {\bibfnamefont {Y.}~\bibnamefont {Zhang}},
  \bibinfo {author} {\bibfnamefont {F.}~\bibnamefont {Fan}}, \bibinfo {author}
  {\bibfnamefont {T.}~\bibnamefont {Taniguchi}}, \bibinfo {author}
  {\bibfnamefont {K.}~\bibnamefont {Watanabe}}, \bibinfo {author}
  {\bibfnamefont {Y.}~\bibnamefont {Ran}}, \bibinfo {author} {\bibfnamefont
  {T.}~\bibnamefont {Cao}}, \bibinfo {author} {\bibfnamefont {L.}~\bibnamefont
  {Fu}}, \bibinfo {author} {\bibfnamefont {D.}~\bibnamefont {Xiao}}, \bibinfo
  {author} {\bibfnamefont {W.}~\bibnamefont {Yao}},\ and\ \bibinfo {author}
  {\bibfnamefont {X.}~\bibnamefont {Xu}},\ }\bibfield  {title} {\bibinfo
  {title} {Signatures of fractional quantum anomalous {{{H}all}} states in
  twisted {{{M}o{T}e$_2$}}},\ }\href
  {https://doi.org/10.1038/s41586-023-06289-w} {\bibfield  {journal} {\bibinfo
  {journal} {Nature}\ }\textbf {\bibinfo {volume} {622}},\ \bibinfo {pages}
  {63} (\bibinfo {year} {2023})}\BibitemShut {NoStop}%
\bibitem [{\citenamefont {Zeng}\ \emph {et~al.}(2023)\citenamefont {Zeng},
  \citenamefont {Xia}, \citenamefont {Kang}, \citenamefont {Zhu}, \citenamefont
  {Kn{\"u}ppel}, \citenamefont {Vaswani}, \citenamefont {Watanabe},
  \citenamefont {Taniguchi}, \citenamefont {Mak},\ and\ \citenamefont
  {Shan}}]{zeng2023thermodynamic}%
  \BibitemOpen
  \bibfield  {author} {\bibinfo {author} {\bibfnamefont {Y.}~\bibnamefont
  {Zeng}}, \bibinfo {author} {\bibfnamefont {Z.}~\bibnamefont {Xia}}, \bibinfo
  {author} {\bibfnamefont {K.}~\bibnamefont {Kang}}, \bibinfo {author}
  {\bibfnamefont {J.}~\bibnamefont {Zhu}}, \bibinfo {author} {\bibfnamefont
  {P.}~\bibnamefont {Kn{\"u}ppel}}, \bibinfo {author} {\bibfnamefont
  {C.}~\bibnamefont {Vaswani}}, \bibinfo {author} {\bibfnamefont
  {K.}~\bibnamefont {Watanabe}}, \bibinfo {author} {\bibfnamefont
  {T.}~\bibnamefont {Taniguchi}}, \bibinfo {author} {\bibfnamefont {K.~F.}\
  \bibnamefont {Mak}},\ and\ \bibinfo {author} {\bibfnamefont {J.}~\bibnamefont
  {Shan}},\ }\bibfield  {title} {\bibinfo {title} {Thermodynamic evidence of
  fractional {{{C}hern}} insulator in moir{\'e} {{{M}o{T}e$_2$}}},\ }\href
  {https://doi.org/10.1038/s41586-023-06452-3} {\bibfield  {journal} {\bibinfo
  {journal} {Nature}\ }\textbf {\bibinfo {volume} {622}},\ \bibinfo {pages}
  {69} (\bibinfo {year} {2023})}\BibitemShut {NoStop}%
\bibitem [{\citenamefont {Park}\ \emph {et~al.}(2023)\citenamefont {Park},
  \citenamefont {Cai}, \citenamefont {Anderson}, \citenamefont {Zhang},
  \citenamefont {Zhu}, \citenamefont {Liu}, \citenamefont {Wang}, \citenamefont
  {Holtzmann}, \citenamefont {Hu}, \citenamefont {Liu}, \citenamefont
  {Taniguchi}, \citenamefont {Watanabe}, \citenamefont {Chu}, \citenamefont
  {Cao}, \citenamefont {Fu}, \citenamefont {Yao}, \citenamefont {Chang},
  \citenamefont {Cobden}, \citenamefont {Xiao},\ and\ \citenamefont
  {Xu}}]{park2023observation}%
  \BibitemOpen
  \bibfield  {author} {\bibinfo {author} {\bibfnamefont {H.}~\bibnamefont
  {Park}}, \bibinfo {author} {\bibfnamefont {J.}~\bibnamefont {Cai}}, \bibinfo
  {author} {\bibfnamefont {E.}~\bibnamefont {Anderson}}, \bibinfo {author}
  {\bibfnamefont {Y.}~\bibnamefont {Zhang}}, \bibinfo {author} {\bibfnamefont
  {J.}~\bibnamefont {Zhu}}, \bibinfo {author} {\bibfnamefont {X.}~\bibnamefont
  {Liu}}, \bibinfo {author} {\bibfnamefont {C.}~\bibnamefont {Wang}}, \bibinfo
  {author} {\bibfnamefont {W.}~\bibnamefont {Holtzmann}}, \bibinfo {author}
  {\bibfnamefont {C.}~\bibnamefont {Hu}}, \bibinfo {author} {\bibfnamefont
  {Z.}~\bibnamefont {Liu}}, \bibinfo {author} {\bibfnamefont {T.}~\bibnamefont
  {Taniguchi}}, \bibinfo {author} {\bibfnamefont {K.}~\bibnamefont {Watanabe}},
  \bibinfo {author} {\bibfnamefont {J.-H.}\ \bibnamefont {Chu}}, \bibinfo
  {author} {\bibfnamefont {T.}~\bibnamefont {Cao}}, \bibinfo {author}
  {\bibfnamefont {L.}~\bibnamefont {Fu}}, \bibinfo {author} {\bibfnamefont
  {W.}~\bibnamefont {Yao}}, \bibinfo {author} {\bibfnamefont {C.-Z.}\
  \bibnamefont {Chang}}, \bibinfo {author} {\bibfnamefont {D.}~\bibnamefont
  {Cobden}}, \bibinfo {author} {\bibfnamefont {D.}~\bibnamefont {Xiao}},\ and\
  \bibinfo {author} {\bibfnamefont {X.}~\bibnamefont {Xu}},\ }\bibfield
  {title} {\bibinfo {title} {Observation of fractionally quantized anomalous
  {{{H}all}} effect},\ }\href {https://doi.org/10.1038/s41586-023-06536-0}
  {\bibfield  {journal} {\bibinfo  {journal} {Nature}\ }\textbf {\bibinfo
  {volume} {622}},\ \bibinfo {pages} {74} (\bibinfo {year} {2023})}\BibitemShut
  {NoStop}%
\bibitem [{\citenamefont {Xu}\ \emph {et~al.}(2023)\citenamefont {Xu},
  \citenamefont {Sun}, \citenamefont {Jia}, \citenamefont {Liu}, \citenamefont
  {Xu}, \citenamefont {Li}, \citenamefont {Gu}, \citenamefont {Watanabe},
  \citenamefont {Taniguchi}, \citenamefont {Tong}, \citenamefont {Jia},
  \citenamefont {Shi}, \citenamefont {Jiang}, \citenamefont {Zhang},
  \citenamefont {Liu},\ and\ \citenamefont {Li}}]{xu2023observation}%
  \BibitemOpen
  \bibfield  {author} {\bibinfo {author} {\bibfnamefont {F.}~\bibnamefont
  {Xu}}, \bibinfo {author} {\bibfnamefont {Z.}~\bibnamefont {Sun}}, \bibinfo
  {author} {\bibfnamefont {T.}~\bibnamefont {Jia}}, \bibinfo {author}
  {\bibfnamefont {C.}~\bibnamefont {Liu}}, \bibinfo {author} {\bibfnamefont
  {C.}~\bibnamefont {Xu}}, \bibinfo {author} {\bibfnamefont {C.}~\bibnamefont
  {Li}}, \bibinfo {author} {\bibfnamefont {Y.}~\bibnamefont {Gu}}, \bibinfo
  {author} {\bibfnamefont {K.}~\bibnamefont {Watanabe}}, \bibinfo {author}
  {\bibfnamefont {T.}~\bibnamefont {Taniguchi}}, \bibinfo {author}
  {\bibfnamefont {B.}~\bibnamefont {Tong}}, \bibinfo {author} {\bibfnamefont
  {J.}~\bibnamefont {Jia}}, \bibinfo {author} {\bibfnamefont {Z.}~\bibnamefont
  {Shi}}, \bibinfo {author} {\bibfnamefont {S.}~\bibnamefont {Jiang}}, \bibinfo
  {author} {\bibfnamefont {Y.}~\bibnamefont {Zhang}}, \bibinfo {author}
  {\bibfnamefont {X.}~\bibnamefont {Liu}},\ and\ \bibinfo {author}
  {\bibfnamefont {T.}~\bibnamefont {Li}},\ }\bibfield  {title} {\bibinfo
  {title} {Observation of {{Integer}} and {{Fractional Quantum Anomalous {H}all
  Effects}} in {{Twisted Bilayer {M}o{T}e}} $_2$},\ }\href
  {https://doi.org/10.1103/PhysRevX.13.031037} {\bibfield  {journal} {\bibinfo
  {journal} {Phys. Rev. X}\ }\textbf {\bibinfo {volume} {13}},\ \bibinfo
  {pages} {031037} (\bibinfo {year} {2023})}\BibitemShut {NoStop}%
\bibitem [{\citenamefont {Lu}\ \emph {et~al.}(2024)\citenamefont {Lu},
  \citenamefont {Han}, \citenamefont {Yao}, \citenamefont {Reddy},
  \citenamefont {Yang}, \citenamefont {Seo}, \citenamefont {Watanabe},
  \citenamefont {Taniguchi}, \citenamefont {Fu},\ and\ \citenamefont
  {Ju}}]{lu2024fractional}%
  \BibitemOpen
  \bibfield  {author} {\bibinfo {author} {\bibfnamefont {Z.}~\bibnamefont
  {Lu}}, \bibinfo {author} {\bibfnamefont {T.}~\bibnamefont {Han}}, \bibinfo
  {author} {\bibfnamefont {Y.}~\bibnamefont {Yao}}, \bibinfo {author}
  {\bibfnamefont {A.~P.}\ \bibnamefont {Reddy}}, \bibinfo {author}
  {\bibfnamefont {J.}~\bibnamefont {Yang}}, \bibinfo {author} {\bibfnamefont
  {J.}~\bibnamefont {Seo}}, \bibinfo {author} {\bibfnamefont {K.}~\bibnamefont
  {Watanabe}}, \bibinfo {author} {\bibfnamefont {T.}~\bibnamefont {Taniguchi}},
  \bibinfo {author} {\bibfnamefont {L.}~\bibnamefont {Fu}},\ and\ \bibinfo
  {author} {\bibfnamefont {L.}~\bibnamefont {Ju}},\ }\bibfield  {title}
  {\bibinfo {title} {Fractional quantum anomalous {{{H}all}} effect in
  multilayer graphene},\ }\href {https://doi.org/10.1038/s41586-023-07010-7}
  {\bibfield  {journal} {\bibinfo  {journal} {Nature}\ }\textbf {\bibinfo
  {volume} {626}},\ \bibinfo {pages} {759} (\bibinfo {year}
  {2024})}\BibitemShut {NoStop}%
\bibitem [{\citenamefont {Xu}\ \emph {et~al.}(2025)\citenamefont {Xu},
  \citenamefont {Sun}, \citenamefont {Li}, \citenamefont {Zheng}, \citenamefont
  {Xu}, \citenamefont {Gao}, \citenamefont {Jia}, \citenamefont {Watanabe},
  \citenamefont {Taniguchi}, \citenamefont {Tong}, \citenamefont {Lu},
  \citenamefont {Jia}, \citenamefont {Shi}, \citenamefont {Jiang},
  \citenamefont {Zhang}, \citenamefont {Zhang}, \citenamefont {Lei},
  \citenamefont {Liu},\ and\ \citenamefont {Li}}]{xu2025superconductivity}%
  \BibitemOpen
  \bibfield  {author} {\bibinfo {author} {\bibfnamefont {F.}~\bibnamefont
  {Xu}}, \bibinfo {author} {\bibfnamefont {Z.}~\bibnamefont {Sun}}, \bibinfo
  {author} {\bibfnamefont {J.}~\bibnamefont {Li}}, \bibinfo {author}
  {\bibfnamefont {C.}~\bibnamefont {Zheng}}, \bibinfo {author} {\bibfnamefont
  {C.}~\bibnamefont {Xu}}, \bibinfo {author} {\bibfnamefont {J.}~\bibnamefont
  {Gao}}, \bibinfo {author} {\bibfnamefont {T.}~\bibnamefont {Jia}}, \bibinfo
  {author} {\bibfnamefont {K.}~\bibnamefont {Watanabe}}, \bibinfo {author}
  {\bibfnamefont {T.}~\bibnamefont {Taniguchi}}, \bibinfo {author}
  {\bibfnamefont {B.}~\bibnamefont {Tong}}, \bibinfo {author} {\bibfnamefont
  {L.}~\bibnamefont {Lu}}, \bibinfo {author} {\bibfnamefont {J.}~\bibnamefont
  {Jia}}, \bibinfo {author} {\bibfnamefont {Z.}~\bibnamefont {Shi}}, \bibinfo
  {author} {\bibfnamefont {S.}~\bibnamefont {Jiang}}, \bibinfo {author}
  {\bibfnamefont {Y.}~\bibnamefont {Zhang}}, \bibinfo {author} {\bibfnamefont
  {Y.}~\bibnamefont {Zhang}}, \bibinfo {author} {\bibfnamefont
  {S.}~\bibnamefont {Lei}}, \bibinfo {author} {\bibfnamefont {X.}~\bibnamefont
  {Liu}},\ and\ \bibinfo {author} {\bibfnamefont {T.}~\bibnamefont {Li}},\
  }\href {https://arxiv.org/abs/2504.06972} {\bibinfo {title} {Signatures of
  unconventional superconductivity near reentrant and fractional quantum
  anomalous {H}all insulators}} (\bibinfo {year} {2025}),\ \Eprint
  {https://arxiv.org/abs/2504.06972} {arXiv:2504.06972 [cond-mat.mes-{H}all]}
  \BibitemShut {NoStop}%
\bibitem [{\citenamefont {Sun}\ \emph {et~al.}(2026)\citenamefont {Sun},
  \citenamefont {Xu}, \citenamefont {Li}, \citenamefont {Jiang}, \citenamefont
  {Gao}, \citenamefont {Xu}, \citenamefont {Jia}, \citenamefont {Cheng},
  \citenamefont {Zhang}, \citenamefont {Tian}, \citenamefont {Watanabe},
  \citenamefont {Taniguchi}, \citenamefont {Jia}, \citenamefont {Jiang},
  \citenamefont {Zhang}, \citenamefont {Zhang}, \citenamefont {Lei},
  \citenamefont {Liu},\ and\ \citenamefont {Li}}]{sun2026twistangle}%
  \BibitemOpen
  \bibfield  {author} {\bibinfo {author} {\bibfnamefont {Z.}~\bibnamefont
  {Sun}}, \bibinfo {author} {\bibfnamefont {F.}~\bibnamefont {Xu}}, \bibinfo
  {author} {\bibfnamefont {J.}~\bibnamefont {Li}}, \bibinfo {author}
  {\bibfnamefont {Y.}~\bibnamefont {Jiang}}, \bibinfo {author} {\bibfnamefont
  {J.}~\bibnamefont {Gao}}, \bibinfo {author} {\bibfnamefont {C.}~\bibnamefont
  {Xu}}, \bibinfo {author} {\bibfnamefont {T.}~\bibnamefont {Jia}}, \bibinfo
  {author} {\bibfnamefont {K.}~\bibnamefont {Cheng}}, \bibinfo {author}
  {\bibfnamefont {J.}~\bibnamefont {Zhang}}, \bibinfo {author} {\bibfnamefont
  {W.}~\bibnamefont {Tian}}, \bibinfo {author} {\bibfnamefont {K.}~\bibnamefont
  {Watanabe}}, \bibinfo {author} {\bibfnamefont {T.}~\bibnamefont {Taniguchi}},
  \bibinfo {author} {\bibfnamefont {J.}~\bibnamefont {Jia}}, \bibinfo {author}
  {\bibfnamefont {S.}~\bibnamefont {Jiang}}, \bibinfo {author} {\bibfnamefont
  {Y.}~\bibnamefont {Zhang}}, \bibinfo {author} {\bibfnamefont
  {Y.}~\bibnamefont {Zhang}}, \bibinfo {author} {\bibfnamefont
  {S.}~\bibnamefont {Lei}}, \bibinfo {author} {\bibfnamefont {X.}~\bibnamefont
  {Liu}},\ and\ \bibinfo {author} {\bibfnamefont {T.}~\bibnamefont {Li}},\
  }\href {https://arxiv.org/abs/2603.16412} {\bibinfo {title} {Twist-angle
  evolution from valley-polarized fractional topological phases to
  valley-degenerate superconductivity in twisted bilayer {M}o{T}e$_2$}}
  (\bibinfo {year} {2026}),\ \Eprint {https://arxiv.org/abs/2603.16412}
  {arXiv:2603.16412 [cond-mat.mes-{H}all]} \BibitemShut {NoStop}%
\bibitem [{\citenamefont {Lu}\ \emph {et~al.}(2025)\citenamefont {Lu},
  \citenamefont {Han}, \citenamefont {Yao}, \citenamefont {Hadjri},
  \citenamefont {Yang}, \citenamefont {Seo}, \citenamefont {Shi}, \citenamefont
  {Ye}, \citenamefont {Watanabe}, \citenamefont {Taniguchi},\ and\
  \citenamefont {Ju}}]{Lu2025Extended}%
  \BibitemOpen
  \bibfield  {author} {\bibinfo {author} {\bibfnamefont {Z.}~\bibnamefont
  {Lu}}, \bibinfo {author} {\bibfnamefont {T.}~\bibnamefont {Han}}, \bibinfo
  {author} {\bibfnamefont {Y.}~\bibnamefont {Yao}}, \bibinfo {author}
  {\bibfnamefont {Z.}~\bibnamefont {Hadjri}}, \bibinfo {author} {\bibfnamefont
  {J.}~\bibnamefont {Yang}}, \bibinfo {author} {\bibfnamefont {J.}~\bibnamefont
  {Seo}}, \bibinfo {author} {\bibfnamefont {L.}~\bibnamefont {Shi}}, \bibinfo
  {author} {\bibfnamefont {S.}~\bibnamefont {Ye}}, \bibinfo {author}
  {\bibfnamefont {K.}~\bibnamefont {Watanabe}}, \bibinfo {author}
  {\bibfnamefont {T.}~\bibnamefont {Taniguchi}},\ and\ \bibinfo {author}
  {\bibfnamefont {L.}~\bibnamefont {Ju}},\ }\bibfield  {title} {\bibinfo
  {title} {Extended quantum anomalous {H}all states in graphene/h{BN} moir{\'e}
  superlattices},\ }\href {https://doi.org/10.1038/s41586-024-08470-1}
  {\bibfield  {journal} {\bibinfo  {journal} {Nature}\ }\textbf {\bibinfo
  {volume} {637}},\ \bibinfo {pages} {1090} (\bibinfo {year}
  {2025})}\BibitemShut {NoStop}%
\bibitem [{\citenamefont {Aronson}\ \emph {et~al.}(2025)\citenamefont
  {Aronson}, \citenamefont {Han}, \citenamefont {Lu}, \citenamefont {Yao},
  \citenamefont {Butler}, \citenamefont {Watanabe}, \citenamefont {Taniguchi},
  \citenamefont {Ju},\ and\ \citenamefont {Ashoori}}]{Aronson2025Displacement}%
  \BibitemOpen
  \bibfield  {author} {\bibinfo {author} {\bibfnamefont {S.~H.}\ \bibnamefont
  {Aronson}}, \bibinfo {author} {\bibfnamefont {T.}~\bibnamefont {Han}},
  \bibinfo {author} {\bibfnamefont {Z.}~\bibnamefont {Lu}}, \bibinfo {author}
  {\bibfnamefont {Y.}~\bibnamefont {Yao}}, \bibinfo {author} {\bibfnamefont
  {J.~P.}\ \bibnamefont {Butler}}, \bibinfo {author} {\bibfnamefont
  {K.}~\bibnamefont {Watanabe}}, \bibinfo {author} {\bibfnamefont
  {T.}~\bibnamefont {Taniguchi}}, \bibinfo {author} {\bibfnamefont
  {L.}~\bibnamefont {Ju}},\ and\ \bibinfo {author} {\bibfnamefont {R.~C.}\
  \bibnamefont {Ashoori}},\ }\bibfield  {title} {\bibinfo {title} {Displacement
  field-controlled fractional {C}hern insulators and charge density waves in a
  graphene/h{BN} moir\'e superlattice},\ }\href
  {https://doi.org/10.1103/75gl-jzl6} {\bibfield  {journal} {\bibinfo
  {journal} {Phys. Rev. X}\ }\textbf {\bibinfo {volume} {15}},\ \bibinfo
  {pages} {031026} (\bibinfo {year} {2025})}\BibitemShut {NoStop}%
\bibitem [{\citenamefont {Roy}(2014)}]{roy2014band}%
  \BibitemOpen
  \bibfield  {author} {\bibinfo {author} {\bibfnamefont {R.}~\bibnamefont
  {Roy}},\ }\bibfield  {title} {\bibinfo {title} {Band geometry of fractional
  topological insulators},\ }\href {https://doi.org/10.1103/PhysRevB.90.165139}
  {\bibfield  {journal} {\bibinfo  {journal} {Phys. Rev. B}\ }\textbf {\bibinfo
  {volume} {90}},\ \bibinfo {pages} {165139} (\bibinfo {year}
  {2014})}\BibitemShut {NoStop}%
\bibitem [{\citenamefont {Jackson}\ \emph {et~al.}(2015)\citenamefont
  {Jackson}, \citenamefont {M{\"o}ller},\ and\ \citenamefont
  {Roy}}]{jackson2015geometric}%
  \BibitemOpen
  \bibfield  {author} {\bibinfo {author} {\bibfnamefont {T.~S.}\ \bibnamefont
  {Jackson}}, \bibinfo {author} {\bibfnamefont {G.}~\bibnamefont
  {M{\"o}ller}},\ and\ \bibinfo {author} {\bibfnamefont {R.}~\bibnamefont
  {Roy}},\ }\bibfield  {title} {\bibinfo {title} {Geometric stability of
  topological lattice phases},\ }\href {https://doi.org/10.1038/ncomms9629}
  {\bibfield  {journal} {\bibinfo  {journal} {Nat Commun}\ }\textbf {\bibinfo
  {volume} {6}},\ \bibinfo {pages} {8629} (\bibinfo {year} {2015})}\BibitemShut
  {NoStop}%
\bibitem [{\citenamefont {Parameswaran}\ \emph {et~al.}(2013)\citenamefont
  {Parameswaran}, \citenamefont {Roy},\ and\ \citenamefont
  {Sondhi}}]{Parameswaran2013Fractional}%
  \BibitemOpen
  \bibfield  {author} {\bibinfo {author} {\bibfnamefont {S.~A.}\ \bibnamefont
  {Parameswaran}}, \bibinfo {author} {\bibfnamefont {R.}~\bibnamefont {Roy}},\
  and\ \bibinfo {author} {\bibfnamefont {S.~L.}\ \bibnamefont {Sondhi}},\
  }\bibfield  {title} {\bibinfo {title} {Fractional quantum {Hall} physics in
  topological flat bands},\ }\href {https://doi.org/10.1016/j.crhy.2013.04.003}
  {\bibfield  {journal} {\bibinfo  {journal} {Comptes Rendus. Physique}\
  }\textbf {\bibinfo {volume} {14}},\ \bibinfo {pages} {816} (\bibinfo {year}
  {2013})}\BibitemShut {NoStop}%
\bibitem [{\citenamefont {Ledwith}\ \emph {et~al.}(2023)\citenamefont
  {Ledwith}, \citenamefont {Vishwanath},\ and\ \citenamefont
  {Parker}}]{ledwith2023vortexability}%
  \BibitemOpen
  \bibfield  {author} {\bibinfo {author} {\bibfnamefont {P.~J.}\ \bibnamefont
  {Ledwith}}, \bibinfo {author} {\bibfnamefont {A.}~\bibnamefont
  {Vishwanath}},\ and\ \bibinfo {author} {\bibfnamefont {D.~E.}\ \bibnamefont
  {Parker}},\ }\bibfield  {title} {\bibinfo {title} {Vortexability: {{A}}
  unifying criterion for ideal fractional {{{C}hern}} insulators},\ }\href
  {https://doi.org/10.1103/PhysRevB.108.205144} {\bibfield  {journal} {\bibinfo
   {journal} {Phys. Rev. B}\ }\textbf {\bibinfo {volume} {108}},\ \bibinfo
  {pages} {205144} (\bibinfo {year} {2023})}\BibitemShut {NoStop}%
\bibitem [{\citenamefont {Ledwith}\ \emph {et~al.}(2020)\citenamefont
  {Ledwith}, \citenamefont {Tarnopolsky}, \citenamefont {Khalaf},\ and\
  \citenamefont {Vishwanath}}]{ledwith2020fractional}%
  \BibitemOpen
  \bibfield  {author} {\bibinfo {author} {\bibfnamefont {P.~J.}\ \bibnamefont
  {Ledwith}}, \bibinfo {author} {\bibfnamefont {G.}~\bibnamefont
  {Tarnopolsky}}, \bibinfo {author} {\bibfnamefont {E.}~\bibnamefont
  {Khalaf}},\ and\ \bibinfo {author} {\bibfnamefont {A.}~\bibnamefont
  {Vishwanath}},\ }\bibfield  {title} {\bibinfo {title} {Fractional {{{C}hern}}
  insulator states in twisted bilayer graphene: {{An}} analytical approach},\
  }\href {https://doi.org/10.1103/PhysRevResearch.2.023237} {\bibfield
  {journal} {\bibinfo  {journal} {Phys. Rev. Research}\ }\textbf {\bibinfo
  {volume} {2}},\ \bibinfo {pages} {023237} (\bibinfo {year}
  {2020})}\BibitemShut {NoStop}%
\bibitem [{\citenamefont {Wang}\ \emph {et~al.}(2021)\citenamefont {Wang},
  \citenamefont {Cano}, \citenamefont {Millis}, \citenamefont {Liu},\ and\
  \citenamefont {Yang}}]{wang2021exact}%
  \BibitemOpen
  \bibfield  {author} {\bibinfo {author} {\bibfnamefont {J.}~\bibnamefont
  {Wang}}, \bibinfo {author} {\bibfnamefont {J.}~\bibnamefont {Cano}}, \bibinfo
  {author} {\bibfnamefont {A.~J.}\ \bibnamefont {Millis}}, \bibinfo {author}
  {\bibfnamefont {Z.}~\bibnamefont {Liu}},\ and\ \bibinfo {author}
  {\bibfnamefont {B.}~\bibnamefont {Yang}},\ }\bibfield  {title} {\bibinfo
  {title} {Exact {{{L}andau Level Description}} of {{Geometry}} and
  {{Interaction}} in a {{Flatband}}},\ }\href
  {https://doi.org/10.1103/PhysRevLett.127.246403} {\bibfield  {journal}
  {\bibinfo  {journal} {Phys. Rev. Lett.}\ }\textbf {\bibinfo {volume} {127}},\
  \bibinfo {pages} {246403} (\bibinfo {year} {2021})}\BibitemShut {NoStop}%
\bibitem [{\citenamefont {Cano}\ and\ \citenamefont
  {Wang}(2026)}]{cano2026ideal}%
  \BibitemOpen
  \bibfield  {author} {\bibinfo {author} {\bibfnamefont {J.}~\bibnamefont
  {Cano}}\ and\ \bibinfo {author} {\bibfnamefont {J.}~\bibnamefont {Wang}},\
  }\bibfield  {title} {\bibinfo {title} {Ideal quantum geometry for fractional
  chern insulators},\ }\href@noop {} {\bibfield  {journal} {\bibinfo  {journal}
  {arXiv preprint arXiv:2606.05496}\ } (\bibinfo {year} {2026})}\BibitemShut
  {NoStop}%
\bibitem [{\citenamefont {Valenti}\ \emph
  {et~al.}(2025{\natexlab{a}})\citenamefont {Valenti}, \citenamefont {Vituri},
  \citenamefont {Yang}, \citenamefont {Parker}, \citenamefont {Soejima},
  \citenamefont {Dong}, \citenamefont {Morales}, \citenamefont {Vishwanath},
  \citenamefont {Berg},\ and\ \citenamefont {Zhang}}]{valenti2025quantum}%
  \BibitemOpen
  \bibfield  {author} {\bibinfo {author} {\bibfnamefont {A.}~\bibnamefont
  {Valenti}}, \bibinfo {author} {\bibfnamefont {Y.}~\bibnamefont {Vituri}},
  \bibinfo {author} {\bibfnamefont {Y.}~\bibnamefont {Yang}}, \bibinfo {author}
  {\bibfnamefont {D.~E.}\ \bibnamefont {Parker}}, \bibinfo {author}
  {\bibfnamefont {T.}~\bibnamefont {Soejima}}, \bibinfo {author} {\bibfnamefont
  {J.}~\bibnamefont {Dong}}, \bibinfo {author} {\bibfnamefont {M.~A.}\
  \bibnamefont {Morales}}, \bibinfo {author} {\bibfnamefont {A.}~\bibnamefont
  {Vishwanath}}, \bibinfo {author} {\bibfnamefont {E.}~\bibnamefont {Berg}},\
  and\ \bibinfo {author} {\bibfnamefont {S.}~\bibnamefont {Zhang}},\ }\href
  {https://arxiv.org/abs/2512.07947} {\bibinfo {title} {Quantum geometry driven
  crystallization: A neural-network variational {M}onte {C}arlo study}}
  (\bibinfo {year} {2025}{\natexlab{a}}),\ \Eprint
  {https://arxiv.org/abs/2512.07947} {arXiv:2512.07947 [cond-mat.str-el]}
  \BibitemShut {NoStop}%
\bibitem [{\citenamefont {Joy}\ \emph {et~al.}(2025)\citenamefont {Joy},
  \citenamefont {Levitov},\ and\ \citenamefont
  {Skinner}}]{Joy2026ChiralWigner}%
  \BibitemOpen
  \bibfield  {author} {\bibinfo {author} {\bibfnamefont {S.}~\bibnamefont
  {Joy}}, \bibinfo {author} {\bibfnamefont {L.}~\bibnamefont {Levitov}},\ and\
  \bibinfo {author} {\bibfnamefont {B.}~\bibnamefont {Skinner}},\ }\bibfield
  {title} {\bibinfo {title} {Chiral {W}igner crystal phases induced by {B}erry
  curvature},\ }\href {https://doi.org/10.1103/h5hy-jh6m} {\bibfield  {journal}
  {\bibinfo  {journal} {Phys. Rev. Lett.}\ }\textbf {\bibinfo {volume} {135}},\
  \bibinfo {pages} {256502} (\bibinfo {year} {2025})}\BibitemShut {NoStop}%
\bibitem [{\citenamefont {Kim}(2026)}]{kim2026exchange}%
  \BibitemOpen
  \bibfield  {author} {\bibinfo {author} {\bibfnamefont {K.-S.}\ \bibnamefont
  {Kim}},\ }\bibfield  {title} {\bibinfo {title} {Exchange interactions of a
  {W}igner crystal in a magnetic field and {B}erry curvature: Multiparticle
  tunneling through complex trajectories},\ }\href
  {https://doi.org/10.1103/lj9b-3myv} {\bibfield  {journal} {\bibinfo
  {journal} {Phys. Rev. B}\ }\textbf {\bibinfo {volume} {113}},\ \bibinfo
  {pages} {144434} (\bibinfo {year} {2026})}\BibitemShut {NoStop}%
\bibitem [{\citenamefont {Zverevich}\ \emph {et~al.}(2026)\citenamefont
  {Zverevich}, \citenamefont {Levchenko},\ and\ \citenamefont
  {Esterlis}}]{Zverevich2026Spin}%
  \BibitemOpen
  \bibfield  {author} {\bibinfo {author} {\bibfnamefont {D.}~\bibnamefont
  {Zverevich}}, \bibinfo {author} {\bibfnamefont {A.}~\bibnamefont
  {Levchenko}},\ and\ \bibinfo {author} {\bibfnamefont {I.}~\bibnamefont
  {Esterlis}},\ }\bibfield  {title} {\bibinfo {title} {Spin-triplet paired
  {W}igner crystal stabilized by quantum geometry},\ }\href
  {https://doi.org/10.1103/ksxq-1l4q} {\bibfield  {journal} {\bibinfo
  {journal} {Phys. Rev. Lett.}\ }\textbf {\bibinfo {volume} {137}},\ \bibinfo
  {pages} {036506} (\bibinfo {year} {2026})}\BibitemShut {NoStop}%
\bibitem [{\citenamefont {Zhou}\ and\ \citenamefont
  {Zhang}(2025)}]{Zhou2025NewClasses}%
  \BibitemOpen
  \bibfield  {author} {\bibinfo {author} {\bibfnamefont {B.}~\bibnamefont
  {Zhou}}\ and\ \bibinfo {author} {\bibfnamefont {Y.-H.}\ \bibnamefont
  {Zhang}},\ }\bibfield  {title} {\bibinfo {title} {New classes of quantum
  anomalous {H}all crystals in multilayer graphene},\ }\href
  {https://doi.org/10.1103/26pl-gkh7} {\bibfield  {journal} {\bibinfo
  {journal} {Phys. Rev. Lett.}\ }\textbf {\bibinfo {volume} {135}},\ \bibinfo
  {pages} {036501} (\bibinfo {year} {2025})}\BibitemShut {NoStop}%
\bibitem [{\citenamefont {Dong}\ \emph
  {et~al.}(2024{\natexlab{a}})\citenamefont {Dong}, \citenamefont {Wang},
  \citenamefont {Wang}, \citenamefont {Soejima}, \citenamefont {Zaletel},
  \citenamefont {Vishwanath},\ and\ \citenamefont
  {Parker}}]{dong2024anomalous}%
  \BibitemOpen
  \bibfield  {author} {\bibinfo {author} {\bibfnamefont {J.}~\bibnamefont
  {Dong}}, \bibinfo {author} {\bibfnamefont {T.}~\bibnamefont {Wang}}, \bibinfo
  {author} {\bibfnamefont {T.}~\bibnamefont {Wang}}, \bibinfo {author}
  {\bibfnamefont {T.}~\bibnamefont {Soejima}}, \bibinfo {author} {\bibfnamefont
  {M.~P.}\ \bibnamefont {Zaletel}}, \bibinfo {author} {\bibfnamefont
  {A.}~\bibnamefont {Vishwanath}},\ and\ \bibinfo {author} {\bibfnamefont
  {D.~E.}\ \bibnamefont {Parker}},\ }\bibfield  {title} {\bibinfo {title}
  {Anomalous {H}all crystals in rhombohedral multilayer graphene. {I}.
  interaction-driven {C}hern bands and fractional quantum {H}all states at zero
  magnetic field},\ }\href@noop {} {\bibfield  {journal} {\bibinfo  {journal}
  {Physical Review Letters}\ }\textbf {\bibinfo {volume} {133}},\ \bibinfo
  {pages} {206503} (\bibinfo {year} {2024}{\natexlab{a}})}\BibitemShut
  {NoStop}%
\bibitem [{\citenamefont {Zeng}\ \emph {et~al.}(2024)\citenamefont {Zeng},
  \citenamefont {Guerci}, \citenamefont {Cr{\'e}pel}, \citenamefont {Millis},\
  and\ \citenamefont {Cano}}]{zeng2024sublattice}%
  \BibitemOpen
  \bibfield  {author} {\bibinfo {author} {\bibfnamefont {Y.}~\bibnamefont
  {Zeng}}, \bibinfo {author} {\bibfnamefont {D.}~\bibnamefont {Guerci}},
  \bibinfo {author} {\bibfnamefont {V.}~\bibnamefont {Cr{\'e}pel}}, \bibinfo
  {author} {\bibfnamefont {A.~J.}\ \bibnamefont {Millis}},\ and\ \bibinfo
  {author} {\bibfnamefont {J.}~\bibnamefont {Cano}},\ }\bibfield  {title}
  {\bibinfo {title} {Sublattice structure and topology in spontaneously
  crystallized electronic states},\ }\href@noop {} {\bibfield  {journal}
  {\bibinfo  {journal} {Physical Review Letters}\ }\textbf {\bibinfo {volume}
  {132}},\ \bibinfo {pages} {236601} (\bibinfo {year} {2024})}\BibitemShut
  {NoStop}%
\bibitem [{\citenamefont {Cr{\'e}pel}\ and\ \citenamefont
  {Cano}(2025)}]{crepel2025efficient}%
  \BibitemOpen
  \bibfield  {author} {\bibinfo {author} {\bibfnamefont {V.}~\bibnamefont
  {Cr{\'e}pel}}\ and\ \bibinfo {author} {\bibfnamefont {J.}~\bibnamefont
  {Cano}},\ }\bibfield  {title} {\bibinfo {title} {Efficient prediction of
  superlattice and anomalous miniband topology from quantum geometry},\
  }\href@noop {} {\bibfield  {journal} {\bibinfo  {journal} {Physical Review
  X}\ }\textbf {\bibinfo {volume} {15}},\ \bibinfo {pages} {011004} (\bibinfo
  {year} {2025})}\BibitemShut {NoStop}%
\bibitem [{\citenamefont {Soejima}\ \emph {et~al.}(2025)\citenamefont
  {Soejima}, \citenamefont {Dong}, \citenamefont {Vishwanath},\ and\
  \citenamefont {Parker}}]{Soejima2025Lambda}%
  \BibitemOpen
  \bibfield  {author} {\bibinfo {author} {\bibfnamefont {T.}~\bibnamefont
  {Soejima}}, \bibinfo {author} {\bibfnamefont {J.}~\bibnamefont {Dong}},
  \bibinfo {author} {\bibfnamefont {A.}~\bibnamefont {Vishwanath}},\ and\
  \bibinfo {author} {\bibfnamefont {D.~E.}\ \bibnamefont {Parker}},\ }\bibfield
   {title} {\bibinfo {title} {$\ensuremath{\lambda}$-jellium model for the
  anomalous {H}all crystal},\ }\href {https://doi.org/10.1103/x53d-12s6}
  {\bibfield  {journal} {\bibinfo  {journal} {Phys. Rev. Lett.}\ }\textbf
  {\bibinfo {volume} {135}},\ \bibinfo {pages} {186505} (\bibinfo {year}
  {2025})}\BibitemShut {NoStop}%
\bibitem [{\citenamefont {Desrochers}\ \emph
  {et~al.}(2026{\natexlab{a}})\citenamefont {Desrochers}, \citenamefont
  {Huxford}, \citenamefont {Hirsbrunner},\ and\ \citenamefont
  {Kim}}]{Desrochers2026Electronic}%
  \BibitemOpen
  \bibfield  {author} {\bibinfo {author} {\bibfnamefont {F.}~\bibnamefont
  {Desrochers}}, \bibinfo {author} {\bibfnamefont {J.}~\bibnamefont {Huxford}},
  \bibinfo {author} {\bibfnamefont {M.~R.}\ \bibnamefont {Hirsbrunner}},\ and\
  \bibinfo {author} {\bibfnamefont {Y.~B.}\ \bibnamefont {Kim}},\ }\bibfield
  {title} {\bibinfo {title} {Electronic crystal phases in the presence of
  nonuniform {B}erry curvature and tunable {B}erry flux: The
  ${\ensuremath{\lambda}}_{N}$-jellium model},\ }\href
  {https://doi.org/10.1103/5w7g-j4cf} {\bibfield  {journal} {\bibinfo
  {journal} {Phys. Rev. B}\ }\textbf {\bibinfo {volume} {113}},\ \bibinfo
  {pages} {045148} (\bibinfo {year} {2026}{\natexlab{a}})}\BibitemShut
  {NoStop}%
\bibitem [{\citenamefont {Dong}\ \emph
  {et~al.}(2024{\natexlab{b}})\citenamefont {Dong}, \citenamefont {Patri},\
  and\ \citenamefont {Senthil}}]{Dong2024Stability}%
  \BibitemOpen
  \bibfield  {author} {\bibinfo {author} {\bibfnamefont {Z.}~\bibnamefont
  {Dong}}, \bibinfo {author} {\bibfnamefont {A.~S.}\ \bibnamefont {Patri}},\
  and\ \bibinfo {author} {\bibfnamefont {T.}~\bibnamefont {Senthil}},\
  }\bibfield  {title} {\bibinfo {title} {Stability of anomalous {H}all crystals
  in multilayer rhombohedral graphene},\ }\href
  {https://doi.org/10.1103/PhysRevB.110.205130} {\bibfield  {journal} {\bibinfo
   {journal} {Phys. Rev. B}\ }\textbf {\bibinfo {volume} {110}},\ \bibinfo
  {pages} {205130} (\bibinfo {year} {2024}{\natexlab{b}})}\BibitemShut
  {NoStop}%
\bibitem [{\citenamefont {Patri}\ \emph {et~al.}(2024)\citenamefont {Patri},
  \citenamefont {Dong},\ and\ \citenamefont {Senthil}}]{Patri2024Extended}%
  \BibitemOpen
  \bibfield  {author} {\bibinfo {author} {\bibfnamefont {A.~S.}\ \bibnamefont
  {Patri}}, \bibinfo {author} {\bibfnamefont {Z.}~\bibnamefont {Dong}},\ and\
  \bibinfo {author} {\bibfnamefont {T.}~\bibnamefont {Senthil}},\ }\bibfield
  {title} {\bibinfo {title} {Extended quantum anomalous {H}all effect in
  moir\'e structures: Phase transitions and transport},\ }\href
  {https://doi.org/10.1103/PhysRevB.110.245115} {\bibfield  {journal} {\bibinfo
   {journal} {Phys. Rev. B}\ }\textbf {\bibinfo {volume} {110}},\ \bibinfo
  {pages} {245115} (\bibinfo {year} {2024})}\BibitemShut {NoStop}%
\bibitem [{\citenamefont {Tan}\ and\ \citenamefont
  {Devakul}(2024)}]{Tan2024Parent}%
  \BibitemOpen
  \bibfield  {author} {\bibinfo {author} {\bibfnamefont {T.}~\bibnamefont
  {Tan}}\ and\ \bibinfo {author} {\bibfnamefont {T.}~\bibnamefont {Devakul}},\
  }\bibfield  {title} {\bibinfo {title} {Parent {B}erry curvature and the ideal
  anomalous {H}all crystal},\ }\href
  {https://doi.org/10.1103/PhysRevX.14.041040} {\bibfield  {journal} {\bibinfo
  {journal} {Phys. Rev. X}\ }\textbf {\bibinfo {volume} {14}},\ \bibinfo
  {pages} {041040} (\bibinfo {year} {2024})}\BibitemShut {NoStop}%
\bibitem [{\citenamefont {Tan}\ \emph {et~al.}(2025{\natexlab{a}})\citenamefont
  {Tan}, \citenamefont {May-Mann},\ and\ \citenamefont
  {Devakul}}]{Tan2025Variational}%
  \BibitemOpen
  \bibfield  {author} {\bibinfo {author} {\bibfnamefont {T.}~\bibnamefont
  {Tan}}, \bibinfo {author} {\bibfnamefont {J.}~\bibnamefont {May-Mann}},\ and\
  \bibinfo {author} {\bibfnamefont {T.}~\bibnamefont {Devakul}},\ }\bibfield
  {title} {\bibinfo {title} {Variational wave-function analysis of the
  fractional anomalous {H}all crystal},\ }\href
  {https://doi.org/10.1103/dd2d-kk3w} {\bibfield  {journal} {\bibinfo
  {journal} {Phys. Rev. Lett.}\ }\textbf {\bibinfo {volume} {135}},\ \bibinfo
  {pages} {036604} (\bibinfo {year} {2025}{\natexlab{a}})}\BibitemShut
  {NoStop}%
\bibitem [{\citenamefont {Soejima}\ \emph {et~al.}(2024)\citenamefont
  {Soejima}, \citenamefont {Dong}, \citenamefont {Wang}, \citenamefont {Wang},
  \citenamefont {Zaletel}, \citenamefont {Vishwanath},\ and\ \citenamefont
  {Parker}}]{Soejima2024Anomalous}%
  \BibitemOpen
  \bibfield  {author} {\bibinfo {author} {\bibfnamefont {T.}~\bibnamefont
  {Soejima}}, \bibinfo {author} {\bibfnamefont {J.}~\bibnamefont {Dong}},
  \bibinfo {author} {\bibfnamefont {T.}~\bibnamefont {Wang}}, \bibinfo {author}
  {\bibfnamefont {T.}~\bibnamefont {Wang}}, \bibinfo {author} {\bibfnamefont
  {M.~P.}\ \bibnamefont {Zaletel}}, \bibinfo {author} {\bibfnamefont
  {A.}~\bibnamefont {Vishwanath}},\ and\ \bibinfo {author} {\bibfnamefont
  {D.~E.}\ \bibnamefont {Parker}},\ }\bibfield  {title} {\bibinfo {title}
  {Anomalous {H}all crystals in rhombohedral multilayer graphene. {II}. general
  mechanism and a minimal model},\ }\href
  {https://doi.org/10.1103/PhysRevB.110.205124} {\bibfield  {journal} {\bibinfo
   {journal} {Phys. Rev. B}\ }\textbf {\bibinfo {volume} {110}},\ \bibinfo
  {pages} {205124} (\bibinfo {year} {2024})}\BibitemShut {NoStop}%
\bibitem [{\citenamefont {Zhou}\ \emph {et~al.}(2024)\citenamefont {Zhou},
  \citenamefont {Yang},\ and\ \citenamefont {Zhang}}]{Zhou2024Fractional}%
  \BibitemOpen
  \bibfield  {author} {\bibinfo {author} {\bibfnamefont {B.}~\bibnamefont
  {Zhou}}, \bibinfo {author} {\bibfnamefont {H.}~\bibnamefont {Yang}},\ and\
  \bibinfo {author} {\bibfnamefont {Y.-H.}\ \bibnamefont {Zhang}},\ }\bibfield
  {title} {\bibinfo {title} {Fractional quantum anomalous {H}all effect in
  rhombohedral multilayer graphene in the moir\'eless limit},\ }\href
  {https://doi.org/10.1103/PhysRevLett.133.206504} {\bibfield  {journal}
  {\bibinfo  {journal} {Phys. Rev. Lett.}\ }\textbf {\bibinfo {volume} {133}},\
  \bibinfo {pages} {206504} (\bibinfo {year} {2024})}\BibitemShut {NoStop}%
\bibitem [{\citenamefont {Su}\ \emph {et~al.}(2025)\citenamefont {Su},
  \citenamefont {Waters}, \citenamefont {Zhou}, \citenamefont {Watanabe},
  \citenamefont {Taniguchi}, \citenamefont {Zhang}, \citenamefont {Yankowitz},\
  and\ \citenamefont {Folk}}]{Su2025Moire}%
  \BibitemOpen
  \bibfield  {author} {\bibinfo {author} {\bibfnamefont {R.}~\bibnamefont
  {Su}}, \bibinfo {author} {\bibfnamefont {D.}~\bibnamefont {Waters}}, \bibinfo
  {author} {\bibfnamefont {B.}~\bibnamefont {Zhou}}, \bibinfo {author}
  {\bibfnamefont {K.}~\bibnamefont {Watanabe}}, \bibinfo {author}
  {\bibfnamefont {T.}~\bibnamefont {Taniguchi}}, \bibinfo {author}
  {\bibfnamefont {Y.-H.}\ \bibnamefont {Zhang}}, \bibinfo {author}
  {\bibfnamefont {M.}~\bibnamefont {Yankowitz}},\ and\ \bibinfo {author}
  {\bibfnamefont {J.}~\bibnamefont {Folk}},\ }\bibfield  {title} {\bibinfo
  {title} {Moir{\'e}-driven topological electronic crystals in twisted
  graphene},\ }\href {https://doi.org/10.1038/s41586-024-08239-6} {\bibfield
  {journal} {\bibinfo  {journal} {Nature}\ }\textbf {\bibinfo {volume} {637}},\
  \bibinfo {pages} {1084} (\bibinfo {year} {2025})}\BibitemShut {NoStop}%
\bibitem [{\citenamefont {Drummond}\ and\ \citenamefont
  {Needs}(2009)}]{Drummond2009WC}%
  \BibitemOpen
  \bibfield  {author} {\bibinfo {author} {\bibfnamefont {N.~D.}\ \bibnamefont
  {Drummond}}\ and\ \bibinfo {author} {\bibfnamefont {R.~J.}\ \bibnamefont
  {Needs}},\ }\bibfield  {title} {\bibinfo {title} {Phase diagram of the
  low-density two-dimensional homogeneous electron gas},\ }\href
  {https://doi.org/10.1103/PhysRevLett.102.126402} {\bibfield  {journal}
  {\bibinfo  {journal} {Phys. Rev. Lett.}\ }\textbf {\bibinfo {volume} {102}},\
  \bibinfo {pages} {126402} (\bibinfo {year} {2009})}\BibitemShut {NoStop}%
\bibitem [{\citenamefont {Smith}\ \emph {et~al.}(2024)\citenamefont {Smith},
  \citenamefont {Chen}, \citenamefont {Levy}, \citenamefont {Yang},
  \citenamefont {Morales},\ and\ \citenamefont {Zhang}}]{Smith2024WC}%
  \BibitemOpen
  \bibfield  {author} {\bibinfo {author} {\bibfnamefont {C.}~\bibnamefont
  {Smith}}, \bibinfo {author} {\bibfnamefont {Y.}~\bibnamefont {Chen}},
  \bibinfo {author} {\bibfnamefont {R.}~\bibnamefont {Levy}}, \bibinfo {author}
  {\bibfnamefont {Y.}~\bibnamefont {Yang}}, \bibinfo {author} {\bibfnamefont
  {M.~A.}\ \bibnamefont {Morales}},\ and\ \bibinfo {author} {\bibfnamefont
  {S.}~\bibnamefont {Zhang}},\ }\bibfield  {title} {\bibinfo {title} {Unified
  variational approach description of ground-state phases of the
  two-dimensional electron gas},\ }\href
  {https://doi.org/10.1103/PhysRevLett.133.266504} {\bibfield  {journal}
  {\bibinfo  {journal} {Phys. Rev. Lett.}\ }\textbf {\bibinfo {volume} {133}},\
  \bibinfo {pages} {266504} (\bibinfo {year} {2024})}\BibitemShut {NoStop}%
\bibitem [{\citenamefont {Kim}\ \emph {et~al.}(2022)\citenamefont {Kim},
  \citenamefont {Murthy}, \citenamefont {Pandey},\ and\ \citenamefont
  {Kivelson}}]{Kim2022Interstitials}%
  \BibitemOpen
  \bibfield  {author} {\bibinfo {author} {\bibfnamefont {K.-S.}\ \bibnamefont
  {Kim}}, \bibinfo {author} {\bibfnamefont {C.}~\bibnamefont {Murthy}},
  \bibinfo {author} {\bibfnamefont {A.}~\bibnamefont {Pandey}},\ and\ \bibinfo
  {author} {\bibfnamefont {S.~A.}\ \bibnamefont {Kivelson}},\ }\bibfield
  {title} {\bibinfo {title} {Interstitial-induced ferromagnetism in a
  two-dimensional {W}igner crystal},\ }\href
  {https://doi.org/10.1103/PhysRevLett.129.227202} {\bibfield  {journal}
  {\bibinfo  {journal} {Phys. Rev. Lett.}\ }\textbf {\bibinfo {volume} {129}},\
  \bibinfo {pages} {227202} (\bibinfo {year} {2022})}\BibitemShut {NoStop}%
\bibitem [{\citenamefont {Kim}\ \emph {et~al.}(2024)\citenamefont {Kim},
  \citenamefont {Esterlis}, \citenamefont {Murthy},\ and\ \citenamefont
  {Kivelson}}]{Kim2024Dynamical}%
  \BibitemOpen
  \bibfield  {author} {\bibinfo {author} {\bibfnamefont {K.-S.}\ \bibnamefont
  {Kim}}, \bibinfo {author} {\bibfnamefont {I.}~\bibnamefont {Esterlis}},
  \bibinfo {author} {\bibfnamefont {C.}~\bibnamefont {Murthy}},\ and\ \bibinfo
  {author} {\bibfnamefont {S.~A.}\ \bibnamefont {Kivelson}},\ }\bibfield
  {title} {\bibinfo {title} {Dynamical defects in a two-dimensional {W}igner
  crystal: Self-doping and kinetic magnetism},\ }\href
  {https://doi.org/10.1103/PhysRevB.109.235130} {\bibfield  {journal} {\bibinfo
   {journal} {Phys. Rev. B}\ }\textbf {\bibinfo {volume} {109}},\ \bibinfo
  {pages} {235130} (\bibinfo {year} {2024})}\BibitemShut {NoStop}%
\bibitem [{\citenamefont {Valenti}\ \emph
  {et~al.}(2025{\natexlab{b}})\citenamefont {Valenti}, \citenamefont {Calvera},
  \citenamefont {Yang}, \citenamefont {Morales}, \citenamefont {Kivelson},
  \citenamefont {Esterlis},\ and\ \citenamefont {Zhang}}]{Valenti2025Gate}%
  \BibitemOpen
  \bibfield  {author} {\bibinfo {author} {\bibfnamefont {A.}~\bibnamefont
  {Valenti}}, \bibinfo {author} {\bibfnamefont {V.}~\bibnamefont {Calvera}},
  \bibinfo {author} {\bibfnamefont {Y.}~\bibnamefont {Yang}}, \bibinfo {author}
  {\bibfnamefont {M.~A.}\ \bibnamefont {Morales}}, \bibinfo {author}
  {\bibfnamefont {S.~A.}\ \bibnamefont {Kivelson}}, \bibinfo {author}
  {\bibfnamefont {I.}~\bibnamefont {Esterlis}},\ and\ \bibinfo {author}
  {\bibfnamefont {S.}~\bibnamefont {Zhang}},\ }\bibfield  {title} {\bibinfo
  {title} {Critical gate distance for {W}igner crystallization in the
  two-dimensional electron gas},\ }\href {https://doi.org/10.1103/2qgp-v27h}
  {\bibfield  {journal} {\bibinfo  {journal} {Phys. Rev. Lett.}\ }\textbf
  {\bibinfo {volume} {135}},\ \bibinfo {pages} {166501} (\bibinfo {year}
  {2025}{\natexlab{b}})}\BibitemShut {NoStop}%
\bibitem [{\citenamefont {Esterlis}\ \emph {et~al.}(2025)\citenamefont
  {Esterlis}, \citenamefont {Zverevich}, \citenamefont {Zhuang},\ and\
  \citenamefont {Levchenko}}]{Esterlis2025Bilayer}%
  \BibitemOpen
  \bibfield  {author} {\bibinfo {author} {\bibfnamefont {I.}~\bibnamefont
  {Esterlis}}, \bibinfo {author} {\bibfnamefont {D.}~\bibnamefont {Zverevich}},
  \bibinfo {author} {\bibfnamefont {Z.}~\bibnamefont {Zhuang}},\ and\ \bibinfo
  {author} {\bibfnamefont {A.}~\bibnamefont {Levchenko}},\ }\bibfield  {title}
  {\bibinfo {title} {Magnetism of the bilayer {W}igner crystal},\ }\href
  {https://doi.org/10.1103/PhysRevB.111.075159} {\bibfield  {journal} {\bibinfo
   {journal} {Phys. Rev. B}\ }\textbf {\bibinfo {volume} {111}},\ \bibinfo
  {pages} {075159} (\bibinfo {year} {2025})}\BibitemShut {NoStop}%
\bibitem [{\citenamefont {Esterlis}\ and\ \citenamefont
  {Levchenko}(2025)}]{Esterlis2025MagnetismMoire}%
  \BibitemOpen
  \bibfield  {author} {\bibinfo {author} {\bibfnamefont {I.}~\bibnamefont
  {Esterlis}}\ and\ \bibinfo {author} {\bibfnamefont {A.}~\bibnamefont
  {Levchenko}},\ }\bibfield  {title} {\bibinfo {title} {Magnetism from
  multiparticle ring exchange in moir\'e {W}igner crystals},\ }\href
  {https://doi.org/10.1103/PhysRevB.111.L201115} {\bibfield  {journal}
  {\bibinfo  {journal} {Phys. Rev. B}\ }\textbf {\bibinfo {volume} {111}},\
  \bibinfo {pages} {L201115} (\bibinfo {year} {2025})}\BibitemShut {NoStop}%
\bibitem [{\citenamefont {Zhou}\ \emph {et~al.}(2021)\citenamefont {Zhou},
  \citenamefont {Sung}, \citenamefont {Brutschea}, \citenamefont {Esterlis},
  \citenamefont {Wang}, \citenamefont {Scuri}, \citenamefont {Gelly},
  \citenamefont {Heo}, \citenamefont {Taniguchi}, \citenamefont {Watanabe},
  \citenamefont {Zar{\'a}nd}, \citenamefont {Lukin}, \citenamefont {Kim},
  \citenamefont {Demler},\ and\ \citenamefont {Park}}]{Zhou2021BilayerCrystal}%
  \BibitemOpen
  \bibfield  {author} {\bibinfo {author} {\bibfnamefont {Y.}~\bibnamefont
  {Zhou}}, \bibinfo {author} {\bibfnamefont {J.}~\bibnamefont {Sung}}, \bibinfo
  {author} {\bibfnamefont {E.}~\bibnamefont {Brutschea}}, \bibinfo {author}
  {\bibfnamefont {I.}~\bibnamefont {Esterlis}}, \bibinfo {author}
  {\bibfnamefont {Y.}~\bibnamefont {Wang}}, \bibinfo {author} {\bibfnamefont
  {G.}~\bibnamefont {Scuri}}, \bibinfo {author} {\bibfnamefont {R.~J.}\
  \bibnamefont {Gelly}}, \bibinfo {author} {\bibfnamefont {H.}~\bibnamefont
  {Heo}}, \bibinfo {author} {\bibfnamefont {T.}~\bibnamefont {Taniguchi}},
  \bibinfo {author} {\bibfnamefont {K.}~\bibnamefont {Watanabe}}, \bibinfo
  {author} {\bibfnamefont {G.}~\bibnamefont {Zar{\'a}nd}}, \bibinfo {author}
  {\bibfnamefont {M.~D.}\ \bibnamefont {Lukin}}, \bibinfo {author}
  {\bibfnamefont {P.}~\bibnamefont {Kim}}, \bibinfo {author} {\bibfnamefont
  {E.}~\bibnamefont {Demler}},\ and\ \bibinfo {author} {\bibfnamefont
  {H.}~\bibnamefont {Park}},\ }\bibfield  {title} {\bibinfo {title} {Bilayer
  {W}igner crystals in a transition metal dichalcogenide heterostructure},\
  }\href {https://doi.org/10.1038/s41586-021-03560-w} {\bibfield  {journal}
  {\bibinfo  {journal} {Nature}\ }\textbf {\bibinfo {volume} {595}},\ \bibinfo
  {pages} {48} (\bibinfo {year} {2021})}\BibitemShut {NoStop}%
\bibitem [{\citenamefont {Smole{\'{n}}ski}\ \emph {et~al.}(2021)\citenamefont
  {Smole{\'{n}}ski}, \citenamefont {Dolgirev}, \citenamefont {Kuhlenkamp},
  \citenamefont {Popert}, \citenamefont {Shimazaki}, \citenamefont {Back},
  \citenamefont {Lu}, \citenamefont {Kroner}, \citenamefont {Watanabe},
  \citenamefont {Taniguchi}, \citenamefont {Esterlis}, \citenamefont {Demler},\
  and\ \citenamefont {Imamo{\u{g}}lu}}]{Smolenski2021WignerCrystal}%
  \BibitemOpen
  \bibfield  {author} {\bibinfo {author} {\bibfnamefont {T.}~\bibnamefont
  {Smole{\'{n}}ski}}, \bibinfo {author} {\bibfnamefont {P.~E.}\ \bibnamefont
  {Dolgirev}}, \bibinfo {author} {\bibfnamefont {C.}~\bibnamefont
  {Kuhlenkamp}}, \bibinfo {author} {\bibfnamefont {A.}~\bibnamefont {Popert}},
  \bibinfo {author} {\bibfnamefont {Y.}~\bibnamefont {Shimazaki}}, \bibinfo
  {author} {\bibfnamefont {P.}~\bibnamefont {Back}}, \bibinfo {author}
  {\bibfnamefont {X.}~\bibnamefont {Lu}}, \bibinfo {author} {\bibfnamefont
  {M.}~\bibnamefont {Kroner}}, \bibinfo {author} {\bibfnamefont
  {K.}~\bibnamefont {Watanabe}}, \bibinfo {author} {\bibfnamefont
  {T.}~\bibnamefont {Taniguchi}}, \bibinfo {author} {\bibfnamefont
  {I.}~\bibnamefont {Esterlis}}, \bibinfo {author} {\bibfnamefont
  {E.}~\bibnamefont {Demler}},\ and\ \bibinfo {author} {\bibfnamefont
  {A.}~\bibnamefont {Imamo{\u{g}}lu}},\ }\bibfield  {title} {\bibinfo {title}
  {Signatures of {W}igner crystal of electrons in a monolayer semiconductor},\
  }\href {https://doi.org/10.1038/s41586-021-03590-4} {\bibfield  {journal}
  {\bibinfo  {journal} {Nature}\ }\textbf {\bibinfo {volume} {595}},\ \bibinfo
  {pages} {53} (\bibinfo {year} {2021})}\BibitemShut {NoStop}%
\bibitem [{\citenamefont {Sung}\ \emph {et~al.}(2025)\citenamefont {Sung},
  \citenamefont {Wang}, \citenamefont {Esterlis}, \citenamefont {Volkov},
  \citenamefont {Scuri}, \citenamefont {Zhou}, \citenamefont {Brutschea},
  \citenamefont {Taniguchi}, \citenamefont {Watanabe}, \citenamefont {Yang},
  \citenamefont {Morales}, \citenamefont {Zhang}, \citenamefont {Millis},
  \citenamefont {Lukin}, \citenamefont {Kim}, \citenamefont {Demler},\ and\
  \citenamefont {Park}}]{Sung2025Electronic}%
  \BibitemOpen
  \bibfield  {author} {\bibinfo {author} {\bibfnamefont {J.}~\bibnamefont
  {Sung}}, \bibinfo {author} {\bibfnamefont {J.}~\bibnamefont {Wang}}, \bibinfo
  {author} {\bibfnamefont {I.}~\bibnamefont {Esterlis}}, \bibinfo {author}
  {\bibfnamefont {P.~A.}\ \bibnamefont {Volkov}}, \bibinfo {author}
  {\bibfnamefont {G.}~\bibnamefont {Scuri}}, \bibinfo {author} {\bibfnamefont
  {Y.}~\bibnamefont {Zhou}}, \bibinfo {author} {\bibfnamefont {E.}~\bibnamefont
  {Brutschea}}, \bibinfo {author} {\bibfnamefont {T.}~\bibnamefont
  {Taniguchi}}, \bibinfo {author} {\bibfnamefont {K.}~\bibnamefont {Watanabe}},
  \bibinfo {author} {\bibfnamefont {Y.}~\bibnamefont {Yang}}, \bibinfo {author}
  {\bibfnamefont {M.~A.}\ \bibnamefont {Morales}}, \bibinfo {author}
  {\bibfnamefont {S.}~\bibnamefont {Zhang}}, \bibinfo {author} {\bibfnamefont
  {A.~J.}\ \bibnamefont {Millis}}, \bibinfo {author} {\bibfnamefont {M.~D.}\
  \bibnamefont {Lukin}}, \bibinfo {author} {\bibfnamefont {P.}~\bibnamefont
  {Kim}}, \bibinfo {author} {\bibfnamefont {E.}~\bibnamefont {Demler}},\ and\
  \bibinfo {author} {\bibfnamefont {H.}~\bibnamefont {Park}},\ }\bibfield
  {title} {\bibinfo {title} {An electronic microemulsion phase emerging from a
  quantum crystal-to-liquid transition},\ }\href
  {https://doi.org/10.1038/s41567-024-02759-8} {\bibfield  {journal} {\bibinfo
  {journal} {Nature Physics}\ }\textbf {\bibinfo {volume} {21}},\ \bibinfo
  {pages} {437} (\bibinfo {year} {2025})}\BibitemShut {NoStop}%
\bibitem [{\citenamefont {Xu}\ \emph {et~al.}(2020)\citenamefont {Xu},
  \citenamefont {Liu}, \citenamefont {Rhodes}, \citenamefont {Watanabe},
  \citenamefont {Taniguchi}, \citenamefont {Hone}, \citenamefont {Elser},
  \citenamefont {Mak},\ and\ \citenamefont {Shan}}]{CornellWigner}%
  \BibitemOpen
  \bibfield  {author} {\bibinfo {author} {\bibfnamefont {Y.}~\bibnamefont
  {Xu}}, \bibinfo {author} {\bibfnamefont {S.}~\bibnamefont {Liu}}, \bibinfo
  {author} {\bibfnamefont {D.~A.}\ \bibnamefont {Rhodes}}, \bibinfo {author}
  {\bibfnamefont {K.}~\bibnamefont {Watanabe}}, \bibinfo {author}
  {\bibfnamefont {T.}~\bibnamefont {Taniguchi}}, \bibinfo {author}
  {\bibfnamefont {J.}~\bibnamefont {Hone}}, \bibinfo {author} {\bibfnamefont
  {V.}~\bibnamefont {Elser}}, \bibinfo {author} {\bibfnamefont {K.~F.}\
  \bibnamefont {Mak}},\ and\ \bibinfo {author} {\bibfnamefont {J.}~\bibnamefont
  {Shan}},\ }\bibfield  {title} {\bibinfo {title} {Correlated insulating states
  at fractional fillings of moir{\'e} superlattices},\ }\href
  {https://doi.org/10.1038/s41586-020-2868-6} {\bibfield  {journal} {\bibinfo
  {journal} {Nature}\ }\textbf {\bibinfo {volume} {587}},\ \bibinfo {pages}
  {214} (\bibinfo {year} {2020})}\BibitemShut {NoStop}%
\bibitem [{\citenamefont {Huang}\ \emph {et~al.}(2021)\citenamefont {Huang},
  \citenamefont {Wang}, \citenamefont {Miao}, \citenamefont {Wang},
  \citenamefont {Li}, \citenamefont {Lian}, \citenamefont {Taniguchi},
  \citenamefont {Watanabe}, \citenamefont {Okamoto}, \citenamefont {Xiao},
  \citenamefont {Shi},\ and\ \citenamefont {Cui}}]{CaliforniaWigner}%
  \BibitemOpen
  \bibfield  {author} {\bibinfo {author} {\bibfnamefont {X.}~\bibnamefont
  {Huang}}, \bibinfo {author} {\bibfnamefont {T.}~\bibnamefont {Wang}},
  \bibinfo {author} {\bibfnamefont {S.}~\bibnamefont {Miao}}, \bibinfo {author}
  {\bibfnamefont {C.}~\bibnamefont {Wang}}, \bibinfo {author} {\bibfnamefont
  {Z.}~\bibnamefont {Li}}, \bibinfo {author} {\bibfnamefont {Z.}~\bibnamefont
  {Lian}}, \bibinfo {author} {\bibfnamefont {T.}~\bibnamefont {Taniguchi}},
  \bibinfo {author} {\bibfnamefont {K.}~\bibnamefont {Watanabe}}, \bibinfo
  {author} {\bibfnamefont {S.}~\bibnamefont {Okamoto}}, \bibinfo {author}
  {\bibfnamefont {D.}~\bibnamefont {Xiao}}, \bibinfo {author} {\bibfnamefont
  {S.-F.}\ \bibnamefont {Shi}},\ and\ \bibinfo {author} {\bibfnamefont {Y.-T.}\
  \bibnamefont {Cui}},\ }\bibfield  {title} {\bibinfo {title} {Correlated
  insulating states at fractional fillings of the {WS}$_2$/{WS}e$_2$ moir{\'e}
  lattice},\ }\bibfield  {journal} {\bibinfo  {journal} {Nature Physics}\
  }\href {https://doi.org/10.1038/s41567-021-01171-w}
  {10.1038/s41567-021-01171-w} (\bibinfo {year} {2021})\BibitemShut {NoStop}%
\bibitem [{\citenamefont {Regan}\ \emph {et~al.}(2020)\citenamefont {Regan},
  \citenamefont {Wang}, \citenamefont {Jin}, \citenamefont {Bakti~Utama},
  \citenamefont {Gao}, \citenamefont {Wei}, \citenamefont {Zhao}, \citenamefont
  {Zhao}, \citenamefont {Zhang}, \citenamefont {Yumigeta}, \citenamefont
  {Blei}, \citenamefont {Carlstr{\"o}m}, \citenamefont {Watanabe},
  \citenamefont {Taniguchi}, \citenamefont {Tongay}, \citenamefont {Crommie},
  \citenamefont {Zettl},\ and\ \citenamefont {Wang}}]{Berkeley}%
  \BibitemOpen
  \bibfield  {author} {\bibinfo {author} {\bibfnamefont {E.~C.}\ \bibnamefont
  {Regan}}, \bibinfo {author} {\bibfnamefont {D.}~\bibnamefont {Wang}},
  \bibinfo {author} {\bibfnamefont {C.}~\bibnamefont {Jin}}, \bibinfo {author}
  {\bibfnamefont {M.~I.}\ \bibnamefont {Bakti~Utama}}, \bibinfo {author}
  {\bibfnamefont {B.}~\bibnamefont {Gao}}, \bibinfo {author} {\bibfnamefont
  {X.}~\bibnamefont {Wei}}, \bibinfo {author} {\bibfnamefont {S.}~\bibnamefont
  {Zhao}}, \bibinfo {author} {\bibfnamefont {W.}~\bibnamefont {Zhao}}, \bibinfo
  {author} {\bibfnamefont {Z.}~\bibnamefont {Zhang}}, \bibinfo {author}
  {\bibfnamefont {K.}~\bibnamefont {Yumigeta}}, \bibinfo {author}
  {\bibfnamefont {M.}~\bibnamefont {Blei}}, \bibinfo {author} {\bibfnamefont
  {J.~D.}\ \bibnamefont {Carlstr{\"o}m}}, \bibinfo {author} {\bibfnamefont
  {K.}~\bibnamefont {Watanabe}}, \bibinfo {author} {\bibfnamefont
  {T.}~\bibnamefont {Taniguchi}}, \bibinfo {author} {\bibfnamefont
  {S.}~\bibnamefont {Tongay}}, \bibinfo {author} {\bibfnamefont
  {M.}~\bibnamefont {Crommie}}, \bibinfo {author} {\bibfnamefont
  {A.}~\bibnamefont {Zettl}},\ and\ \bibinfo {author} {\bibfnamefont
  {F.}~\bibnamefont {Wang}},\ }\bibfield  {title} {\bibinfo {title} {Mott and
  generalized {W}igner crystal states in {WS}e$_2$/{WS}$_2$ moir{\'e}
  superlattices},\ }\href {https://doi.org/10.1038/s41586-020-2092-4}
  {\bibfield  {journal} {\bibinfo  {journal} {Nature}\ }\textbf {\bibinfo
  {volume} {579}},\ \bibinfo {pages} {359} (\bibinfo {year}
  {2020})}\BibitemShut {NoStop}%
\bibitem [{\citenamefont {Jin}\ \emph {et~al.}(2021)\citenamefont {Jin},
  \citenamefont {Tao}, \citenamefont {Li}, \citenamefont {Xu}, \citenamefont
  {Tang}, \citenamefont {Zhu}, \citenamefont {Liu}, \citenamefont {Watanabe},
  \citenamefont {Taniguchi}, \citenamefont {Hone}, \citenamefont {Fu},
  \citenamefont {Shan},\ and\ \citenamefont {Mak}}]{CornellWignerStripe}%
  \BibitemOpen
  \bibfield  {author} {\bibinfo {author} {\bibfnamefont {C.}~\bibnamefont
  {Jin}}, \bibinfo {author} {\bibfnamefont {Z.}~\bibnamefont {Tao}}, \bibinfo
  {author} {\bibfnamefont {T.}~\bibnamefont {Li}}, \bibinfo {author}
  {\bibfnamefont {Y.}~\bibnamefont {Xu}}, \bibinfo {author} {\bibfnamefont
  {Y.}~\bibnamefont {Tang}}, \bibinfo {author} {\bibfnamefont {J.}~\bibnamefont
  {Zhu}}, \bibinfo {author} {\bibfnamefont {S.}~\bibnamefont {Liu}}, \bibinfo
  {author} {\bibfnamefont {K.}~\bibnamefont {Watanabe}}, \bibinfo {author}
  {\bibfnamefont {T.}~\bibnamefont {Taniguchi}}, \bibinfo {author}
  {\bibfnamefont {J.~C.}\ \bibnamefont {Hone}}, \bibinfo {author}
  {\bibfnamefont {L.}~\bibnamefont {Fu}}, \bibinfo {author} {\bibfnamefont
  {J.}~\bibnamefont {Shan}},\ and\ \bibinfo {author} {\bibfnamefont {K.~F.}\
  \bibnamefont {Mak}},\ }\bibfield  {title} {\bibinfo {title} {Stripe phases in
  {WS}e$_2$/{WS}$_2$ moir{\'e} superlattices},\ }\bibfield  {journal} {\bibinfo
   {journal} {Nature Materials}\ }\href
  {https://doi.org/10.1038/s41563-021-00959-8} {10.1038/s41563-021-00959-8}
  (\bibinfo {year} {2021})\BibitemShut {NoStop}%
\bibitem [{\citenamefont {Li}\ \emph {et~al.}(2021)\citenamefont {Li},
  \citenamefont {Li}, \citenamefont {Regan}, \citenamefont {Wang},
  \citenamefont {Zhao}, \citenamefont {Kahn}, \citenamefont {Yumigeta},
  \citenamefont {Blei}, \citenamefont {Taniguchi}, \citenamefont {Watanabe},
  \citenamefont {Tongay}, \citenamefont {Zettl}, \citenamefont {Crommie},\ and\
  \citenamefont {Wang}}]{STM_Wigner}%
  \BibitemOpen
  \bibfield  {author} {\bibinfo {author} {\bibfnamefont {H.}~\bibnamefont
  {Li}}, \bibinfo {author} {\bibfnamefont {S.}~\bibnamefont {Li}}, \bibinfo
  {author} {\bibfnamefont {E.~C.}\ \bibnamefont {Regan}}, \bibinfo {author}
  {\bibfnamefont {D.}~\bibnamefont {Wang}}, \bibinfo {author} {\bibfnamefont
  {W.}~\bibnamefont {Zhao}}, \bibinfo {author} {\bibfnamefont {S.}~\bibnamefont
  {Kahn}}, \bibinfo {author} {\bibfnamefont {K.}~\bibnamefont {Yumigeta}},
  \bibinfo {author} {\bibfnamefont {M.}~\bibnamefont {Blei}}, \bibinfo {author}
  {\bibfnamefont {T.}~\bibnamefont {Taniguchi}}, \bibinfo {author}
  {\bibfnamefont {K.}~\bibnamefont {Watanabe}}, \bibinfo {author}
  {\bibfnamefont {S.}~\bibnamefont {Tongay}}, \bibinfo {author} {\bibfnamefont
  {A.}~\bibnamefont {Zettl}}, \bibinfo {author} {\bibfnamefont {M.~F.}\
  \bibnamefont {Crommie}},\ and\ \bibinfo {author} {\bibfnamefont
  {F.}~\bibnamefont {Wang}},\ }\bibfield  {title} {\bibinfo {title} {Imaging
  two-dimensional generalized {W}igner crystals},\ }\href
  {https://doi.org/10.1038/s41586-021-03874-9} {\bibfield  {journal} {\bibinfo
  {journal} {Nature}\ }\textbf {\bibinfo {volume} {597}},\ \bibinfo {pages}
  {650} (\bibinfo {year} {2021})}\BibitemShut {NoStop}%
\bibitem [{\citenamefont {Tang}\ \emph {et~al.}(2022)\citenamefont {Tang},
  \citenamefont {Gu}, \citenamefont {Liu}, \citenamefont {Watanabe},
  \citenamefont {Taniguchi}, \citenamefont {Hone}, \citenamefont {Mak},\ and\
  \citenamefont {Shan}}]{ContinuousWigner}%
  \BibitemOpen
  \bibfield  {author} {\bibinfo {author} {\bibfnamefont {Y.}~\bibnamefont
  {Tang}}, \bibinfo {author} {\bibfnamefont {J.}~\bibnamefont {Gu}}, \bibinfo
  {author} {\bibfnamefont {S.}~\bibnamefont {Liu}}, \bibinfo {author}
  {\bibfnamefont {K.}~\bibnamefont {Watanabe}}, \bibinfo {author}
  {\bibfnamefont {T.}~\bibnamefont {Taniguchi}}, \bibinfo {author}
  {\bibfnamefont {J.~C.}\ \bibnamefont {Hone}}, \bibinfo {author}
  {\bibfnamefont {K.~F.}\ \bibnamefont {Mak}},\ and\ \bibinfo {author}
  {\bibfnamefont {J.}~\bibnamefont {Shan}},\ }\bibfield  {title} {\bibinfo
  {title} {Dielectric catastrophe at the {W}igner-{M}ott transition in a
  moir{\'e} superlattice},\ }\href {https://doi.org/10.1038/s41467-022-32037-1}
  {\bibfield  {journal} {\bibinfo  {journal} {Nature Communications}\ }\textbf
  {\bibinfo {volume} {13}},\ \bibinfo {pages} {4271} (\bibinfo {year}
  {2022})}\BibitemShut {NoStop}%
\bibitem [{\citenamefont {Yoshioka}\ and\ \citenamefont
  {Fukuyama}(1979)}]{Yoshioka_Fukuyama}%
  \BibitemOpen
  \bibfield  {author} {\bibinfo {author} {\bibfnamefont {D.}~\bibnamefont
  {Yoshioka}}\ and\ \bibinfo {author} {\bibfnamefont {H.}~\bibnamefont
  {Fukuyama}},\ }\bibfield  {title} {\bibinfo {title} {Charge density wave
  state of two-dimensional electrons in strong magnetic fields},\ }\href
  {https://doi.org/10.1143/JPSJ.47.394} {\bibfield  {journal} {\bibinfo
  {journal} {Journal of the Physical Society of Japan}\ }\textbf {\bibinfo
  {volume} {47}},\ \bibinfo {pages} {394} (\bibinfo {year} {1979})},\ \Eprint
  {https://arxiv.org/abs/https://doi.org/10.1143/JPSJ.47.394}
  {https://doi.org/10.1143/JPSJ.47.394} \BibitemShut {NoStop}%
\bibitem [{\citenamefont {Yoshioka}\ and\ \citenamefont
  {Lee}(1983)}]{Yoshioka_Lee}%
  \BibitemOpen
  \bibfield  {author} {\bibinfo {author} {\bibfnamefont {D.}~\bibnamefont
  {Yoshioka}}\ and\ \bibinfo {author} {\bibfnamefont {P.~A.}\ \bibnamefont
  {Lee}},\ }\bibfield  {title} {\bibinfo {title} {Ground-state energy of a
  two-dimensional charge-density-wave state in a strong magnetic field},\
  }\href {https://doi.org/10.1103/PhysRevB.27.4986} {\bibfield  {journal}
  {\bibinfo  {journal} {Phys. Rev. B}\ }\textbf {\bibinfo {volume} {27}},\
  \bibinfo {pages} {4986} (\bibinfo {year} {1983})}\BibitemShut {NoStop}%
\bibitem [{\citenamefont {Maki}\ and\ \citenamefont
  {Zotos}(1983)}]{maki1983static}%
  \BibitemOpen
  \bibfield  {author} {\bibinfo {author} {\bibfnamefont {K.}~\bibnamefont
  {Maki}}\ and\ \bibinfo {author} {\bibfnamefont {X.}~\bibnamefont {Zotos}},\
  }\bibfield  {title} {\bibinfo {title} {Static and dynamic properties of a
  two-dimensional {W}igner crystal in a strong magnetic field},\ }\href@noop {}
  {\bibfield  {journal} {\bibinfo  {journal} {Physical Review B}\ }\textbf
  {\bibinfo {volume} {28}},\ \bibinfo {pages} {4349} (\bibinfo {year}
  {1983})}\BibitemShut {NoStop}%
\bibitem [{\citenamefont {MacDonald}\ and\ \citenamefont
  {Murray}(1985)}]{MacDonald1985Broken}%
  \BibitemOpen
  \bibfield  {author} {\bibinfo {author} {\bibfnamefont {A.~H.}\ \bibnamefont
  {MacDonald}}\ and\ \bibinfo {author} {\bibfnamefont {D.~B.}\ \bibnamefont
  {Murray}},\ }\bibfield  {title} {\bibinfo {title} {Broken symmetry states for
  two-dimensional electrons in a strong magnetic field},\ }\href
  {https://doi.org/10.1103/PhysRevB.32.2291} {\bibfield  {journal} {\bibinfo
  {journal} {Phys. Rev. B}\ }\textbf {\bibinfo {volume} {32}},\ \bibinfo
  {pages} {2291} (\bibinfo {year} {1985})}\BibitemShut {NoStop}%
\bibitem [{\citenamefont {Platzman}\ and\ \citenamefont
  {Price}(1993)}]{Platzman1993QuantumFreezing}%
  \BibitemOpen
  \bibfield  {author} {\bibinfo {author} {\bibfnamefont {P.~M.}\ \bibnamefont
  {Platzman}}\ and\ \bibinfo {author} {\bibfnamefont {R.}~\bibnamefont
  {Price}},\ }\bibfield  {title} {\bibinfo {title} {Quantum freezing of the
  fractional quantum {H}all liquid},\ }\href
  {https://doi.org/10.1103/PhysRevLett.70.3487} {\bibfield  {journal} {\bibinfo
   {journal} {Phys. Rev. Lett.}\ }\textbf {\bibinfo {volume} {70}},\ \bibinfo
  {pages} {3487} (\bibinfo {year} {1993})}\BibitemShut {NoStop}%
\bibitem [{\citenamefont {Aharonov}\ and\ \citenamefont
  {Casher}(1979)}]{aharonov1979ground}%
  \BibitemOpen
  \bibfield  {author} {\bibinfo {author} {\bibfnamefont {Y.}~\bibnamefont
  {Aharonov}}\ and\ \bibinfo {author} {\bibfnamefont {A.}~\bibnamefont
  {Casher}},\ }\bibfield  {title} {\bibinfo {title} {Ground state of a
  spin-{$\frac{1}{2}$} charged particle in a two-dimensional magnetic field},\
  }\href {https://doi.org/10.1103/PhysRevA.19.2461} {\bibfield  {journal}
  {\bibinfo  {journal} {Phys. Rev. A}\ }\textbf {\bibinfo {volume} {19}},\
  \bibinfo {pages} {2461} (\bibinfo {year} {1979})}\BibitemShut {NoStop}%
\bibitem [{\citenamefont {Morales-Dur\'an}\ \emph {et~al.}(2024)\citenamefont
  {Morales-Dur\'an}, \citenamefont {Wei}, \citenamefont {Shi},\ and\
  \citenamefont {MacDonald}}]{Duran2024Magic}%
  \BibitemOpen
  \bibfield  {author} {\bibinfo {author} {\bibfnamefont {N.}~\bibnamefont
  {Morales-Dur\'an}}, \bibinfo {author} {\bibfnamefont {N.}~\bibnamefont
  {Wei}}, \bibinfo {author} {\bibfnamefont {J.}~\bibnamefont {Shi}},\ and\
  \bibinfo {author} {\bibfnamefont {A.~H.}\ \bibnamefont {MacDonald}},\
  }\bibfield  {title} {\bibinfo {title} {Magic angles and fractional {C}hern
  insulators in twisted homobilayer transition metal dichalcogenides},\ }\href
  {https://doi.org/10.1103/PhysRevLett.132.096602} {\bibfield  {journal}
  {\bibinfo  {journal} {Phys. Rev. Lett.}\ }\textbf {\bibinfo {volume} {132}},\
  \bibinfo {pages} {096602} (\bibinfo {year} {2024})}\BibitemShut {NoStop}%
\bibitem [{\citenamefont {Shi}\ \emph {et~al.}(2024)\citenamefont {Shi},
  \citenamefont {Morales-Dur\'an}, \citenamefont {Khalaf},\ and\ \citenamefont
  {MacDonald}}]{Shi2024Adiabatic}%
  \BibitemOpen
  \bibfield  {author} {\bibinfo {author} {\bibfnamefont {J.}~\bibnamefont
  {Shi}}, \bibinfo {author} {\bibfnamefont {N.}~\bibnamefont
  {Morales-Dur\'an}}, \bibinfo {author} {\bibfnamefont {E.}~\bibnamefont
  {Khalaf}},\ and\ \bibinfo {author} {\bibfnamefont {A.~H.}\ \bibnamefont
  {MacDonald}},\ }\bibfield  {title} {\bibinfo {title} {Adiabatic approximation
  and {A}haronov-{C}asher bands in twisted homobilayer transition metal
  dichalcogenides},\ }\href {https://doi.org/10.1103/PhysRevB.110.035130}
  {\bibfield  {journal} {\bibinfo  {journal} {Phys. Rev. B}\ }\textbf {\bibinfo
  {volume} {110}},\ \bibinfo {pages} {035130} (\bibinfo {year}
  {2024})}\BibitemShut {NoStop}%
\bibitem [{\citenamefont {Shi}\ \emph {et~al.}(2026)\citenamefont {Shi},
  \citenamefont {Cano},\ and\ \citenamefont
  {Morales-Dur{\'a}n}}]{shi2026effects}%
  \BibitemOpen
  \bibfield  {author} {\bibinfo {author} {\bibfnamefont {J.}~\bibnamefont
  {Shi}}, \bibinfo {author} {\bibfnamefont {J.}~\bibnamefont {Cano}},\ and\
  \bibinfo {author} {\bibfnamefont {N.}~\bibnamefont {Morales-Dur{\'a}n}},\
  }\bibfield  {title} {\bibinfo {title} {Effects of {B}erry curvature on ideal
  fractional {C}hern insulator many-body gaps},\ }\href@noop {} {\bibfield
  {journal} {\bibinfo  {journal} {Physical Review Research}\ }\textbf {\bibinfo
  {volume} {8}},\ \bibinfo {pages} {L022045} (\bibinfo {year}
  {2026})}\BibitemShut {NoStop}%
\bibitem [{\citenamefont {Wu}\ \emph {et~al.}(2012)\citenamefont {Wu},
  \citenamefont {Jain},\ and\ \citenamefont {Sun}}]{Wu2012Adiabatic}%
  \BibitemOpen
  \bibfield  {author} {\bibinfo {author} {\bibfnamefont {Y.-H.}\ \bibnamefont
  {Wu}}, \bibinfo {author} {\bibfnamefont {J.~K.}\ \bibnamefont {Jain}},\ and\
  \bibinfo {author} {\bibfnamefont {K.}~\bibnamefont {Sun}},\ }\bibfield
  {title} {\bibinfo {title} {Adiabatic continuity between {H}ofstadter and
  {C}hern insulator states},\ }\href
  {https://doi.org/10.1103/PhysRevB.86.165129} {\bibfield  {journal} {\bibinfo
  {journal} {Phys. Rev. B}\ }\textbf {\bibinfo {volume} {86}},\ \bibinfo
  {pages} {165129} (\bibinfo {year} {2012})}\BibitemShut {NoStop}%
\bibitem [{\citenamefont {Moitra}\ and\ \citenamefont
  {Sodemann~Villadiego}(2026)}]{Moitra2026Instability}%
  \BibitemOpen
  \bibfield  {author} {\bibinfo {author} {\bibfnamefont {S.}~\bibnamefont
  {Moitra}}\ and\ \bibinfo {author} {\bibfnamefont {I.}~\bibnamefont
  {Sodemann~Villadiego}},\ }\bibfield  {title} {\bibinfo {title} {Instability
  of {L}aughlin fractional quantum {H}all liquids into gapless power-law
  correlated states with continuous exponents in ideal {C}hern bands: Rigorous
  results from plasma mapping},\ }\href {https://doi.org/10.1103/3y87-96dw}
  {\bibfield  {journal} {\bibinfo  {journal} {Phys. Rev. B}\ }\textbf {\bibinfo
  {volume} {113}},\ \bibinfo {pages} {L161102} (\bibinfo {year}
  {2026})}\BibitemShut {NoStop}%
\bibitem [{\citenamefont {Wu}\ \emph {et~al.}(2019)\citenamefont {Wu},
  \citenamefont {Lovorn}, \citenamefont {Tutuc}, \citenamefont {Martin},\ and\
  \citenamefont {MacDonald}}]{Wu2019Topological}%
  \BibitemOpen
  \bibfield  {author} {\bibinfo {author} {\bibfnamefont {F.}~\bibnamefont
  {Wu}}, \bibinfo {author} {\bibfnamefont {T.}~\bibnamefont {Lovorn}}, \bibinfo
  {author} {\bibfnamefont {E.}~\bibnamefont {Tutuc}}, \bibinfo {author}
  {\bibfnamefont {I.}~\bibnamefont {Martin}},\ and\ \bibinfo {author}
  {\bibfnamefont {A.~H.}\ \bibnamefont {MacDonald}},\ }\bibfield  {title}
  {\bibinfo {title} {Topological insulators in twisted transition metal
  dichalcogenide homobilayers},\ }\href
  {https://doi.org/10.1103/PhysRevLett.122.086402} {\bibfield  {journal}
  {\bibinfo  {journal} {Phys. Rev. Lett.}\ }\textbf {\bibinfo {volume} {122}},\
  \bibinfo {pages} {086402} (\bibinfo {year} {2019})}\BibitemShut {NoStop}%
\bibitem [{\citenamefont {Yu}\ \emph {et~al.}(2020)\citenamefont {Yu},
  \citenamefont {Chen},\ and\ \citenamefont {Yao}}]{Yu2020Giant}%
  \BibitemOpen
  \bibfield  {author} {\bibinfo {author} {\bibfnamefont {H.}~\bibnamefont
  {Yu}}, \bibinfo {author} {\bibfnamefont {M.}~\bibnamefont {Chen}},\ and\
  \bibinfo {author} {\bibfnamefont {W.}~\bibnamefont {Yao}},\ }\bibfield
  {title} {\bibinfo {title} {Giant magnetic field from moiré induced {B}erry
  phase in homobilayer semiconductors},\ }\href
  {https://doi.org/10.1093/nsr/nwz117} {\bibfield  {journal} {\bibinfo
  {journal} {National Science Review}\ }\textbf {\bibinfo {volume} {7}},\
  \bibinfo {pages} {12} (\bibinfo {year} {2020})}\BibitemShut {NoStop}%
\bibitem [{\citenamefont {Zhang}\ \emph {et~al.}(2025)\citenamefont {Zhang},
  \citenamefont {Morales-Dur{\'a}n}, \citenamefont {Li}, \citenamefont {Yao},
  \citenamefont {Su}, \citenamefont {Lin}, \citenamefont {Dong}, \citenamefont
  {Liu}, \citenamefont {Chen}, \citenamefont {Kim}, \citenamefont {Watanabe},
  \citenamefont {Taniguchi}, \citenamefont {Li}, \citenamefont {Robinson},
  \citenamefont {Macdonald},\ and\ \citenamefont
  {Shih}}]{Zhang2025Experimental}%
  \BibitemOpen
  \bibfield  {author} {\bibinfo {author} {\bibfnamefont {F.}~\bibnamefont
  {Zhang}}, \bibinfo {author} {\bibfnamefont {N.}~\bibnamefont
  {Morales-Dur{\'a}n}}, \bibinfo {author} {\bibfnamefont {Y.}~\bibnamefont
  {Li}}, \bibinfo {author} {\bibfnamefont {W.}~\bibnamefont {Yao}}, \bibinfo
  {author} {\bibfnamefont {J.-J.}\ \bibnamefont {Su}}, \bibinfo {author}
  {\bibfnamefont {Y.-C.}\ \bibnamefont {Lin}}, \bibinfo {author} {\bibfnamefont
  {C.}~\bibnamefont {Dong}}, \bibinfo {author} {\bibfnamefont {X.}~\bibnamefont
  {Liu}}, \bibinfo {author} {\bibfnamefont {F.-X.~R.}\ \bibnamefont {Chen}},
  \bibinfo {author} {\bibfnamefont {H.}~\bibnamefont {Kim}}, \bibinfo {author}
  {\bibfnamefont {K.}~\bibnamefont {Watanabe}}, \bibinfo {author}
  {\bibfnamefont {T.}~\bibnamefont {Taniguchi}}, \bibinfo {author}
  {\bibfnamefont {X.}~\bibnamefont {Li}}, \bibinfo {author} {\bibfnamefont
  {J.~A.}\ \bibnamefont {Robinson}}, \bibinfo {author} {\bibfnamefont {A.~H.}\
  \bibnamefont {Macdonald}},\ and\ \bibinfo {author} {\bibfnamefont {C.-K.}\
  \bibnamefont {Shih}},\ }\bibfield  {title} {\bibinfo {title} {Experimental
  signature of layer skyrmions and implications for band topology in twisted
  {WS}e$_2$ bilayers},\ }\href {https://doi.org/10.1038/s41567-025-02876-y}
  {\bibfield  {journal} {\bibinfo  {journal} {Nature Physics}\ }\textbf
  {\bibinfo {volume} {21}},\ \bibinfo {pages} {1217} (\bibinfo {year}
  {2025})}\BibitemShut {NoStop}%
\bibitem [{\citenamefont {Thompson}\ \emph {et~al.}(2025)\citenamefont
  {Thompson}, \citenamefont {Chu}, \citenamefont {Mesple}, \citenamefont
  {Zhang}, \citenamefont {Hu}, \citenamefont {Zhao}, \citenamefont {Park},
  \citenamefont {Cai}, \citenamefont {Anderson}, \citenamefont {Watanabe},
  \citenamefont {Taniguchi}, \citenamefont {Yang}, \citenamefont {Chu},
  \citenamefont {Xu}, \citenamefont {Cao}, \citenamefont {Xiao},\ and\
  \citenamefont {Yankowitz}}]{Thompson2025Microscopic}%
  \BibitemOpen
  \bibfield  {author} {\bibinfo {author} {\bibfnamefont {E.}~\bibnamefont
  {Thompson}}, \bibinfo {author} {\bibfnamefont {K.~T.}\ \bibnamefont {Chu}},
  \bibinfo {author} {\bibfnamefont {F.}~\bibnamefont {Mesple}}, \bibinfo
  {author} {\bibfnamefont {X.-W.}\ \bibnamefont {Zhang}}, \bibinfo {author}
  {\bibfnamefont {C.}~\bibnamefont {Hu}}, \bibinfo {author} {\bibfnamefont
  {Y.}~\bibnamefont {Zhao}}, \bibinfo {author} {\bibfnamefont {H.}~\bibnamefont
  {Park}}, \bibinfo {author} {\bibfnamefont {J.}~\bibnamefont {Cai}}, \bibinfo
  {author} {\bibfnamefont {E.}~\bibnamefont {Anderson}}, \bibinfo {author}
  {\bibfnamefont {K.}~\bibnamefont {Watanabe}}, \bibinfo {author}
  {\bibfnamefont {T.}~\bibnamefont {Taniguchi}}, \bibinfo {author}
  {\bibfnamefont {J.}~\bibnamefont {Yang}}, \bibinfo {author} {\bibfnamefont
  {J.-H.}\ \bibnamefont {Chu}}, \bibinfo {author} {\bibfnamefont
  {X.}~\bibnamefont {Xu}}, \bibinfo {author} {\bibfnamefont {T.}~\bibnamefont
  {Cao}}, \bibinfo {author} {\bibfnamefont {D.}~\bibnamefont {Xiao}},\ and\
  \bibinfo {author} {\bibfnamefont {M.}~\bibnamefont {Yankowitz}},\ }\bibfield
  {title} {\bibinfo {title} {Microscopic signatures of topology in twisted
  {M}o{T}e$_2$},\ }\href {https://doi.org/10.1038/s41567-025-02877-x}
  {\bibfield  {journal} {\bibinfo  {journal} {Nature Physics}\ }\textbf
  {\bibinfo {volume} {21}},\ \bibinfo {pages} {1224} (\bibinfo {year}
  {2025})}\BibitemShut {NoStop}%
\bibitem [{\citenamefont {Devakul}\ \emph {et~al.}(2021)\citenamefont
  {Devakul}, \citenamefont {Cr{\'e}pel}, \citenamefont {Zhang},\ and\
  \citenamefont {Fu}}]{devakul2021magic}%
  \BibitemOpen
  \bibfield  {author} {\bibinfo {author} {\bibfnamefont {T.}~\bibnamefont
  {Devakul}}, \bibinfo {author} {\bibfnamefont {V.}~\bibnamefont {Cr{\'e}pel}},
  \bibinfo {author} {\bibfnamefont {Y.}~\bibnamefont {Zhang}},\ and\ \bibinfo
  {author} {\bibfnamefont {L.}~\bibnamefont {Fu}},\ }\bibfield  {title}
  {\bibinfo {title} {Magic in twisted transition metal dichalcogenide
  bilayers},\ }\href {https://doi.org/10.1038/s41467-021-27042-9} {\bibfield
  {journal} {\bibinfo  {journal} {Nat Commun}\ }\textbf {\bibinfo {volume}
  {12}},\ \bibinfo {pages} {6730} (\bibinfo {year} {2021})}\BibitemShut
  {NoStop}%
\bibitem [{\citenamefont {Morales-Dur\'an}\ \emph {et~al.}(2023)\citenamefont
  {Morales-Dur\'an}, \citenamefont {Wang}, \citenamefont {Schleder},
  \citenamefont {Angeli}, \citenamefont {Zhu}, \citenamefont {Kaxiras},
  \citenamefont {Repellin},\ and\ \citenamefont {Cano}}]{Duran2023Pressure}%
  \BibitemOpen
  \bibfield  {author} {\bibinfo {author} {\bibfnamefont {N.}~\bibnamefont
  {Morales-Dur\'an}}, \bibinfo {author} {\bibfnamefont {J.}~\bibnamefont
  {Wang}}, \bibinfo {author} {\bibfnamefont {G.~R.}\ \bibnamefont {Schleder}},
  \bibinfo {author} {\bibfnamefont {M.}~\bibnamefont {Angeli}}, \bibinfo
  {author} {\bibfnamefont {Z.}~\bibnamefont {Zhu}}, \bibinfo {author}
  {\bibfnamefont {E.}~\bibnamefont {Kaxiras}}, \bibinfo {author} {\bibfnamefont
  {C.}~\bibnamefont {Repellin}},\ and\ \bibinfo {author} {\bibfnamefont
  {J.}~\bibnamefont {Cano}},\ }\bibfield  {title} {\bibinfo {title}
  {Pressure-enhanced fractional {C}hern insulators along a magic line in
  moir\'e transition metal dichalcogenides},\ }\href
  {https://doi.org/10.1103/PhysRevResearch.5.L032022} {\bibfield  {journal}
  {\bibinfo  {journal} {Phys. Rev. Res.}\ }\textbf {\bibinfo {volume} {5}},\
  \bibinfo {pages} {L032022} (\bibinfo {year} {2023})}\BibitemShut {NoStop}%
\bibitem [{\citenamefont {Tan}\ \emph {et~al.}(2025{\natexlab{b}})\citenamefont
  {Tan}, \citenamefont {Ledwith},\ and\ \citenamefont
  {Devakul}}]{tan2025ideal}%
  \BibitemOpen
  \bibfield  {author} {\bibinfo {author} {\bibfnamefont {T.}~\bibnamefont
  {Tan}}, \bibinfo {author} {\bibfnamefont {P.~J.}\ \bibnamefont {Ledwith}},\
  and\ \bibinfo {author} {\bibfnamefont {T.}~\bibnamefont {Devakul}},\ }\href
  {https://arxiv.org/abs/2511.07402} {\bibinfo {title} {The ideal limit of
  rhombohedral graphene: Interaction-induced layer-skyrmion lattices and their
  collective excitations}} (\bibinfo {year} {2025}{\natexlab{b}}),\ \Eprint
  {https://arxiv.org/abs/2511.07402} {arXiv:2511.07402 [cond-mat.mes-hall]}
  \BibitemShut {NoStop}%
\bibitem [{\citenamefont {MacDonald}(1984)}]{MacDonald1984Influence}%
  \BibitemOpen
  \bibfield  {author} {\bibinfo {author} {\bibfnamefont {A.~H.}\ \bibnamefont
  {MacDonald}},\ }\bibfield  {title} {\bibinfo {title} {Influence of
  {L}andau-level mixing on the charge-density-wave state of a two-dimensional
  electron gas in a strong magnetic field},\ }\href
  {https://doi.org/10.1103/PhysRevB.30.4392} {\bibfield  {journal} {\bibinfo
  {journal} {Phys. Rev. B}\ }\textbf {\bibinfo {volume} {30}},\ \bibinfo
  {pages} {4392} (\bibinfo {year} {1984})}\BibitemShut {NoStop}%
\bibitem [{\citenamefont {Lam}\ and\ \citenamefont
  {Girvin}(1984)}]{Girvin_Lam}%
  \BibitemOpen
  \bibfield  {author} {\bibinfo {author} {\bibfnamefont {P.~K.}\ \bibnamefont
  {Lam}}\ and\ \bibinfo {author} {\bibfnamefont {S.~M.}\ \bibnamefont
  {Girvin}},\ }\bibfield  {title} {\bibinfo {title} {Liquid-solid transition
  and the fractional quantum-{H}all effect},\ }\href
  {https://doi.org/10.1103/PhysRevB.30.473} {\bibfield  {journal} {\bibinfo
  {journal} {Phys. Rev. B}\ }\textbf {\bibinfo {volume} {30}},\ \bibinfo
  {pages} {473(R)} (\bibinfo {year} {1984})}\BibitemShut {NoStop}%
\bibitem [{\citenamefont {Yi}\ and\ \citenamefont
  {Fertig}(1998)}]{Yi1998Laughlin}%
  \BibitemOpen
  \bibfield  {author} {\bibinfo {author} {\bibfnamefont {H.}~\bibnamefont
  {Yi}}\ and\ \bibinfo {author} {\bibfnamefont {H.~A.}\ \bibnamefont
  {Fertig}},\ }\bibfield  {title} {\bibinfo {title}
  {Laughlin-{J}astrow-correlated {W}igner crystal in a strong magnetic field},\
  }\href {https://doi.org/10.1103/PhysRevB.58.4019} {\bibfield  {journal}
  {\bibinfo  {journal} {Phys. Rev. B}\ }\textbf {\bibinfo {volume} {58}},\
  \bibinfo {pages} {4019} (\bibinfo {year} {1998})}\BibitemShut {NoStop}%
\bibitem [{\citenamefont {DaSilva}\ \emph {et~al.}(2016)\citenamefont
  {DaSilva}, \citenamefont {Jung},\ and\ \citenamefont
  {MacDonald}}]{DaSilva2016Fractional}%
  \BibitemOpen
  \bibfield  {author} {\bibinfo {author} {\bibfnamefont {A.~M.}\ \bibnamefont
  {DaSilva}}, \bibinfo {author} {\bibfnamefont {J.}~\bibnamefont {Jung}},\ and\
  \bibinfo {author} {\bibfnamefont {A.~H.}\ \bibnamefont {MacDonald}},\
  }\bibfield  {title} {\bibinfo {title} {Fractional {H}ofstadter states in
  graphene on hexagonal boron nitride},\ }\href
  {https://doi.org/10.1103/PhysRevLett.117.036802} {\bibfield  {journal}
  {\bibinfo  {journal} {Phys. Rev. Lett.}\ }\textbf {\bibinfo {volume} {117}},\
  \bibinfo {pages} {036802} (\bibinfo {year} {2016})}\BibitemShut {NoStop}%
\bibitem [{\citenamefont {Levesque}\ \emph {et~al.}(1984)\citenamefont
  {Levesque}, \citenamefont {Weis},\ and\ \citenamefont
  {MacDonald}}]{Levesque1984Crystallization}%
  \BibitemOpen
  \bibfield  {author} {\bibinfo {author} {\bibfnamefont {D.}~\bibnamefont
  {Levesque}}, \bibinfo {author} {\bibfnamefont {J.~J.}\ \bibnamefont {Weis}},\
  and\ \bibinfo {author} {\bibfnamefont {A.~H.}\ \bibnamefont {MacDonald}},\
  }\bibfield  {title} {\bibinfo {title} {Crystallization of the incompressible
  quantum-fluid state of a two-dimensional electron gas in a strong magnetic
  field},\ }\href {https://doi.org/10.1103/PhysRevB.30.1056} {\bibfield
  {journal} {\bibinfo  {journal} {Phys. Rev. B}\ }\textbf {\bibinfo {volume}
  {30}},\ \bibinfo {pages} {1056} (\bibinfo {year} {1984})}\BibitemShut
  {NoStop}%
\bibitem [{\citenamefont {Laughlin}(1983)}]{Laughlin1983Anomalous}%
  \BibitemOpen
  \bibfield  {author} {\bibinfo {author} {\bibfnamefont {R.~B.}\ \bibnamefont
  {Laughlin}},\ }\bibfield  {title} {\bibinfo {title} {Anomalous quantum {H}all
  effect: An incompressible quantum fluid with fractionally charged
  excitations},\ }\href {https://doi.org/10.1103/PhysRevLett.50.1395}
  {\bibfield  {journal} {\bibinfo  {journal} {Phys. Rev. Lett.}\ }\textbf
  {\bibinfo {volume} {50}},\ \bibinfo {pages} {1395} (\bibinfo {year}
  {1983})}\BibitemShut {NoStop}%
\bibitem [{\citenamefont {Girvin}\ \emph {et~al.}(1986)\citenamefont {Girvin},
  \citenamefont {MacDonald},\ and\ \citenamefont
  {Platzman}}]{girvin1986magneto}%
  \BibitemOpen
  \bibfield  {author} {\bibinfo {author} {\bibfnamefont {S.}~\bibnamefont
  {Girvin}}, \bibinfo {author} {\bibfnamefont {A.}~\bibnamefont {MacDonald}},\
  and\ \bibinfo {author} {\bibfnamefont {P.}~\bibnamefont {Platzman}},\
  }\bibfield  {title} {\bibinfo {title} {Magneto-roton theory of collective
  excitations in the fractional quantum {H}all effect},\ }\href@noop {}
  {\bibfield  {journal} {\bibinfo  {journal} {Physical Review B}\ }\textbf
  {\bibinfo {volume} {33}},\ \bibinfo {pages} {2481} (\bibinfo {year}
  {1986})}\BibitemShut {NoStop}%
\bibitem [{\citenamefont {Wolf}\ \emph {et~al.}(2025)\citenamefont {Wolf},
  \citenamefont {Chao}, \citenamefont {MacDonald},\ and\ \citenamefont
  {Su}}]{Wolf2025Intraband}%
  \BibitemOpen
  \bibfield  {author} {\bibinfo {author} {\bibfnamefont {T.}~\bibnamefont
  {Wolf}}, \bibinfo {author} {\bibfnamefont {Y.-C.}\ \bibnamefont {Chao}},
  \bibinfo {author} {\bibfnamefont {A.~H.}\ \bibnamefont {MacDonald}},\ and\
  \bibinfo {author} {\bibfnamefont {J.-J.}\ \bibnamefont {Su}},\ }\bibfield
  {title} {\bibinfo {title} {Intraband collective excitations and spatial
  correlations in fractional {C}hern insulators},\ }\href
  {https://doi.org/10.1103/PhysRevLett.134.116501} {\bibfield  {journal}
  {\bibinfo  {journal} {Phys. Rev. Lett.}\ }\textbf {\bibinfo {volume} {134}},\
  \bibinfo {pages} {116501} (\bibinfo {year} {2025})}\BibitemShut {NoStop}%
\bibitem [{\citenamefont {Kousa}\ \emph {et~al.}(2025)\citenamefont {Kousa},
  \citenamefont {Morales-Dur\'an}, \citenamefont {Wolf}, \citenamefont
  {Khalaf},\ and\ \citenamefont {MacDonald}}]{Kousa2025Magnetoroton}%
  \BibitemOpen
  \bibfield  {author} {\bibinfo {author} {\bibfnamefont {B.~M.}\ \bibnamefont
  {Kousa}}, \bibinfo {author} {\bibfnamefont {N.}~\bibnamefont
  {Morales-Dur\'an}}, \bibinfo {author} {\bibfnamefont {T.~M.~R.}\ \bibnamefont
  {Wolf}}, \bibinfo {author} {\bibfnamefont {E.}~\bibnamefont {Khalaf}},\ and\
  \bibinfo {author} {\bibfnamefont {A.~H.}\ \bibnamefont {MacDonald}},\
  }\bibfield  {title} {\bibinfo {title} {Theory of magnetoroton bands in
  moir\'e materials},\ }\href {https://doi.org/10.1103/w57n-q4xn} {\bibfield
  {journal} {\bibinfo  {journal} {Phys. Rev. Lett.}\ }\textbf {\bibinfo
  {volume} {135}},\ \bibinfo {pages} {246604} (\bibinfo {year}
  {2025})}\BibitemShut {NoStop}%
\bibitem [{\citenamefont {Ferrari}(1990)}]{ferrari1990twodimensional}%
  \BibitemOpen
  \bibfield  {author} {\bibinfo {author} {\bibfnamefont {R.}~\bibnamefont
  {Ferrari}},\ }\bibfield  {title} {\bibinfo {title} {Two-dimensional electrons
  in a strong magnetic field: {{A}} basis for single-particle states},\ }\href
  {https://doi.org/10.1103/PhysRevB.42.4598} {\bibfield  {journal} {\bibinfo
  {journal} {Phys. Rev. B}\ }\textbf {\bibinfo {volume} {42}},\ \bibinfo
  {pages} {4598} (\bibinfo {year} {1990})}\BibitemShut {NoStop}%
\bibitem [{\citenamefont {Haldane}(2018)}]{haldane2018modularinvariant}%
  \BibitemOpen
  \bibfield  {author} {\bibinfo {author} {\bibfnamefont {F.~D.~M.}\
  \bibnamefont {Haldane}},\ }\bibfield  {title} {\bibinfo {title} {A
  modular-invariant modified {{Weierstrass}} sigma-function as a building block
  for lowest-{{Landau-level}} wavefunctions on the torus},\ }\href
  {https://doi.org/10.1063/1.5042618} {\bibfield  {journal} {\bibinfo
  {journal} {Journal of Mathematical Physics}\ }\textbf {\bibinfo {volume}
  {59}},\ \bibinfo {pages} {071901} (\bibinfo {year} {2018})}\BibitemShut
  {NoStop}%
\bibitem [{\citenamefont {Price}\ \emph {et~al.}(2015)\citenamefont {Price},
  \citenamefont {Ozawa}, \citenamefont {Cooper},\ and\ \citenamefont
  {Carusotto}}]{Price2015Artificial}%
  \BibitemOpen
  \bibfield  {author} {\bibinfo {author} {\bibfnamefont {H.~M.}\ \bibnamefont
  {Price}}, \bibinfo {author} {\bibfnamefont {T.}~\bibnamefont {Ozawa}},
  \bibinfo {author} {\bibfnamefont {N.~R.}\ \bibnamefont {Cooper}},\ and\
  \bibinfo {author} {\bibfnamefont {I.}~\bibnamefont {Carusotto}},\ }\bibfield
  {title} {\bibinfo {title} {Artificial magnetic fields in momentum space in
  spin-orbit-coupled systems},\ }\href
  {https://doi.org/10.1103/PhysRevA.91.033606} {\bibfield  {journal} {\bibinfo
  {journal} {Phys. Rev. A}\ }\textbf {\bibinfo {volume} {91}},\ \bibinfo
  {pages} {033606} (\bibinfo {year} {2015})}\BibitemShut {NoStop}%
\bibitem [{\citenamefont {Claassen}\ \emph {et~al.}(2015)\citenamefont
  {Claassen}, \citenamefont {Lee}, \citenamefont {Thomale}, \citenamefont
  {Qi},\ and\ \citenamefont {Devereaux}}]{claassen2015position}%
  \BibitemOpen
  \bibfield  {author} {\bibinfo {author} {\bibfnamefont {M.}~\bibnamefont
  {Claassen}}, \bibinfo {author} {\bibfnamefont {C.~H.}\ \bibnamefont {Lee}},
  \bibinfo {author} {\bibfnamefont {R.}~\bibnamefont {Thomale}}, \bibinfo
  {author} {\bibfnamefont {X.-L.}\ \bibnamefont {Qi}},\ and\ \bibinfo {author}
  {\bibfnamefont {T.~P.}\ \bibnamefont {Devereaux}},\ }\bibfield  {title}
  {\bibinfo {title} {Position-momentum duality and fractional quantum {H}all
  effect in {C}hern insulators},\ }\href@noop {} {\bibfield  {journal}
  {\bibinfo  {journal} {Physical review letters}\ }\textbf {\bibinfo {volume}
  {114}},\ \bibinfo {pages} {236802} (\bibinfo {year} {2015})}\BibitemShut
  {NoStop}%
\bibitem [{\citenamefont {Okuma}(2024)}]{okuma2024constructing}%
  \BibitemOpen
  \bibfield  {author} {\bibinfo {author} {\bibfnamefont {N.}~\bibnamefont
  {Okuma}},\ }\bibfield  {title} {\bibinfo {title} {Constructing vortex
  functions and basis states of {C}hern insulators: Ideal condition, inequality
  from index theorem, and coherentlike states on the von {N}eumann lattice},\
  }\href@noop {} {\bibfield  {journal} {\bibinfo  {journal} {Physical Review
  B}\ }\textbf {\bibinfo {volume} {110}},\ \bibinfo {pages} {245112} (\bibinfo
  {year} {2024})}\BibitemShut {NoStop}%
\bibitem [{\citenamefont {Li}\ \emph {et~al.}(2024)\citenamefont {Li},
  \citenamefont {Dong}, \citenamefont {Ledwith},\ and\ \citenamefont
  {Khalaf}}]{li2024constraints}%
  \BibitemOpen
  \bibfield  {author} {\bibinfo {author} {\bibfnamefont {Q.}~\bibnamefont
  {Li}}, \bibinfo {author} {\bibfnamefont {J.}~\bibnamefont {Dong}}, \bibinfo
  {author} {\bibfnamefont {P.~J.}\ \bibnamefont {Ledwith}},\ and\ \bibinfo
  {author} {\bibfnamefont {E.}~\bibnamefont {Khalaf}},\ }\href
  {https://arxiv.org/abs/2407.02561} {\bibinfo {title} {Constraints on real
  space representations of {C}hern bands}} (\bibinfo {year} {2024}),\ \Eprint
  {https://arxiv.org/abs/2407.02561} {arXiv:2407.02561 [cond-mat.str-el]}
  \BibitemShut {NoStop}%
\bibitem [{\citenamefont {Mishra}\ \emph {et~al.}(2026)\citenamefont {Mishra},
  \citenamefont {Wolf},\ and\ \citenamefont {MacDonald}}]{mishra2026charge}%
  \BibitemOpen
  \bibfield  {author} {\bibinfo {author} {\bibfnamefont {S.}~\bibnamefont
  {Mishra}}, \bibinfo {author} {\bibfnamefont {T.~M.~R.}\ \bibnamefont
  {Wolf}},\ and\ \bibinfo {author} {\bibfnamefont {A.~H.}\ \bibnamefont
  {MacDonald}},\ }\href {https://arxiv.org/abs/2603.05714} {\bibinfo {title}
  {Charge-ordered states in twisted {M}o{T}e$_2$}} (\bibinfo {year} {2026}),\
  \Eprint {https://arxiv.org/abs/2603.05714} {arXiv:2603.05714
  [cond-mat.str-el]} \BibitemShut {NoStop}%
\bibitem [{\citenamefont {Ishikawa}\ and\ \citenamefont
  {Maeda}(1997)}]{ishikawa1997flux}%
  \BibitemOpen
  \bibfield  {author} {\bibinfo {author} {\bibfnamefont {K.}~\bibnamefont
  {Ishikawa}}\ and\ \bibinfo {author} {\bibfnamefont {N.}~\bibnamefont
  {Maeda}},\ }\bibfield  {title} {\bibinfo {title} {Flux state in von {N}eumann
  lattices and the fractional {H}all effect},\ }\href@noop {} {\bibfield
  {journal} {\bibinfo  {journal} {Progress of Theoretical Physics}\ }\textbf
  {\bibinfo {volume} {97}},\ \bibinfo {pages} {507} (\bibinfo {year}
  {1997})}\BibitemShut {NoStop}%
\bibitem [{\citenamefont {Ishikawa}\ \emph {et~al.}(1998)\citenamefont
  {Ishikawa}, \citenamefont {Maeda}, \citenamefont {Ochiai},\ and\
  \citenamefont {Suzuki}}]{ishikawa1998duality}%
  \BibitemOpen
  \bibfield  {author} {\bibinfo {author} {\bibfnamefont {K.}~\bibnamefont
  {Ishikawa}}, \bibinfo {author} {\bibfnamefont {N.}~\bibnamefont {Maeda}},
  \bibinfo {author} {\bibfnamefont {T.}~\bibnamefont {Ochiai}},\ and\ \bibinfo
  {author} {\bibfnamefont {H.}~\bibnamefont {Suzuki}},\ }\bibfield  {title}
  {\bibinfo {title} {Duality relation among periodic-potential problems in the
  lowest {L}andau level},\ }\href@noop {} {\bibfield  {journal} {\bibinfo
  {journal} {Physical Review B}\ }\textbf {\bibinfo {volume} {58}},\ \bibinfo
  {pages} {1088} (\bibinfo {year} {1998})}\BibitemShut {NoStop}%
\bibitem [{\citenamefont {Lindemann}(1910)}]{Lindemann1910}%
  \BibitemOpen
  \bibfield  {author} {\bibinfo {author} {\bibfnamefont {F.~A.}\ \bibnamefont
  {Lindemann}},\ }\bibfield  {title} {\bibinfo {title} {The calculation of
  molecular vibration frequencies},\ }\href@noop {} {\bibfield  {journal}
  {\bibinfo  {journal} {Phys. Z}\ }\textbf {\bibinfo {volume} {11}},\ \bibinfo
  {pages} {609} (\bibinfo {year} {1910})}\BibitemShut {NoStop}%
\bibitem [{\citenamefont {Khrapak}(2020)}]{Khrapak2020Lindemann}%
  \BibitemOpen
  \bibfield  {author} {\bibinfo {author} {\bibfnamefont {S.~A.}\ \bibnamefont
  {Khrapak}},\ }\bibfield  {title} {\bibinfo {title} {Lindemann melting
  criterion in two dimensions},\ }\href
  {https://doi.org/10.1103/PhysRevResearch.2.012040} {\bibfield  {journal}
  {\bibinfo  {journal} {Phys. Rev. Res.}\ }\textbf {\bibinfo {volume} {2}},\
  \bibinfo {pages} {012040(R)} (\bibinfo {year} {2020})}\BibitemShut {NoStop}%
\bibitem [{\citenamefont {Bedanov}\ \emph {et~al.}(1985)\citenamefont
  {Bedanov}, \citenamefont {Gadiyak},\ and\ \citenamefont
  {Lozovik}}]{Bedanov1985modified}%
  \BibitemOpen
  \bibfield  {author} {\bibinfo {author} {\bibfnamefont {V.}~\bibnamefont
  {Bedanov}}, \bibinfo {author} {\bibfnamefont {G.}~\bibnamefont {Gadiyak}},\
  and\ \bibinfo {author} {\bibfnamefont {Y.}~\bibnamefont {Lozovik}},\
  }\bibfield  {title} {\bibinfo {title} {On a modified lindemann-like criterion
  for 2d melting},\ }\href
  {https://doi.org/https://doi.org/10.1016/0375-9601(85)90617-6} {\bibfield
  {journal} {\bibinfo  {journal} {Physics Letters A}\ }\textbf {\bibinfo
  {volume} {109}},\ \bibinfo {pages} {289} (\bibinfo {year}
  {1985})}\BibitemShut {NoStop}%
\bibitem [{\citenamefont {Ross}(1969)}]{Ross1969Generalized}%
  \BibitemOpen
  \bibfield  {author} {\bibinfo {author} {\bibfnamefont {M.}~\bibnamefont
  {Ross}},\ }\bibfield  {title} {\bibinfo {title} {Generalized lindemann
  melting law},\ }\href {https://doi.org/10.1103/PhysRev.184.233} {\bibfield
  {journal} {\bibinfo  {journal} {Physical Review}\ }\textbf {\bibinfo {volume}
  {184}},\ \bibinfo {pages} {233} (\bibinfo {year} {1969})}\BibitemShut
  {NoStop}%
\bibitem [{\citenamefont {Paul}\ \emph {et~al.}(2026)\citenamefont {Paul},
  \citenamefont {Abouelkomsan}, \citenamefont {Reddy},\ and\ \citenamefont
  {Fu}}]{Paul2026Shining}%
  \BibitemOpen
  \bibfield  {author} {\bibinfo {author} {\bibfnamefont {N.}~\bibnamefont
  {Paul}}, \bibinfo {author} {\bibfnamefont {A.}~\bibnamefont {Abouelkomsan}},
  \bibinfo {author} {\bibfnamefont {A.}~\bibnamefont {Reddy}},\ and\ \bibinfo
  {author} {\bibfnamefont {L.}~\bibnamefont {Fu}},\ }\bibfield  {title}
  {\bibinfo {title} {Shining light on collective modes in moir\'e fractional
  {C}hern insulators},\ }\href {https://doi.org/10.1103/7sbg-yqhs} {\bibfield
  {journal} {\bibinfo  {journal} {Phys. Rev. Lett.}\ }\textbf {\bibinfo
  {volume} {137}},\ \bibinfo {pages} {096505} (\bibinfo {year}
  {2026})}\BibitemShut {NoStop}%
\bibitem [{\citenamefont {Dong}\ \emph {et~al.}(2025)\citenamefont {Dong},
  \citenamefont {Sommer}, \citenamefont {Soejima}, \citenamefont {Parker},\
  and\ \citenamefont {Vishwanath}}]{Dong2025Phonons}%
  \BibitemOpen
  \bibfield  {author} {\bibinfo {author} {\bibfnamefont {J.}~\bibnamefont
  {Dong}}, \bibinfo {author} {\bibfnamefont {O.~E.}\ \bibnamefont {Sommer}},
  \bibinfo {author} {\bibfnamefont {T.}~\bibnamefont {Soejima}}, \bibinfo
  {author} {\bibfnamefont {D.~E.}\ \bibnamefont {Parker}},\ and\ \bibinfo
  {author} {\bibfnamefont {A.}~\bibnamefont {Vishwanath}},\ }\bibfield  {title}
  {\bibinfo {title} {Phonons in electron crystals with {B}erry curvature},\
  }\href {https://doi.org/10.1073/pnas.2515532122} {\bibfield  {journal}
  {\bibinfo  {journal} {Proceedings of the National Academy of Sciences}\
  }\textbf {\bibinfo {volume} {122}},\ \bibinfo {pages} {e2515532122} (\bibinfo
  {year} {2025})},\ \Eprint
  {https://arxiv.org/abs/https://www.pnas.org/doi/pdf/10.1073/pnas.2515532122}
  {https://www.pnas.org/doi/pdf/10.1073/pnas.2515532122} \BibitemShut {NoStop}%
\bibitem [{\citenamefont {Desrochers}\ \emph
  {et~al.}(2026{\natexlab{b}})\citenamefont {Desrochers}, \citenamefont
  {Hirsbrunner}, \citenamefont {Huxford}, \citenamefont {Patri}, \citenamefont
  {Senthil},\ and\ \citenamefont {Kim}}]{Desrochers2026Elastic}%
  \BibitemOpen
  \bibfield  {author} {\bibinfo {author} {\bibfnamefont {F.}~\bibnamefont
  {Desrochers}}, \bibinfo {author} {\bibfnamefont {M.~R.}\ \bibnamefont
  {Hirsbrunner}}, \bibinfo {author} {\bibfnamefont {J.}~\bibnamefont
  {Huxford}}, \bibinfo {author} {\bibfnamefont {A.~S.}\ \bibnamefont {Patri}},
  \bibinfo {author} {\bibfnamefont {T.}~\bibnamefont {Senthil}},\ and\ \bibinfo
  {author} {\bibfnamefont {Y.~B.}\ \bibnamefont {Kim}},\ }\bibfield  {title}
  {\bibinfo {title} {Elastic response and instabilities of anomalous {H}all
  crystals},\ }\href {https://doi.org/10.1103/vyss-2hkj} {\bibfield  {journal}
  {\bibinfo  {journal} {Phys. Rev. Lett.}\ }\textbf {\bibinfo {volume} {136}},\
  \bibinfo {pages} {166503} (\bibinfo {year} {2026}{\natexlab{b}})}\BibitemShut
  {NoStop}%
\bibitem [{\citenamefont {Pan}\ \emph {et~al.}(2026)\citenamefont {Pan},
  \citenamefont {Yang}, \citenamefont {Wang}, \citenamefont {Cai},
  \citenamefont {Wang}, \citenamefont {Zhao}, \citenamefont {Watanabe},
  \citenamefont {Taniguchi}, \citenamefont {Zhang}, \citenamefont {Liu},
  \citenamefont {Yang},\ and\ \citenamefont {Gao}}]{Pan2026optical}%
  \BibitemOpen
  \bibfield  {author} {\bibinfo {author} {\bibfnamefont {H.}~\bibnamefont
  {Pan}}, \bibinfo {author} {\bibfnamefont {S.}~\bibnamefont {Yang}}, \bibinfo
  {author} {\bibfnamefont {Y.}~\bibnamefont {Wang}}, \bibinfo {author}
  {\bibfnamefont {X.}~\bibnamefont {Cai}}, \bibinfo {author} {\bibfnamefont
  {W.}~\bibnamefont {Wang}}, \bibinfo {author} {\bibfnamefont {Y.}~\bibnamefont
  {Zhao}}, \bibinfo {author} {\bibfnamefont {K.}~\bibnamefont {Watanabe}},
  \bibinfo {author} {\bibfnamefont {T.}~\bibnamefont {Taniguchi}}, \bibinfo
  {author} {\bibfnamefont {L.}~\bibnamefont {Zhang}}, \bibinfo {author}
  {\bibfnamefont {Y.}~\bibnamefont {Liu}}, \bibinfo {author} {\bibfnamefont
  {B.}~\bibnamefont {Yang}},\ and\ \bibinfo {author} {\bibfnamefont
  {W.}~\bibnamefont {Gao}},\ }\bibfield  {title} {\bibinfo {title} {Optical
  signatures of $\ensuremath{-}\frac{1}{3}$ fractional quantum anomalous {H}all
  state in twisted {M}o{T}e$_2$},\ }\href {https://doi.org/10.1103/f4dj-7sts}
  {\bibfield  {journal} {\bibinfo  {journal} {Phys. Rev. Lett.}\ }\textbf
  {\bibinfo {volume} {136}},\ \bibinfo {pages} {056601} (\bibinfo {year}
  {2026})}\BibitemShut {NoStop}%
\bibitem [{\citenamefont {Guerci}\ \emph {et~al.}(2025)\citenamefont {Guerci},
  \citenamefont {Abouelkomsan},\ and\ \citenamefont {Fu}}]{Guerci_AC_SC}%
  \BibitemOpen
  \bibfield  {author} {\bibinfo {author} {\bibfnamefont {D.}~\bibnamefont
  {Guerci}}, \bibinfo {author} {\bibfnamefont {A.}~\bibnamefont
  {Abouelkomsan}},\ and\ \bibinfo {author} {\bibfnamefont {L.}~\bibnamefont
  {Fu}},\ }\bibfield  {title} {\bibinfo {title} {From fractionalization to
  chiral topological superconductivity in a flat {C}hern band},\ }\href
  {https://doi.org/10.1103/zm39-dstj} {\bibfield  {journal} {\bibinfo
  {journal} {Phys. Rev. Lett.}\ }\textbf {\bibinfo {volume} {135}},\ \bibinfo
  {pages} {186601} (\bibinfo {year} {2025})}\BibitemShut {NoStop}%
\bibitem [{\citenamefont {Wang}\ and\ \citenamefont
  {Zaletel}(2025)}]{wang2025chiral}%
  \BibitemOpen
  \bibfield  {author} {\bibinfo {author} {\bibfnamefont {T.}~\bibnamefont
  {Wang}}\ and\ \bibinfo {author} {\bibfnamefont {M.~P.}\ \bibnamefont
  {Zaletel}},\ }\href {https://arxiv.org/abs/2507.07921} {\bibinfo {title}
  {Chiral superconductivity near a fractional {C}hern insulator}} (\bibinfo
  {year} {2025}),\ \Eprint {https://arxiv.org/abs/2507.07921} {arXiv:2507.07921
  [cond-mat.str-el]} \BibitemShut {NoStop}%
\bibitem [{\citenamefont {Morales-Durán}\ \emph {et~al.}(2026)\citenamefont
  {Morales-Durán}, \citenamefont {Shi}, \citenamefont {Voinea}, \citenamefont
  {Potasz},\ and\ \citenamefont {Cano}}]{moralesduran2026bandmixing}%
  \BibitemOpen
  \bibfield  {author} {\bibinfo {author} {\bibfnamefont {N.}~\bibnamefont
  {Morales-Durán}}, \bibinfo {author} {\bibfnamefont {J.}~\bibnamefont {Shi}},
  \bibinfo {author} {\bibfnamefont {C.}~\bibnamefont {Voinea}}, \bibinfo
  {author} {\bibfnamefont {P.}~\bibnamefont {Potasz}},\ and\ \bibinfo {author}
  {\bibfnamefont {J.}~\bibnamefont {Cano}},\ }\href
  {https://arxiv.org/abs/2604.16847} {\bibinfo {title} {Band mixing and
  particle-hole asymmetry in moir\'e fractional {C}hern insulators}} (\bibinfo
  {year} {2026}),\ \Eprint {https://arxiv.org/abs/2604.16847} {arXiv:2604.16847
  [cond-mat.str-el]} \BibitemShut {NoStop}%
\bibitem [{\citenamefont {Haldane}\ and\ \citenamefont
  {Rezayi}(1985)}]{Haldane1985Periodic}%
  \BibitemOpen
  \bibfield  {author} {\bibinfo {author} {\bibfnamefont {F.~D.~M.}\
  \bibnamefont {Haldane}}\ and\ \bibinfo {author} {\bibfnamefont {E.~H.}\
  \bibnamefont {Rezayi}},\ }\bibfield  {title} {\bibinfo {title} {Periodic
  laughlin-{J}astrow wave functions for the fractional quantized {H}all
  effect},\ }\href {https://doi.org/10.1103/PhysRevB.31.2529} {\bibfield
  {journal} {\bibinfo  {journal} {Phys. Rev. B}\ }\textbf {\bibinfo {volume}
  {31}},\ \bibinfo {pages} {2529} (\bibinfo {year} {1985})}\BibitemShut
  {NoStop}%
\bibitem [{\citenamefont {Ciftja}\ and\ \citenamefont
  {Wexler}(2003)}]{Ciftja2003MonteCarlo}%
  \BibitemOpen
  \bibfield  {author} {\bibinfo {author} {\bibfnamefont {O.}~\bibnamefont
  {Ciftja}}\ and\ \bibinfo {author} {\bibfnamefont {C.}~\bibnamefont
  {Wexler}},\ }\bibfield  {title} {\bibinfo {title} {Monte {C}arlo simulation
  method for {L}aughlin-like states in a disk geometry},\ }\href
  {https://doi.org/10.1103/PhysRevB.67.075304} {\bibfield  {journal} {\bibinfo
  {journal} {Phys. Rev. B}\ }\textbf {\bibinfo {volume} {67}},\ \bibinfo
  {pages} {075304} (\bibinfo {year} {2003})}\BibitemShut {NoStop}%
\end{thebibliography}%

\clearpage
\appendix
\onecolumngrid

\section{Perturbation theory for the energy of the Laughlin-like state in AC bands}
\label{Appendix:LaughlinPerturbation}

Here we consider an AC band in the limit of weak periodic field modulations, corresponding to $B_1\ll1$. The Laughlin-like state in the AC band can be written as \cite{wang2021exact}
\begin{align}
    \ket{\Psi_{\text{L}}^{\text{AC}}}= e^{\sum_i\chi({\bm r}_i)}\ket{\Psi_{\text{L}}^{\text{LLL}}}=e^X\ket{\Psi_{\text{L}}^{\text{LLL}}},
    \label{eq:Appendix_AC_Laughlin}
\end{align}
where $\ket{\Psi_{\text{L}}^{\text{LLL}}}$ is the Laughlin state in a uniform magnetic field and
\begin{align}
    X\equiv\sum_i\chi({\bm r}_i)=\sum_i \sum_{\bm G}\chi_{\bm G}e^{i {\bm G}\cdot {\bm r}_i}=\sum_{\bm G}\chi_{\bm G}\rho_{\bm G},
\end{align}
where $\chi_{\bm G}=-B_{\bm G}/G^2$ are the harmonics of $\chi({\bm r})$ \cite{Duran2024Magic}. Using Eq.~\eqref{eq:Appendix_AC_Laughlin}, the energy of the AC Laughlin-like state can be expressed in terms of expectation values of operators evaluated on the uniform field Laughlin state
\begin{align}
    E_{\text L}^{\text{AC}}= \frac{\braket{\Psi_{\text{L}}^{AC}|H|\Psi_{\text{L}}^{AC}}}{\braket{\Psi_{\text{L}}^{AC}|\Psi_{\text{L}}^{AC}}}=\frac{\braket{\Psi_\text{L}^{\text{LLL}}|e^XHe^X|\Psi_\text{L}^{\text{LLL}}}}{\braket{\Psi_\text{L}^{\text{LLL}}|e^{2X}|\Psi_\text{L}^{\text{LLL}}}}.
    \label{eq:App_LaughlinE}
\end{align}
In the weak modulation limit, the numerator of Eq.~\eqref{eq:App_LaughlinE} can be Taylor-expanded as
\begin{align}
    \braket{e^X He^X}=&\braket{(1+X+\frac{X^2}{2}+\cdots)\,H\,(1+X+\frac{X^2}{2}+\cdots)}\nonumber \\
    \approx& \braket{H}+\braket{XH}+\braket{HX}+\braket{XHX}+\frac{1}{2}\left(\braket{X^2H}+\braket{HX^2}\right)+\cdots,
\end{align}
where $\braket{O}$ refers to the expectation value of an operator $O$ taken in the uniform field Laughlin ground state $\ket{\Psi_{\text{L}}^{\text{LLL}}}$. The denominator of Eq.~\eqref{eq:App_LaughlinE} is expanded as
\begin{align}
    \frac{1}{\braket{e^{2X}}}=&\braket{1+2X+2X^2+\cdots}^{-1}
    \approx 1-2\braket{X}-2\braket{X^2}+4\braket{X}\braket{X}+\cdots.
\end{align}
Due to the continuous translation symmetry of the uniform Laughlin state, $\braket{\rho_{\bm G}} = 0$ for all non-zero ${\bm G}$. In addition, since $\chi_0 = 0$, $\braket{X}=\braket{\sum_{\bm G}\chi_{\bm G}\rho_{\bm G}}=0$ and $\braket{XH} = 0$.
Thus, we approximate the energy of the AC Laughlin-like state given by Eq.~\eqref{eq:Appendix_AC_Laughlin} as
\begin{align}
    E_{\text{L}}^{\text{AC}}=\frac{\braket{e^X He^X}}{\braket{e^{2X}}}\approx\braket{H}&+\braket{XHX}+\frac{1}{2}\left(\braket{X^2H}+\braket{HX^2}\right)-2\braket{X^2}\braket{H}+\cdots.
\end{align}
In terms of the Fourier coefficients of $X$, the expression for the energy reads
\begin{align}
    E_{\text{L}}^{\text{AC}}&\approx E_{\text{L}}^{\text{LLL}}+\sum_{\bm G}\chi_{\bm G}\chi_{-\bm G}\left( \braket{\bar{\rho}_{-\bm G}\,H\,\bar{\rho}_{\bm G}}-E_{\text{L}}^{\text{LLL}}\braket{\bar{\rho}_{-\bm G}\bar{\rho}_{\bm G}}\right),
    %\nonumber \\ %&=E_{\text{L}}^{\text{LLL}}+\sum_{\bm G}\chi_{\bm G}\chi_{-\bm G}\left(E_{\text{L}}^{\text{LLL}}\,N\,\bar{S}_{\bm G}+N\bar{f}_{\bm G}-E_{\text{L}}^{\text{LLL}}\,N\,\bar{S}_{\bm G}\right),
\end{align}
where $\braket{H}=E_{\text{L}}^{\text{LLL}}$ is the energy of the uniform Laughlin state. The energy per particle can be written simply in terms of the oscillator strength \cite{girvin1986magneto}, $\bar{f}_{\bm G}$, as
\begin{align}
    \frac{E_{\text{L}}^{\text{AC}}}{N}=\frac{E_{\text{L}}^{\text{LLL}}}{N}+\sum_{\bm G}\chi_{\bm G}\chi_{-\bm G}\bar{f}_{\bm G}=\frac{E_{\text{L}}^{\text{LLL}}}{N}+6\chi_1^2\,\bar{f}_{\bm{G}_0},
\end{align}
where in the second equality we use the fact that only the first-shell harmonics of $\chi({\bm r})$ are non-vanishing, and ${\bm G}_0$ is a first-shell reciprocal lattice vector. Finally, the energy per particle of the Laughlin-like state in terms of $B_1$ is
\begin{align}
    \frac{E_{\text{L}}^{\text{AC}}}{N}=\frac{E_{\text{L}}^{\text{LLL}}}{N}+6\left(-\frac{B_1}{G_0^2} \right)^2\,\bar{f}_{\bm{G}_0}=\frac{E_{\text{L}}^{\text{LLL}}}{N}+\left(\frac{9B_1^2}{8\pi^2} \right)\,\bar{f}_{\bm{G}_0},
\end{align}
which is Eq.~\eqref{eq:AC_Laughlin_Energy} in the main text. In the second equality we have used that $B_1$ has units of $\ell^{-2}$ and the relation $2\pi\ell^2=\sqrt{3}a_M^2/2$.

\section{Monte Carlo evaluation of the Laughlin-like state energy}
\label{Appendix:MonteCarlo}

Here we describe the nonperturbative Monte Carlo evaluation of the Laughlin-like-state energy at finite $B_1$, complementing the perturbative treatment of Appendix~\ref{Appendix:LaughlinPerturbation}. The Laughlin-state energies shown in Fig.~\ref{fig:PhaseDiagram}(a) and Fig.~\ref{fig:PhaseDiagram_17_19}(a),(b) are Coulomb expectation values of the state in Eq.~\eqref{eq:Laughlin_State} in the main text, evaluated by sampling the probability density
\begin{align}
    \bigl|\Psi_{\text{L}}^{\text{AC}}(\{{\bm r}_i\})\bigr|^2
    = e^{2\sum_i \chi({\bm r}_i)}\,
    \bigl|\Psi_{\text{L}}^{\text{LLL}}(\{{\bm r}_i\})\bigr|^2.
    \label{eq:MC_measure}
\end{align}
In the plasma analogy \cite{Laughlin1983Anomalous}, this density can be written as the Gibbs weight of a classical two-dimensional Coulomb plasma modified by the one-body K\"ahler factor $\chi({\bm r})$ \cite{Wolf2025Intraband}. We retain only the first reciprocal-lattice shell in $\chi({\bm r})$, with
$\chi_{1}=-B_1/(G_0\ell)^2$ and $(G_0\ell)^2=4\pi/\sqrt{3}$, where $B_1$ is measured in units of $B_0=\ell^{-2}$ throughout. The resulting factor $e^{2\chi({\bm r})}$ is included in the sampling density exactly, without expanding in $B_1$ or truncating the higher harmonics generated by exponentiation.

We evaluate the torus Laughlin wave function $\Psi_{\mathrm{L}}^{\mathrm{LLL}}$ using its Jacobi theta-function representation \cite{Haldane1985Periodic}. We parameterize particle positions by fractional coordinates ${\bm u}\in[0,1)^2$ on a hexagonal torus with modular parameter $\tau=e^{i\pi/3}$ and introduce the complex coordinate $z=u_1+\tau u_2$. The primitive-vector length $L$ is fixed by the flux condition $N_\phi=mN$ ($N$ is the particle number) through
\begin{align}
    L^2\,\mathrm{Im}\,\tau=2\pi N_\phi\ell^2,
    \qquad
    \frac{L}{\ell}
    =\left(\frac{2\pi N_\phi}{\mathrm{Im}\,\tau}\right)^{1/2}.
\end{align}
In these coordinates, the sampling density in Eq.~\eqref{eq:MC_measure} takes the form $e^{-\beta U}$, with
\begin{align}
    \beta U
    = 2m\sum_{i<j}G(z_i-z_j)-2\sum_i\chi({\bm r}_i),
    \qquad
    G(z|\tau)
    = -\ln\bigl|\theta_1(\pi z|\tau)\bigr|
    +\frac{\pi}{\mathrm{Im}\,\tau}
    \bigl(\mathrm{Im}\,z\bigr)^2,
    \label{eq:MC_plasma_weight}
\end{align}
where $\theta_1$ denotes the odd Jacobi theta function and $G$ is the periodic logarithmic Green's function of the two-dimensional plasma. Its quadratic term accounts for the uniform neutralizing background. Eq.~\eqref{eq:MC_plasma_weight} corresponds to the free-center-of-mass form of the torus Laughlin density; throughout, we sample this density rather than a sector-resolved density with an explicit center-of-mass theta factor.\\\\
The simulation torus contains $mN$ moir\'e unit cells, each enclosing one flux quantum. In Table~\ref{table:MonteCarlo} we summarize the different system sizes considered in our study. For each point $(\nu,B_1)$, we use Hamiltonian Monte Carlo with eight independent chains and continue sampling until the autocorrelation-corrected standard error of the energy reaches the target precision.
% , and we discard runs that fail a between/within-chain convergence test (rank-normalized split $\hat{R}$ of the log density above $1.05$ \cite{Vehtari2021RankNormalization}). 
We report the Coulomb energy per particle in units of $e^2/\ell$, evaluated by Ewald summation with a uniform neutralizing background, using the same convention as for the crystal energies. The one-body factor $e^{2\chi}$ modifies only the sampling measure and not the Coulomb interaction itself. \\\\
%The median statistical uncertainties are $(4.0,1.3,1.0)\times10^{-5}\,e^2/\varepsilon\ell$ at $\nu=(1/3,1/7,1/9)$, respectively.\\\\
%
\begin{table}[ht]
    \centering
    \begin{tabular}{|c|c|c|c|}
        \hline
        Filling fraction & Particle number & $E_{\infty}(B_1=0)~[e^2/\ell]$ & $\delta E(B_1=0)~[e^2/\ell]$ \\
        \hline
        $\nu=1/3$    & $N=75$     & $-0.40966(4)$ & $4.0\times10^{-5} $   \\
        \hline
        $\nu=1/7$    & $N=63$     & $-0.28095(1)$ & $1.3\times10^{-5} $    \\
        \hline
        $\nu=1/9$    & $N=49$     & $-0.24996(1)$  & $1.0\times10^{-5} $   \\
        \hline
    \end{tabular}
    \caption{Details on the plasma analogy Monte Carlo sampling. We report the energies and the median statistical uncertainties of the Laughlin-like state for $B_1=0$.}
    \label{table:MonteCarlo}
\end{table}
We control finite-size effects at $B_1=0$ and use this limit to benchmark our calculations. Assuming the leading size dependence to be
$E(N)/N=E_\infty+a/N$, we extrapolate data for $N=27$--$108$ at $\nu=1/3$, $N=28$--$63$ at $\nu=1/7$, and $N=16$--$49$ at $\nu=1/9$. This gives the extrapolated energies reported in Table~\ref{table:MonteCarlo}. At $\nu=1/7$ and $1/9$, these values agree to within $1\times10^{-4}\,e^2/\ell$ with the lowest-Landau-level energies obtained from the interpolation formula of Ref.~\cite{Levesque1984Crystallization}. At $\nu=1/3$, published Laughlin-state energies range from $-0.4094$ to $-0.4100(1)\,e^2/\ell$, depending on geometry and extrapolation procedure \cite{Levesque1984Crystallization,Ciftja2003MonteCarlo}. Our result lies within this range.\\\\
The energy response to the modulation is substantially less sensitive to system size. At $\nu=1/3$ and $B_1=-1.6$, for example, $E(B_1)-E(0)$ ranges from $1.46(6)$ to $1.62(6)\times10^{-3}\,e^2/\ell$ over $N=27$--$108$. We therefore correct the production energies by the constant finite-size offset $E_\infty-E_N(0)$, whose magnitude is at most $5\times10^{-5}\,e^2/\ell$.

\section{Multipole expansion of the direct interaction}
\label{Appendix:Multi_pole}
The direct interaction between two localized orbitals $\psi_{{\bm R}_i}({\bm r})$ and $\psi_{{\bm R}_j}({\bm r})$ is given by
\begin{align}
    D_{{\bm R_i},{\bm R_j}}=e^2\int \frac{d^2 r_1d^2 r_2}{|{\bm r}_1-{\bm r}_2|}|\psi_{{\bm R}_i}({\bm r}_1)|^2|\psi_{{\bm R}_j}({\bm r}_2)|^2.
\end{align}
%
%We will assume that the orbitals are either $C_3$- or $C_6$-symmetric, motivated by the symmetry of the optimal AC band orbitals. 
Let ${\bm X}_i=\int d^2 r\, {\bm r}\,|\psi_{{\bm R}_i}({\bm r})|^2 $ be the orbital's center of mass and define ${\bm R}={\bm X}_i-{\bm X}_j$. By introducing the coordinates relative to the centers of mass: ${\bm u_1}={\bm r}_1-{\bm X}_i$ and ${\bm u_2}={\bm r}_2-{\bm X}_j$, the direct term reads
\begin{align}
    D_{{\bm R_i},{\bm R_j}}=e^2\int \frac{d^2 u_1\,d^2 u_2}{|{\bm u}_1-{\bm u}_2+{\bm R}|}\,|\psi_{{\bm R}_i}({\bm u_1+{\bm X}_i})|^2\,|\psi_{{\bm R}_j}({\bm u_2}+{\bm X}_j)|^2.
    \label{eq:Appendix_Dij}
\end{align}
Noting that the orbital centers $\bm X_i$ are defined so that the dipole moment $\left<\mathbf{r}-\mathbf{X}_i\right>=0$ vanishes, the standard multipole expansion gives,  to order $1/R^5$:
\begin{equation}
 D_{{\bm R_i},{\bm R_j}}=e^2 \left[ \frac{1}{R} + \frac{\left( 3n_\alpha n_\beta - \delta_{\alpha\beta}\right) Q_{\alpha\beta} }{R^3}  + \frac{\left( 35 n_\alpha n_\beta n_\gamma n_\delta - 5\text{Sym}\left[ \delta_{\alpha\beta} n_\gamma n_\delta\right] + \text{Sym}\left[ \delta_{\alpha\beta} \delta_{\gamma\delta} \right]\right)\left( H_{\alpha\beta\gamma\delta} + 3Q_{\alpha\beta}Q_{\gamma\delta} \right)  }{8R^5} \right],
\end{equation}
where $\hat{\mathbf{n}} = \mathbf{R} /R$, and the moments are defined by
\begin{align}
    Q_{\alpha\beta} &= \int d^2u \,u_\alpha u_\beta |\psi_{\mathbf{R}}(\mathbf{X}+\mathbf{u} ) |^2, \qquad \text{and} \qquad
    H_{\alpha\beta\gamma\delta} = \int d^2 u\, u_{\alpha} u_{\beta} u_{\gamma} u_{\delta} |\psi_{\mathbf{R}} \left( \mathbf{X} + \mathbf{u} \right) |^2.
\end{align}
$\text{Sym}$ indicates the term should be symmetrized over all indices, specifically, 
\begin{align}
\text{Sym}\left[ \delta_{\alpha\beta} n_\gamma n_\delta\right] &=  \delta_{\alpha\beta} n_\gamma n_\delta + 
\delta_{\alpha\gamma} n_\beta n_\delta + 
\delta_{\alpha\delta} n_\gamma n_\beta + 
\delta_{\beta\gamma} n_\alpha n_\delta + 
\delta_{\beta\delta} n_\gamma n_\alpha + 
\delta_{\gamma\delta} n_\alpha n_\beta,\\
\text{Sym}\left[\delta_{\alpha\beta} \delta_{\gamma\delta} \right]&= \delta_{\alpha\beta} \delta_{\gamma\delta} + \delta_{\alpha\gamma} \delta_{\beta\delta} + \delta_{\alpha\delta} \delta_{\beta\delta}, 
\end{align}
and the Greek subscripts run over $x$ and $y$.
Making the assumption (valid for three-fold symmetric orbitals) that $Q_{\alpha\beta} = Q\delta_{\alpha\beta} $ and $H_{\alpha\beta\gamma\delta}= \frac{H}{3} \text{Sym} \left[ \delta_{\alpha\beta} \delta_{\gamma\delta}\right]$,
we find
\begin{equation}
    D_{{\bm R_i},{\bm R_j}} = e^2 \left[ \frac{1}{R} + \frac{Q}{R^3} + \frac{ 3 H + 9Q^2}{4R^5}\right].
\end{equation}
We now define the first and second cumulants:
\begin{align}
    \langle r^2\rangle_c &= \int d^2 u |\mathbf{u}|^2 | \psi_{\mathbf{R} }(\mathbf{X} + \mathbf{u} ) | ^2 = 2Q, \\
    \langle r^4 \rangle_c &= \int d^2 u |\mathbf{u}|^4 |\psi_\mathbf{R} (\mathbf{X} + \mathbf{u} )|^2 = 8H/3,
\end{align}
in terms of which the direct interaction is
\begin{equation}
    D_{{\bm R_i},{\bm R_j}} = e^2 \left[ \frac{1}{R} + \frac{\langle r^2\rangle_c}{2R^3} + \frac{ 9 \langle r^4 \rangle_c + 18\langle r^2\rangle_c^2 }{32R^5}\right].
    \label{eq:Appendix_GaussianEnergy}
\end{equation}
Assuming ``Gaussian-like'' orbitals where $\braket{r^4}_c=2\,\braket{r^2}_c^2$, this finally gives 
\begin{equation}
    D_{{\bm R_i},{\bm R_j}} = e^2 \left[ \frac{1}{R} + \frac{\langle r^2\rangle_c}{2R^3} + \frac{ 9\langle r^2\rangle_c^2 }{8R^5}\right],
    \label{eq:Appendix_DirectFinal}
\end{equation}
yielding the expression for the energy of the crystalline state provided in the main text.\\\\
We find that the optimal orbitals obtained from diagonalizing Eq.~\eqref{eq:Effective_H} in the main text are localized at the high-symmetry points within the moiré unit cell. In Fig.~\ref{fig:Analytic_Spread}(a) we show the behavior of the optimal spread with respect to ${\bm \delta}$, for $B_1=\pm2$. We see that for $B_1<0$ the optimal orbital is at $\bm \delta_{\text{AA}}$ and that for $B_1>0$ there are two identical optimal orbitals at $\bm \delta_{\text{AB/BA}}$.

\section{An analytic example of localized orbitals in Aharonov-Casher bands}
\label{Appendix:Gaussian_Orbital}

\begin{figure}
    \centering
    \includegraphics[width=0.95\textwidth]{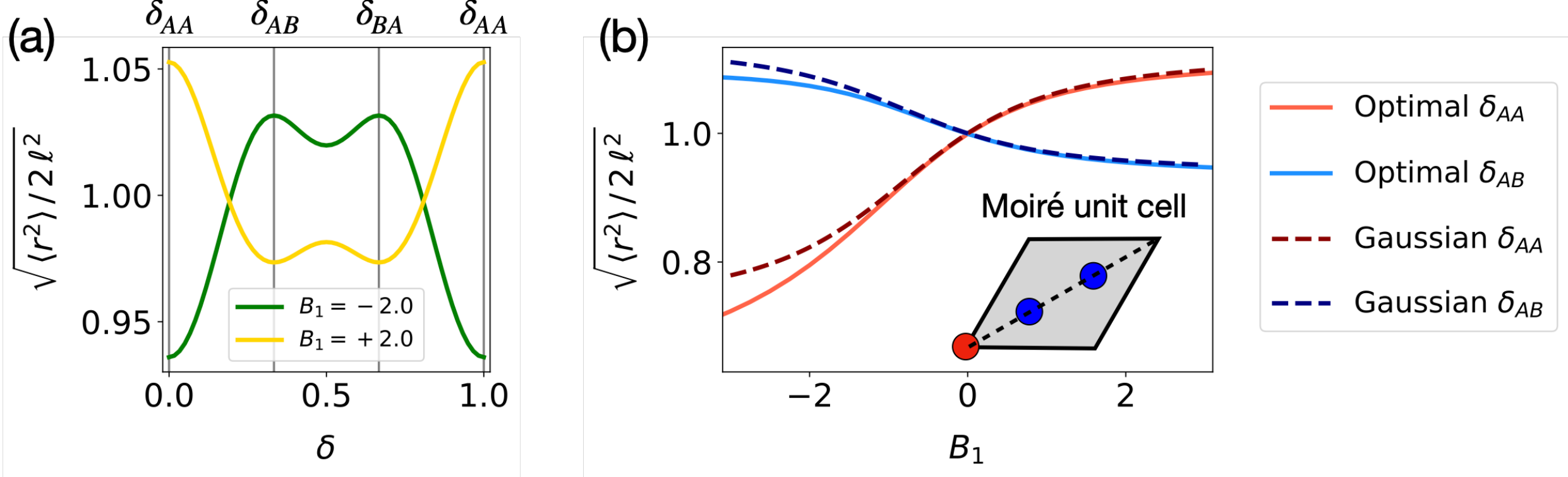}
    
    \caption{(a) Spread of the optimal orbitals, obtained from diagonalizing Eq.~\eqref{eq:Effective_H} in the main text, along a line-cut across the moiré unit cell, for two values of $B_1$. The minimal spread is reached at $\bm \delta_{\text{AA}}$ for $B_1<0$ and at $\bm \delta_{\text{AB}}/\bm \delta_{\text{BA}}$ for $B_1>0$. (b) Comparison between the spread of the analytic orbitals given by Eq.~\eqref{eq:Analytic_Orbital} (dashed lines) and the optimal orbitals used in the main text (solid lines), as a function of $B_1$. The analytic orbitals reproduce the spread of the optimal orbitals very well for $B_1\ge 1$. Below this threshold, the optimal orbitals become more localized by acquiring higher angular momentum terms consistent with the $C_6$-symmetry of the ${\bm \delta}_{\text{AA}}$ centers. The inset indicates the high-symmetry points and the line-cut across the moiré unit cell used for (a).}
    \label{fig:Analytic_Spread}
\end{figure}

Here we use the analytic structure of the AC band to obtain localized orbitals that can be used to construct the many-body wave function ansatz presented in Eq.~\eqref{eq:MZ_Ansatz} in the main text. The analytical orbitals reproduce the main features of the optimal orbitals that we derived by minimizing the spread in the main text. We start from a Gaussian orbital in the LLL
\begin{equation}
    \psi^\text{LLL}_{\bm \delta}(\bm r) = \frac{1}{\sqrt{2\pi \ell^2} }
    e^{-\frac{|{\bm r}-{\bm \delta}|^2}{4\ell^2}+i\frac{({\bm \delta}\times {\bm r})_z}{2\ell^2}}.
    \label{eq:Appendix_GaussianOrbital}
\end{equation}
In a general AC band, the LLL wave function is multiplied by $\exp[\chi({\bm r})]$, along with a normalization factor, yielding
\begin{align}
    \psi^{\text{AC}}_{\bm \delta}({\bm r})
    =\psi^{\text{LLL}}_{\bm \delta}({\bm r})\, \frac{e^{\chi({\bm r})}}{\mathcal{N}_{\bm \delta}}=\left(\frac{1}{2\pi \ell^2\mathcal{N}_{{\bm \delta}}^2}\right)^{1/2}\, e^{-\frac{|{\bm r}-{\bm \delta}|^2}{4\ell^2}+i\frac{({\bm \delta}\times {\bm r})_z}{2\ell^2}+\chi({\bm r})}.
    \label{eq:Analytic_Orbital}
\end{align}
which is Eq.~\eqref{eq:AC_GaussianOrbital} in the main text. The normalization is given by
\begin{align}
    \mathcal{N}^2_{{\bm \delta}}
    =\int d^2 r\,e^{2\chi({\bm r})}|\psi^{\text{LLL}}_{{\bm \delta}}({\bm r})|^2=\sum_{\bm G}\omega_{\bm G}\, e^{i{\bm G}\cdot {\bm \delta}}\,e^{-\frac{ G^2\ell^2}{2}}, 
    \label{eq:AC_GaussianNormalization}
\end{align}
where $G=|{\bm G}|$, and in the second line we have introduced the Fourier expansion 
\begin{align}
    e^{2\chi({\bm r})}=\sum_{{\bm G}}\omega_{\bm G}e^{i {\bm G}\cdot {\bm r}}.
    \label{eq:Khaler_Fourier}
\end{align}
The value of $B_1$ determines the Fourier coefficients $\omega_{\bm G}$. The spread of the orbital defined in Eq.~\eqref{eq:Analytic_Orbital} is given by
\begin{align}
    \braket{(\bm r-\bm \delta)^2}=\int d^2r\,({\bm r}-{\bm \delta})^2 |\psi^{\text{AC}}_{\bm \delta}({\bm r})|^2= 2\ell^2-\ell^4\,\frac{\sum_{\bm G}\omega_{\bm G}\,G^2\, e^{i{\bm G}\cdot {\bm \delta}}\,e^{-\frac{{G}^2\ell^2}{2}}}{\sum_{\bm G}\omega_{\bm G}\, e^{i{\bm G}\cdot {\bm \delta}}\,e^{-\frac{{G}^2\ell^2}{2}}}.
    \label{eq:r2}
\end{align}
For a spatially uniform magnetic field, $B_1=0$, Eq.~\eqref{eq:r2} reproduces the familiar result for a coherent state in the LLL, $\langle r^2 \rangle_c = 2\ell^2$ \cite{maki1983static}. We see that for $B_1\neq 0$ there is a correction, which can be positive or negative, depending on the sign of $B_1$ and on $\bm \delta$. \\\\
In Fig.~\ref{fig:Analytic_Spread}(b) we compare the spread of the analytic orbital given by Eq.~\eqref{eq:Analytic_Orbital} with the spread of the optimal orbitals, which are obtained by diagonalizing the effective Hamiltonian given by Eq.~\eqref{eq:Effective_H} in the main text. The two results for the spread agree well, particularly for $B_1>-1$, supporting the validity of the Gaussian approximation used to derive Eq.~\eqref{eq:Appendix_DirectFinal}. Below $B_1<-1$ the spreads of the analytic and optimal orbitals start to deviate. The reason is that the optimal orbital centered at $\bm \delta_\text{AA}$ has non-vanishing higher angular momentum contributions $m=6, 12, \dots$ that are not captured by the simple analytic expression given in Eq.~\eqref{eq:Appendix_GaussianOrbital}.\\\\
We now compute the energy of the generalized Wigner crystal obtained from using the analytic orbitals given by Eq.~\eqref{eq:Analytic_Orbital} to define the many-body wave function Eq.~\eqref{eq:MZ_Ansatz} in the main text. 
This energy is given by
\begin{align}
    E^{\text{AC}}_{\text{C}}=\frac{\braket{\Psi_{\text{C}}^{\text{AC}}|\frac{1}{2}\sum_{i\neq j}\frac{e^2}{|{\bm r_i-\bm r_j}|}|\Psi_{\text{C}}^{\text{AC}}}}{\braket{\Psi_{\text{C}}^{\text{AC}}|\Psi_{\text{C}}^{\text{AC}}}},
    \label{eq:Appendix_CrystalEnergy}
\end{align}
where we have used the unscreened Coulomb interaction. The numerator of Eq.~\eqref{eq:Appendix_CrystalEnergy} is approximated as \cite{maki1983static}
\begin{align}
    \braket{\Psi_{\text{C}}^{\text{AC}}|\frac{1}{2}\sum_{i\neq j}\frac{1}{|{\bm r_i-\bm r_j}|}|\Psi_{\text{C}}^{\text{AC}}}\approx \sum_{i<j}\int \frac{d{\bm r}_1d{\bm r}_2}{|{\bm r}_1-{\bm r}_2|}\left(|\psi_{{\bm R}_i}({\bm r}_1)|^2|\psi_{{\bm R}_j}({\bm r}_2)|^2 -\psi^*_{{\bm R}_i}({\bm r}_1)\psi^*_{{\bm R}_j}({\bm r}_2)\psi_{{\bm R}_j}({\bm r}_1)\psi_{{\bm R}_i}({\bm r}_2)\right)+\cdots
    \label{eq:Appendix_Numerator}
\end{align}
The two terms in Eq.~\eqref{eq:Appendix_Numerator} correspond to direct and exchange contributions from two orbitals localized at sites ${\bm R}_i$ and ${\bm R}_j$, respectively, and we have neglected terms involving three or more crystal sites. Within the same approximation, the denominator of Eq.~\eqref{eq:Appendix_CrystalEnergy} is  
\begin{align}
    \braket{\Psi_{\text{C}}^{\text{AC}}|\Psi_{\text{C}}^{\text{AC}}}&\approx 1-\frac{1}{2}\sum_{i\neq j} \int d^2 r_1 d^2 r_2\,\psi^*_{\bm R_i}({\bm r}_1)\,\psi^*_{\bm R_j}({\bm r}_2)\,\psi_{\bm R_i}({\bm r}_2)\psi_{\bm R_j}({\bm r}_1)+\cdots =1-\frac{1}{2}\sum_{i\neq j}S({\bm R}_i,{\bm R}_j)+\cdots.
    \label{eq:Appendix_Denominator}
\end{align}
We now evaluate the direct and exchange terms in Eq.~\eqref{eq:Appendix_Numerator} and the overlap in Eq.~\eqref{eq:Appendix_Denominator} explicitly. Given two crystal sites ${\bm R}_i$ and ${\bm R}_j$, we define
\begin{align}
    {\bm R}\equiv{\bm R}_i-{\bm R}_j, \qquad\text{and}\qquad {\bm S}\equiv\frac{{\bm R}_i+{\bm R}_j}{2}.
\end{align}
\underline{{\it Direct term}}: 
By introducing 
\begin{align}
    R_{{\bm G}{\bm G'}}^2=-R^2+\ell^4|{\bm G}+{\bm G'}|^2-2i\ell^2{\bm R}\cdot ({\bm G}+{\bm G'}),
\end{align}
where ${\bm G}$ and ${\bm G}'$ are reciprocal lattice vectors, the direct term can be written as
\begin{align}
    D_{{\bm R}_i,{\bm R}_j}%&=\sum_{{\bm G},{\bm G'}}\,\frac{\omega_{\bm G}\,\omega_{\bm G'}^*}{\mathcal{N}_{\bm \delta}^4}\left[\int_0^{\infty}dq\, I_0(qR_{{\bm G}{\bm G'}})e^{-\ell^2q^2}\right]e^{-\frac{\ell^2}{2}(G^2+G'^2)}e^{i({\bm G}\cdot{\bm R}_i-{\bm G'}\cdot{\bm R}_j)}\nonumber\\
    &=\sum_{{\bm G},{\bm G'}}\,\frac{\omega_{\bm G}\,\omega_{\bm G'}^*}{\mathcal{N}_{\bm \delta}^4}\left[ \frac{\sqrt{\pi}}{2\ell} \exp\left(\frac{R_{{\bm G}{\bm G'}}^2}{8\ell^2}\right)I_0\left(\frac{R_{{\bm G}{\bm G'}}^2}{8\ell^2} \right)\right]e^{-\frac{\ell^2}{2}(G^2+G'^2)}\,e^{i ({\bm G}-{\bm G'})\cdot {\bm \delta}},
    \label{eq:Appendix_Direct}
\end{align}
where $I_0$ is a modified Bessel function and we have used that the crystal site centers can be decomposed as ${\bm R}_{i}={\bm R^0_{i}}+{\bm \delta}$, where ${\bm R^0_{i}}$ is a moiré lattice vector.\\\\
\underline{{\it Exchange term}}: Similarly, we can introduce 
\begin{align}
    \tilde{R}^2_{{\bm G}, {\bm G'}}=R^2+\ell^4|{\bm G}+{\bm G}'|^2-2\ell^2\left[ {\bm R}\times ({\bm G}+{\bm G}')\right]_z,
\end{align}
and the exchange term in Eq.~\eqref{eq:Appendix_Numerator} can be written as
\begin{align}
    X_{{\bm R}_i,{\bm R}_j}%=&\frac{e^{-\frac{R^2}{2\ell^2}}}{\mathcal{N}^4_{\bm \delta}}\sum_{\bm G, \bm G'} \omega_{\bm G}\omega_{\bm G'}^*\left[\int_0^{\infty} dq\,I_0\left( q \tilde{R}_{{\bm G}, {\bm G'}}\right) e^{-\ell^2 q^2}\right]\,e^{-\frac{(G^2+G'^2)\ell^2}{2}} e^{-\frac{({\bm G}+{\bm G'})\times {\bm R}}{2}}\,e^{-i({\bm G}-{\bm G'})\cdot{\bm S}}\nonumber\\
    =&\frac{e^{-\frac{R^2}{2\ell^2}}}{\mathcal{N}^4_{\bm \delta}}\sum_{\bm G, \bm G'} \omega_{\bm G}\omega_{\bm G'}^*\left[ \frac{\sqrt{\pi}}{2\ell}\exp \left(\frac{\tilde{R}^2_{{\bm G}, {\bm G'}}}{8\ell^2}\right)I_0\left(\frac{\tilde{R}^2_{{\bm G}, {\bm G'}}}{8\ell^2} \right)\right]\,e^{-\frac{\ell^2}{2}(G^2+G'^2)} e^{\frac{[{\bm R}\times ({\bm G}+{\bm G'})]_z}{2}}\,e^{-i({\bm G}-{\bm G'})\cdot{\bm S}}.
    \label{eq:Appendix_Exchange}
\end{align}
\begin{figure}
    \centering
    \includegraphics[width=0.9\textwidth]{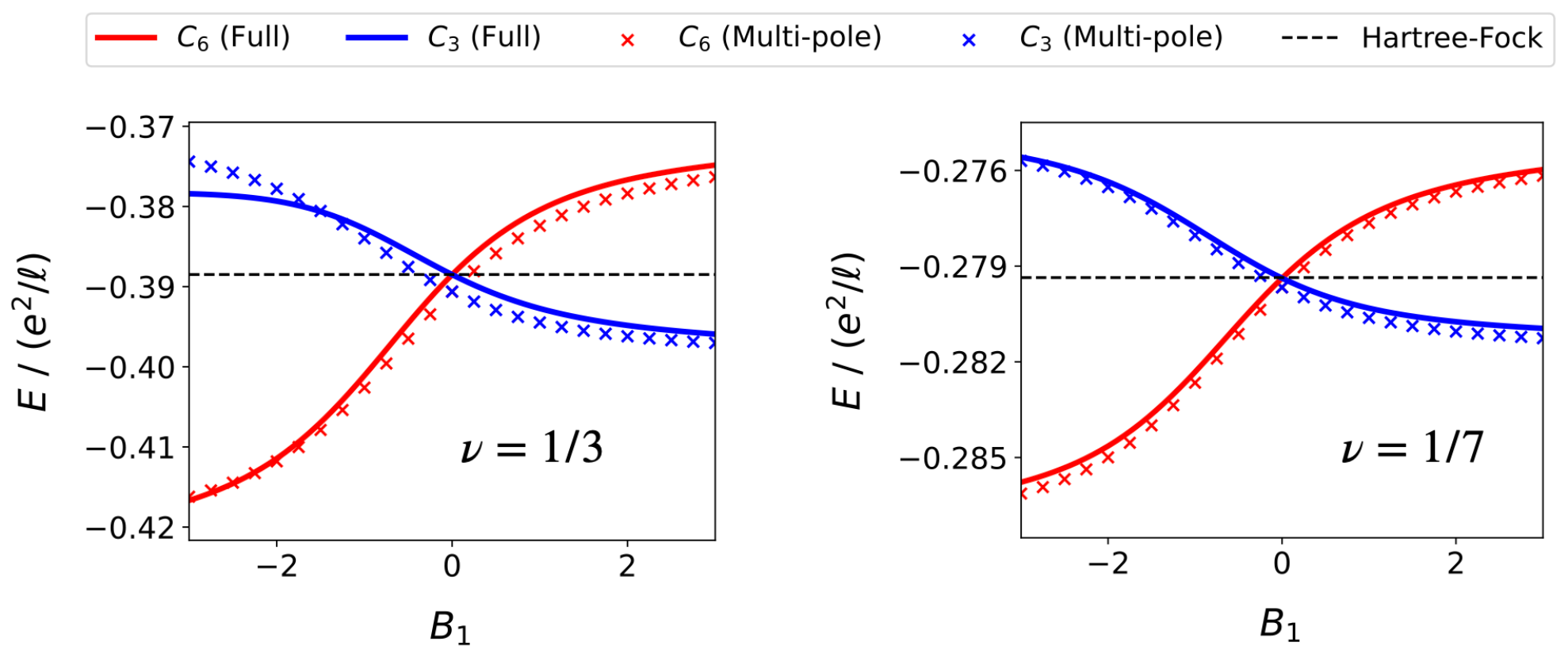}
    
    \caption{Energy of generalized Wigner crystal states on AC bands constructed from the orbitals defined in Eq.~\eqref{eq:Analytic_Orbital} at fillings (a) $\nu=1/3$ and (b) $\nu=1/7$. Results for the $C_6$-crystal are shown in red and results for the $C_3$-crystal are shown in blue. Solid lines correspond to the full crystal energy, Eq.~\eqref{eq:Appendix_Energy_approx}, while the crosses correspond to the multipole expansion Eq.~\eqref{eq:Appendix_Multipole}. The black dashed lines are the Hartree-Fock energies for $B_1=0$.}
    \label{fig:Crystal_Energies}
\end{figure}
{\underline{\it Overlap integral}}: The overlap in Eq.~\eqref{eq:Appendix_Numerator} can be expressed as the square of a matrix element between two Gaussian orbitals in the LLL
\begin{align}
    S_{{\bm R}_i,{\bm R}_j}=\frac{1}{\mathcal{N}^4_{\bm \delta}}|\braket{\psi_{{\bm R}_i}^{\text{LLL}}|e^{2\chi({\bm r})}|\psi_{{\bm R}_j}^{\text{LLL}}}|^2,
    \label{eq:Appendix_S}
\end{align}
which is explicitly evaluated to
\begin{align}
    S_{{\bm R}_i,{\bm R}_j}=\frac{e^{-\frac{R^2}{2\ell^2}}\,}{\mathcal{N}^4_{\bm \delta}}\bigg\rvert \sum_{\bm G}\,\omega_{\bm G}\,e^{i{\bm G}\cdot {\bm S}}\, e^{\frac{-G^2\ell^2}{2}}\, e^{-\frac{({\bm R}\times{\bf G})_z}{2}}\bigg\rvert^2.
    \label{eq:Appendix_Overlap}
\end{align}
Putting Eqs.~\eqref{eq:Appendix_Direct}, \eqref{eq:Appendix_Exchange} and \eqref{eq:Appendix_Overlap} together, the energy of the crystal state is 
\begin{align}
    E_{\text{C}}^{\text{AC}}&=\frac{e^2}{2}\left[ \sum_{i\neq j} D_{{\bm R}_i,{\bm R}_j}-X_{{\bm R}_i,{\bm R}_j}\right]\left[1-\frac{1}{2}\sum_{k\neq l}S_{{\bm R}_k,{\bm R}_l} \right]^{-1}.
    \label{eq:Appendix_Energy_approx}
\end{align}
When $B_1=0$, then $\omega_{\bm G\neq0}=0$, and we recover the expression for the case of a uniform magnetic field \cite{maki1983static}. Note that the exchange and overlap terms include an exponential factor $\exp(-R^2/2\ell^2)$, which will suppress them when the distance between orbitals is much larger than their spread. Therefore, for dilute densities we can approximate the crystal energy by only the direct contribution
\begin{align}
    E_{\text{C}}^{\text{AC}}\approx~\frac{e^2}{2}\sum_{i\neq j} D_{{\bm R}_i,{\bm R}_j}.
\end{align}
Furthermore, the direct term can be subjected to a multipole expansion, as detailed in Appendix~\ref{Appendix:Multi_pole}. Because the analytic orbitals are Gaussian and we choose them to be centered at high-symmetry points with either $C_3$- or $C_6$-symmetry, the energy of the crystal is given by 

\begin{align}
    \frac{E_{\text{C}}^{\text{AC}}}{N}\approx \frac{e^2}{2\ell}\sum_{i}\left[\frac{1}{R_{i}}+\frac{\braket{r^2}_c}{2R_{i}^3}+\frac{9\, \braket{r^2}_c^2}{8R_{i}^5}+\cdots \right],
    \label{eq:Appendix_Multipole}
\end{align}
which is Eq.~\eqref{eq:AC_Crystal_Energy} in the main text. In Fig.~\ref{fig:Crystal_Energies} we compare the energy of the crystal states with analytic orbitals, computed from Eq.~\eqref{eq:Appendix_Energy_approx} and from Eq.~\eqref{eq:Appendix_Multipole}, for filling fractions $\nu=1/3$ and $\nu=1/7$. The good agreement between the two calculations supports the validity of the multipole approximation employed in the main text.

\section{Phase diagram of the AC band for other filling fractions}
\label{Appendix_Other_fillings}

\begin{figure}
    \centering
    \includegraphics[width=0.8\textwidth]{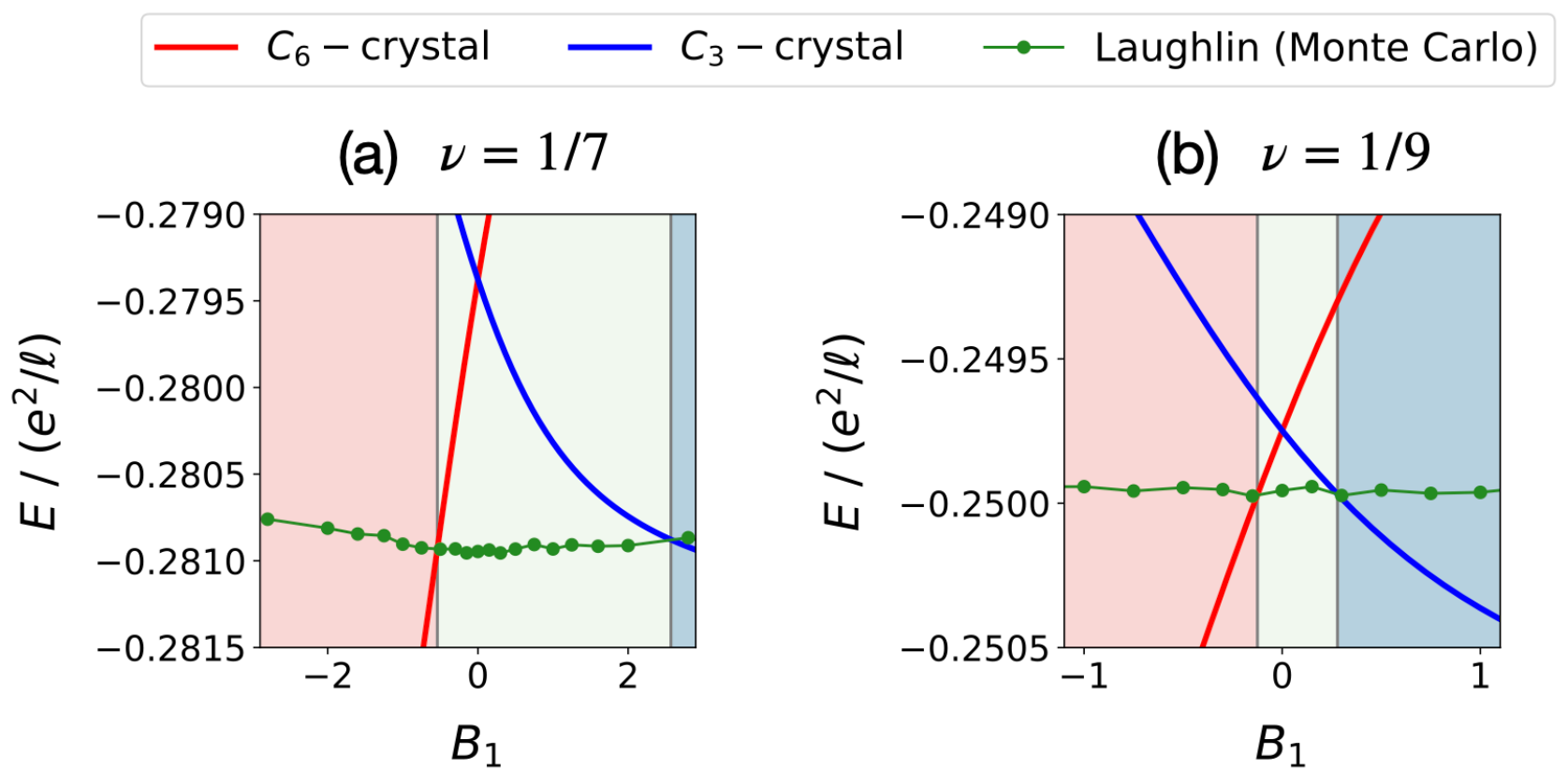}
    
    \caption{Energies per particle of the $C_6$-crystal, the $C_3$-crystal, and the Laughlin liquid at filling fraction (a) $\nu=1/7$ and (b) $\nu=1/9$ of an AC band, as a function of $B_1$.}
    \label{fig:PhaseDiagram_17_19}
\end{figure}

\begin{figure}
    \centering
    \includegraphics[width=0.75\textwidth]{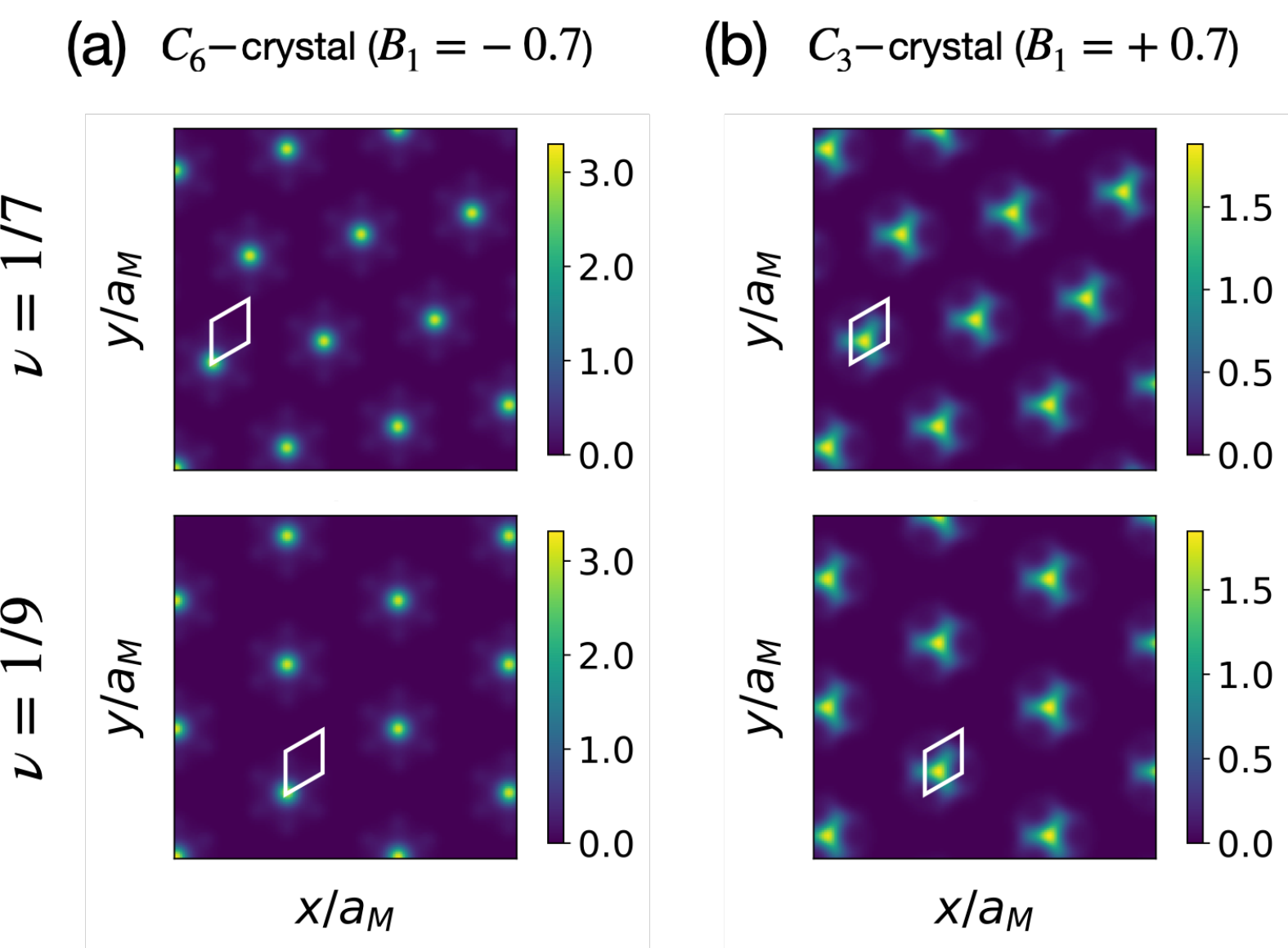}
    
    \caption{Charge densities of the $C_6$- and $C_3$-crystals on AC bands with (a) $B_1=-0.7$ and (b) $B_1=+0.7$, at fillings $\nu=1/7$ (top row) and $\nu=1/9$ (bottom row).}
    \label{fig:Densities_17_19}
\end{figure}

In Fig.~\ref{fig:PhaseDiagram_17_19}(a),(b) we present the energies of the $C_3$- and $C_6$-crystals, together with the energy of the Laughlin-like state as a function of $B_1$, for filling fractions $\nu=1/7$ and $\nu=1/9$, respectively. We use Eq.~\eqref{eq:AC_Crystal_Energy} in the main text to compute the crystal energies, while the Laughlin energy is computed via Monte Carlo. In contrast to the case of $\nu=1/3$, presented in the main text, the Laughlin state is unstable to both crystalline competing phases if the magnitude of the harmonic $|B_1|$ is large enough. The values of $B_1$ where the crystal energies cross the Laughlin energy are used to determine the phase boundaries of the phase diagram presented in Fig.~\ref{fig:PhaseDiagram}(c) in the main text. Finally, in Fig.~\ref{fig:Densities_17_19}(a),(b) we show typical charge densities of the $C_6$- and $C_3$-crystals at $\nu=1/7$ and $\nu=1/9$, respectively.

\end{document}